\documentclass[aps,prb,preprint,showpacs,superscriptaddress,floatfix]{revtex4-2}

\usepackage{graphicx}
\usepackage{xcolor}
\usepackage{amsmath,amssymb,amsfonts}
\usepackage {multirow}
\usepackage{lineno}
\usepackage[normalem]{ulem}

\begin{document}

\title{Tracing the horizon of tetragonal-to-monoclinic distortion in pressurized trilayer nickelate La$_4$Ni$_3$O$_{10}$}

\author{Sitaram Ramakrishnan}
\email{Corresponding author email: sitaram.ramakrishnan@neel.cnrs.fr}
\thanks{Equal contribution}
\affiliation{Institut N$\acute{e}$el CNRS/UGA UPR2940,
25 Rue des Martyrs, 38042 Grenoble, France}

\author{Yingzheng Gao}
\thanks{Equal contribution}
\affiliation{Institut N$\acute{e}$el CNRS/UGA UPR2940,
25 Rue des Martyrs, 38042 Grenoble, France}

\author{Valerio Olevano}
\affiliation{Institut N$\acute{e}$el CNRS/UGA UPR2940,
25 Rue des Martyrs, 38042 Grenoble, France}

\author{Elise Pachoud}
\affiliation{Institut N$\acute{e}$el CNRS/UGA UPR2940,
25 Rue des Martyrs, 38042 Grenoble, France}

\author{Abdellali Hadj-Azzem}
\affiliation{Institut N$\acute{e}$el CNRS/UGA UPR2940,
25 Rue des Martyrs, 38042 Grenoble, France}

\author{Gaston Garbarino}
\affiliation{ESRF -- European Synchrotron Radiation Facility, 38000 Cedex Grenoble, France}

\author{Matteo d’Astuto}
\affiliation{Institut N$\acute{e}$el CNRS/UGA UPR2940,
25 Rue des Martyrs, 38042 Grenoble, France}

\author{Olivier Perez}
\affiliation{Laboratory CRISMAT, UMR 6508 CNRS, ENSICAEN 6 Boulevard du Marechal Juin, F-14050 Caen Cedex 4, France}

\author{Alain Pautrat}
\affiliation{Laboratory CRISMAT, UMR 6508 CNRS, ENSICAEN 6 Boulevard du Marechal Juin, F-14050 Caen Cedex 4, France}

\author{Diego Valenti}
\affiliation{Institut N$\acute{e}$el CNRS/UGA UPR2940,
25 Rue des Martyrs, 38042 Grenoble, France}

\author{Matthieu Qu\'{e}not}
\affiliation{Institut N$\acute{e}$el CNRS/UGA UPR2940,
25 Rue des Martyrs, 38042 Grenoble, France}

\author{S\'{e}bastien Pairis}
\affiliation{Institut N$\acute{e}$el CNRS/UGA UPR2940,
25 Rue des Martyrs, 38042 Grenoble, France}

\author{Dmitry Chernyshov}
\affiliation{Swiss-Norwegian Beamlines, ESRF -- European Synchrotron Radiation Facility, 38000 Cedex Grenoble, France}

\author{Leila Noohinejad}
\affiliation{Deutsches Elektronen-Synchrotron DESY, Notkestrasse 85, 22607 Hamburg, Germany}

\author{Carsten Paulmann}
\affiliation{Deutsches Elektronen-Synchrotron DESY, Notkestrasse 85, 22607 Hamburg, Germany}

\author{Johnathan Bulled}
\affiliation{ESRF -- European Synchrotron Radiation Facility, 38000 Cedex Grenoble, France}

\author{Alexei Bosak}
\affiliation{ESRF -- European Synchrotron Radiation Facility, 38000 Cedex Grenoble, France}

\author{Sander van Smaalen}
\affiliation{Laboratory of Crystallography, Bayerisches Geoinstitut, University of Bayreuth, 95440 Bayreuth, Germany}

\author{Pierre Toulemonde}
\affiliation{Institut N$\acute{e}$el CNRS/UGA UPR2940,
25 Rue des Martyrs, 38042 Grenoble, France}

\author{Marie-Aude M\'{e}asson}
\affiliation{Institut N$\acute{e}$el CNRS/UGA UPR2940,
25 Rue des Martyrs, 38042 Grenoble, France}

\author{Pierre Rodi\`ere}
\email{Corresponding author email: pierre.rodiere@neel.cnrs.fr}
\affiliation{Institut N$\acute{e}$el CNRS/UGA UPR2940,
25 Rue des Martyrs, 38042 Grenoble, France}

\date{\today}

\begin{abstract}
The crux of understanding the superconducting mechanism in pressurized Ruddlesden–Popper nickelates hinges on elucidating their structural phases. Under ambient conditions, the trilayer nickelate La$_4$Ni$_3$O$_{10}$ stabilizes in a twinned monoclinic structure with space group $P2_1/c$. Upon heating, it undergoes a structural transition to the tetragonal $I$4/$mmm$ phase at $T_s$ $\approx$ 1030 K, while a second transition associated with the onset of density-wave (DW) ordering emerges upon cooling below $T_{DW}$ $\approx$ 135 K. Here, from pressure-temperature x-ray diffraction (XRD) on high quality flux-grown single crystals we demonstrate a direct tetragonal-to-monoclinic transition without an intermediate orthorhombic $Bmab$ phase. \textit{Ab initio} density-functional theory calculations as a function of pressure corroborate the experimental observations.
The tetragonal-to-monoclinic transition unfolds as the formation of a two-fold superstructure, as evidenced by  the emergence of commensurate superlattice reflections and can be progressively suppressed from 1030 K down to 20 K under 14 GPa. Notably, from XRD we establish the observation of weak incommensurate satellite reflections associated with the DW ordering in flux-grown samples, as previous findings were confined only to crystals grown by the floating-zone technique. This is further reinforced by Raman spectroscopy that reveal the emergence of additional phonon modes below 130 K, concomitant with the onset of the incommensurate DW state.
\end{abstract}

\maketitle

Groundbreaking scientific discoveries in Cu-based superconductors, where the superconductivity (SC) is believed to arise from the partially filled Cu 3$d$ electron
states in the CuO$_2$ planes have fundamentally shaped our understanding of high temperature SC over the years \cite{Chu1987a, Takagi1988a, Keimer2015a}. Nevertheless, the microscopic mechanisms underlying their unconventional superconductivity remain an enigma. In this context, the search for analogous materials continues to be of paramount importance.  Rare-earth nickelates have emerged as a fertile platform for exploring correlated electron phenomena. Like the cuprates, they host a rich interplay of quantum phases including charge and spin density waves, antiferromagnetism, and superconductivity offering new avenues to probe the intertwined nature of these competing orders \cite{tranquada1994a, Kivelson1998a, Zhang2016d, wang2024f, wang2025h}.

Experimental evidence of SC in the nickelates was first realized on thin films of infinite-layer Nd$_{1-x}$Sr$_x$NiO$_2$ \cite{Li2019a, Li2020a, Zeng2020m}. Interestingly, this structure, composed of NiO$_2$ planes stacked along the $c$ axis, hosts Ni$^{1+}$ ions with a 3$d^9$ electronic configuration, formally isoelectronic to Cu$^{2+}$ in the cuprates. However, superconductivity has not been observed in bulk crystals \cite{Li2020jk, Suyolcu2025}, suggesting a notable influence of the substrate. This breakthrough was shortly followed by the discovery of high temperature SC at $T_c$ = 80 K under hydrostatic pressure in bulk Ruddlesden-Popper bilayer crystal of La$_3$Ni$_2$O$_7$, where the average electronic structure of the Ni is Ni$^{2.5+}$ (d$^{7.5}$) far from the analogs to the cuprates. Interestingly, the SC is accompanied by a change in the symmetry of the crystal structure \cite{Sun2023a, hou2023a, Zhang2024b, Wang2025zx, wang2024a, wang2024au}.

Recently, SC up to 30 K was also discovered both in bulk Ruddlesden-Popper trilayer crystals La$_4$Ni$_3$O$_{10}$ and Pr$_4$Ni$_3$O$_{10}$ grown by the floating zone technique and later by the flux method for the La-analogs \cite{Zhu2024a, zhang2025a, zhang2025b, Shi2025z, Li2025c}. The emergence of superconductivity was evidenced by a drop in the electrical resistance and an anomaly in the magnetic susceptibility around $P_c$ $\approx$ 15 GPa. Notably, the reported critical pressure spans a broader range of 5--40 GPa depending on the pressure transmitting medium and the criteria chosen to define $T_c$ \cite{Zhu2024a}. Pr$_4$Ni$_3$O$_{10}$ at room temperature undergoes a change in symmetry from monoclinic $P2_1/c$ to tetragonal $I$4/$mmm$ at $P_c$ $\approx$ 40 GPa, while for La$_4$Ni$_3$O$_{10}$ the transition occurs at $P_c$ $\approx$ 15 GPa  \cite{Zhu2024a, zhang2025a, zhang2025b, Li2025c}. At ambient pressure the transition towards monoclinic symmetry occurs at $T_S$ $\approx$ 1030 K for La$_4$Ni$_3$O$_{10}$ \cite{nagell2017a, Song2020a}. This suggests that SC emerges once the material has been restored to the higher tetragonal symmetry phase.

In addition under ambient pressure, a density wave (DW) ordering associated with spin and charge manifests in La$_4$Ni$_3$O$_{10}$ at $T_{DW}$ $\approx$ 130 K, characterized by the emergence of incommensurate modulation observed in single crystal by x-ray and neutron diffraction \cite{Zhang2020a}, followed by other probes \cite{xu2025f, Li2025e, khasanov2026a}. This DW ordering marked by an anomaly in resistivity and magnetic susceptibility measurements is remarkably fragile as it collapses under the application of a few GPa \cite{Zhu2024a, Wu2001a}, which presumably is the trigger to promote spin fluctuations. This can be attributed to a pairing mechanism for the formation of the Cooper pairs \cite{Sakakibara24, Huang24, Ouyang25, Chen2024a, Zhang2024h, Zhang2025j}. However, the pressure-dependence of $T_{DW}$ remains enigmatic as optical spectroscopy experiments do not corroborate the transport measurements, thereby stating that the DW competes with the SC \cite{Xu2025a}. Studies involving temperature dependent powder x-ray diffraction (PXRD) on La$_4$Ni$_3$O$_{10}$ reveals a negative thermal expansion of the $b$ axis that could  be potentially linked to structural distortions associated with density-wave ordering \cite{kumar2020a, Zhang2020c, rout2020a}. However, the symmetry in the incommensurate phase and the extent to which it can reshape the 3-$dimensional$ (3-$d$) lattice remains shrouded in uncertainty as later works do not reproduce or probe the modulation through x-ray diffraction.

Further analysis regarding the modulation is preceded by a lingering matter pertaining to the symmetry of La$_4$Ni$_3$O$_{10}$ at ambient conditions, as it also remains in a state of quandary. Earlier, Zhang \textit{et al.}~\cite{Zhang2020c} reported the coexistence of a metastable orthorhombic $Bmab$ phase with the monoclinic $P2_1/c$ phase depending on the crystal growth conditions. A follow up work by Li et al., \cite{Li2025c} propose an experimental phase diagram based on x-ray and transport measurements on single-crystalline La$_4$Ni$_3$O$_{10-\delta}$. However, it is not yet elucidated from a crystallographic point of view. Careful investigation of the diffraction pattern at 11 GPa revealed extremely weak reflections that appear to violate the $B$-centering, an observation that can be ascribed to the formation of twin domains resulting from symmetry breaking, closely resembling our diffraction pattern. However, their interpretation differs from the present work.

Crystal structure is the bedrock of every material, governing its electronic structure and all electronic properties, including superconductivity. Accurate determination of the crystalline symmetry is therefore not merely a formality but an essential prerequisite for any theoretical band-structure calculation aimed at uncovering the microscopic mechanisms of superconductivity and in this paper, we seek to resolve the persistent ambiguities surrounding the symmetries reported in this material by constructing a unified crystallographic framework that connects the two phases. Our crystallographic analysis based exclusively on flux-grown crystals, unveils a direct transition from tetragonal (4/$mmm$) to monoclinic (2/$m$), with the latter emerging as a 2-fold superstructure of the former. This is further reinforced by \textit{ab initio} density-functional theory (DFT) calculations which also unequivocally demonstrates the absence of any intermediate phase between the monoclinic and the tetragonal $I4/mmm$.
The absence of an orthorhombic $Bmab$ phase evidenced by minute deviations of the monoclinic angle from $90^\circ$ and improved fit to the XRD data employing $P2_1$/$c$ symmetry suggests that superconductivity emerges within the tetragonal symmetry under pressure, consistent with previous reports \cite{Zhu2024a, zhang2025a, Li2025c}. In La$_4$Ni$_3$O$_{10}$, a characteristic feature of the tetragonal phase is the absence of tilting of the NiO$_6$ octahedra along \textbf{c}, unlike in the lower-symmetry phases. The linear Ni--O--Ni bond appears favorable for electron hopping and orbital overlap. Furthermore, tetragonal symmetry preserves the equivalence of the in-plane directions, thereby maintaining the symmetry of the $d_{x^2-y^2}$ electronic states believed to be important for superconductivity in the nickelates. Prior investigations employing ARPES have identified contributions from both the $d_{x^2-y^2}$ and $d_{3z^2-r^2}$ orbitals \cite{Li2017m}.

We further track the evolution of the structural transition temperature $T_S$ under hydrostatic pressure, observing a reduction from 1030 K at ambient pressure to 20 K under hydrostatic pressures of 14 GPa. Finally, ambient-pressure XRD on flux-grown crystals reveals incommensurate satellite reflections associated with density-wave (DW) ordering below 130 K. This observation confirms the intrinsic nature of the modulation and demonstrates that it is not dependent on the crystal growth method, as previous reports were limited to floating-zone crystals \cite{Zhang2020a}. These results are consistent with Raman scattering, which shows additional phonon modes around 130 K, indicative of subtle lattice distortions associated with the DW ordering.

\section*{\label{sec:la4ni3o10_results_discussion}%
Results and discussion}

High-temperature powder X-ray diffraction (PXRD) measurements under (See Figures S1--S2 in the the supporting information (SI) \cite{la4ni3o10suppmat2025a}) ambient pressure carried out at beamline BM01 of ESRF in Grenoble, reveal a structural transition from the tetragonal to the monoclinic phase  at $T_S$ $\approx$ 1030 K, in close agreement with \cite{nagell2017a, Song2020a}. As in previous studies \cite{kumar2020a, Zhang2020a} the minute lattice distortion manifested as a small deviation of $\beta$ from 90$^\circ$ by $\approx$ 0.1--0.2$^\circ$ as shown in Figure S1 in the SI \cite{la4ni3o10suppmat2025a} can only be understood by employing non-standard crystallographic settings.

Previously, both Nagell et al. and Song et al. \cite{nagell2017a, Song2020a} reported a monoclinic $P2_1/a$ (alternative setting of standard $P2_1/c$ which involves the permutation of the basis vectors \textbf{a} and \textbf{c}) phase with lattice parameters $a \approx b \approx 5.4$~\AA{} and $c \approx 27$~\AA{}, treating it as a primitive cell, which effectively doubled the number of atoms and refinement parameters. In our work, we employ the non-standard $B2_1/a$ setting, which uses the same lattice parameters, but the $B$-centering makes it equivalent to the reduced primitive $P2_1/a$ cell with $c \approx 14$~\AA{}. Consequently, the number of atoms and independent refinement parameters remains unchanged compared to the reduced cell. This choice allows a cleaner and more reliable refinement while maintaining equivalence with previously reported structural parameters. In addition, the theoretical calculation by D. Puggioni, et.al. \cite{Puggioni2018a} shows that the reported high-temperature phases of La$_4$Ni$_3$O$_{10}$, which have four formula units are unstable at low temperature and transform to a monoclinic $P2_1/a$ phase that can be described using two formula units in
agreement with the angle resolved photoemission spectroscopy \cite{Li2017m}, which corroborates our interpretation.

\begin{figure}[ht]
\includegraphics[width=140mm]{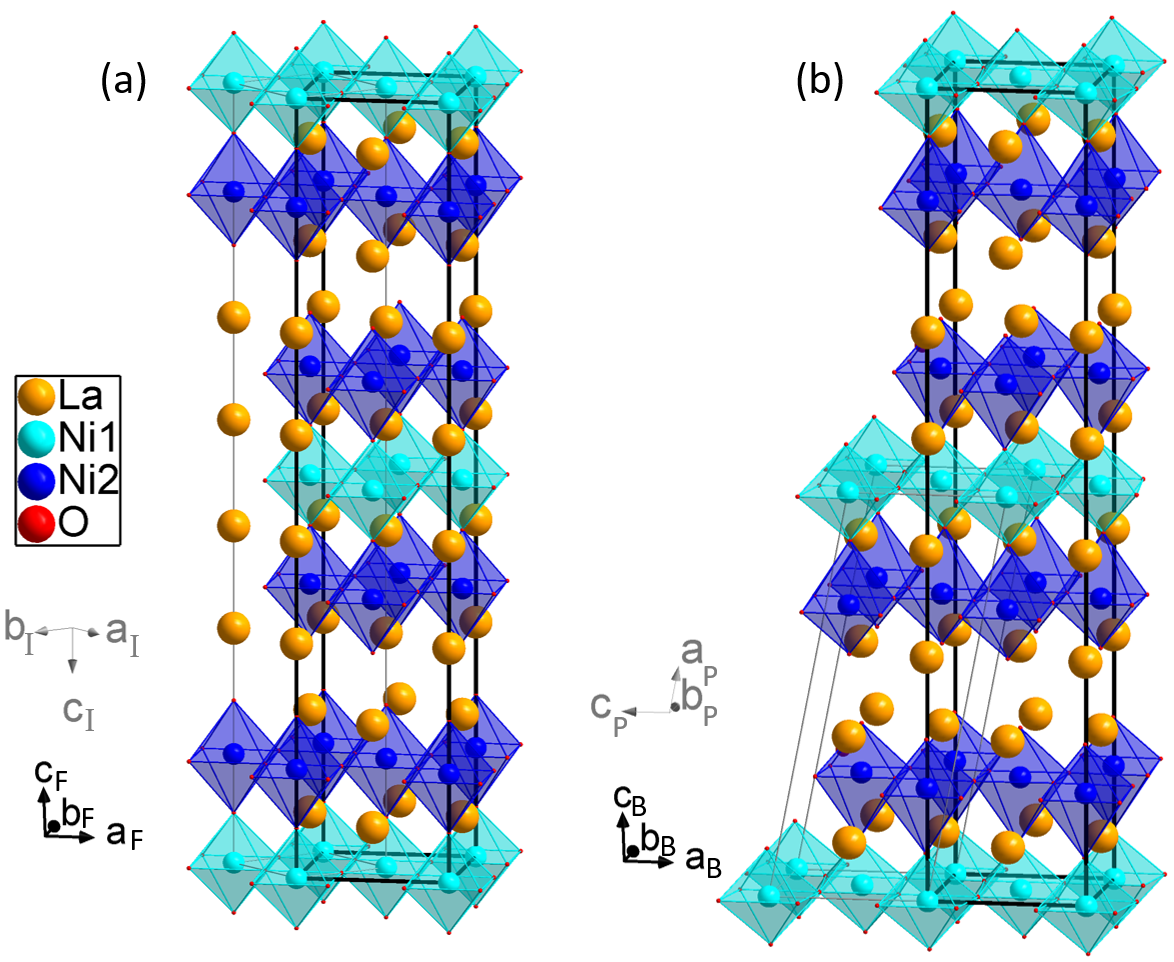}
\caption{\label{fig:UnitCellParamvsT}%
Crystal structures and phase evolution of La$_4$Ni$_3$O$_{10}$.
(a) High-pressure tetragonal phase (15 GPa, RT): the $F4/mmm$ unit cell (thick dark lines) superimposed on standard $I4/mmm$ (thin grey lines).
(b) Monoclinic phase (6 GPa, RT): $B2_1/a$ (thick dark lines) on $P2_1/c$ (thin grey lines) to illustrate the transformation (see equations \ref{eq:transform1} and \ref{eq:transform2}).}
\end{figure}

Moreover, earlier works do not explicitly describe the transformations between standard monoclinic $P2_1/c$ to non-standard $B2_1/a$ and standard tetragonal $I$4/$mmm$ to non-standard $F$4/$mmm$, thereby inadvertently obscuring symmetry relationships. Here as shown in Figure \ref{fig:UnitCellParamvsT} (a)--(b), we clarify these relationships by explicitly showing that the non-standard tetragonal $F4/mmm$ is related to the standard $I4/mmm$ through a transformation (matrix $Q_{Tetra}$) of the basis vectors of the centered cells according to:
\begin{equation}
\label{eq:transform1}
\left( \begin{array}{c}
\mathbf{a}_{F} \\
\mathbf{b}_{F}  \\
\mathbf{c}_{F}
\end{array} \right)=
Q_{Tetra}\textsuperscript{-1,t} \left( \begin{array}{c}
\textbf a_I  \\
\textbf b_I  \\
\textbf c_I
\end{array} \right)=
\begin{pmatrix}
                         1 & -1 & 0 \\
                         -1 & -1 & 0 \\
                         0 & 0 & -1
                       \end{pmatrix} \left(
                       \begin{array}{c}
\textbf a_I  \\
\textbf b_I  \\
\textbf c_I
\end{array} \right)=
\left( \begin{array}{c}
 \mathbf{a}_{I} -\mathbf{b}_{I} \\
 -\mathbf{a}_{I} -\mathbf{b}_{I}         \\
  -\mathbf{c}_{I} \\
\end{array} \right)
\end{equation}
Here, $\mathbf{a}_{F}, \mathbf{b}_{F}, \mathbf{c}_{F}$ are the basis vectors of the $F$-centered cell, while $\mathbf{a}_{I}, \mathbf{b}_{I}, \mathbf{c}_{I}$ correspond to those of the $I$-centered cell.

Similarly, the non-standard monoclinic $B2_1/a$ is related to standard $P2_1/c$ by:

\begin{equation}
\label{eq:transform2}
\left( \begin{array}{c}
\mathbf{a}_{B} \\
\mathbf{b}_{B}  \\
\mathbf{c}_{B}
\end{array} \right)=
Q_{Mono}\textsuperscript{-1,t} \left( \begin{array}{c}
\textbf a_P  \\
\textbf b_P  \\
\textbf c_P
\end{array} \right)=
\begin{pmatrix}
                         0 & 0 & -1 \\
                         0 & 1 & 0 \\
                         2 & 0 & 1
                       \end{pmatrix} \left(
                       \begin{array}{c}
\textbf a_P  \\
\textbf b_P  \\
\textbf c_P
\end{array} \right)=
\left( \begin{array}{c}
 -\mathbf{c}_{P}  \\
 \mathbf{b}_{P}          \\
 2(\mathbf{a}_{P}) + \mathbf{c}_{P} \\
\end{array} \right)
\end{equation}

\begin{figure}[ht]
\includegraphics[width=140mm]{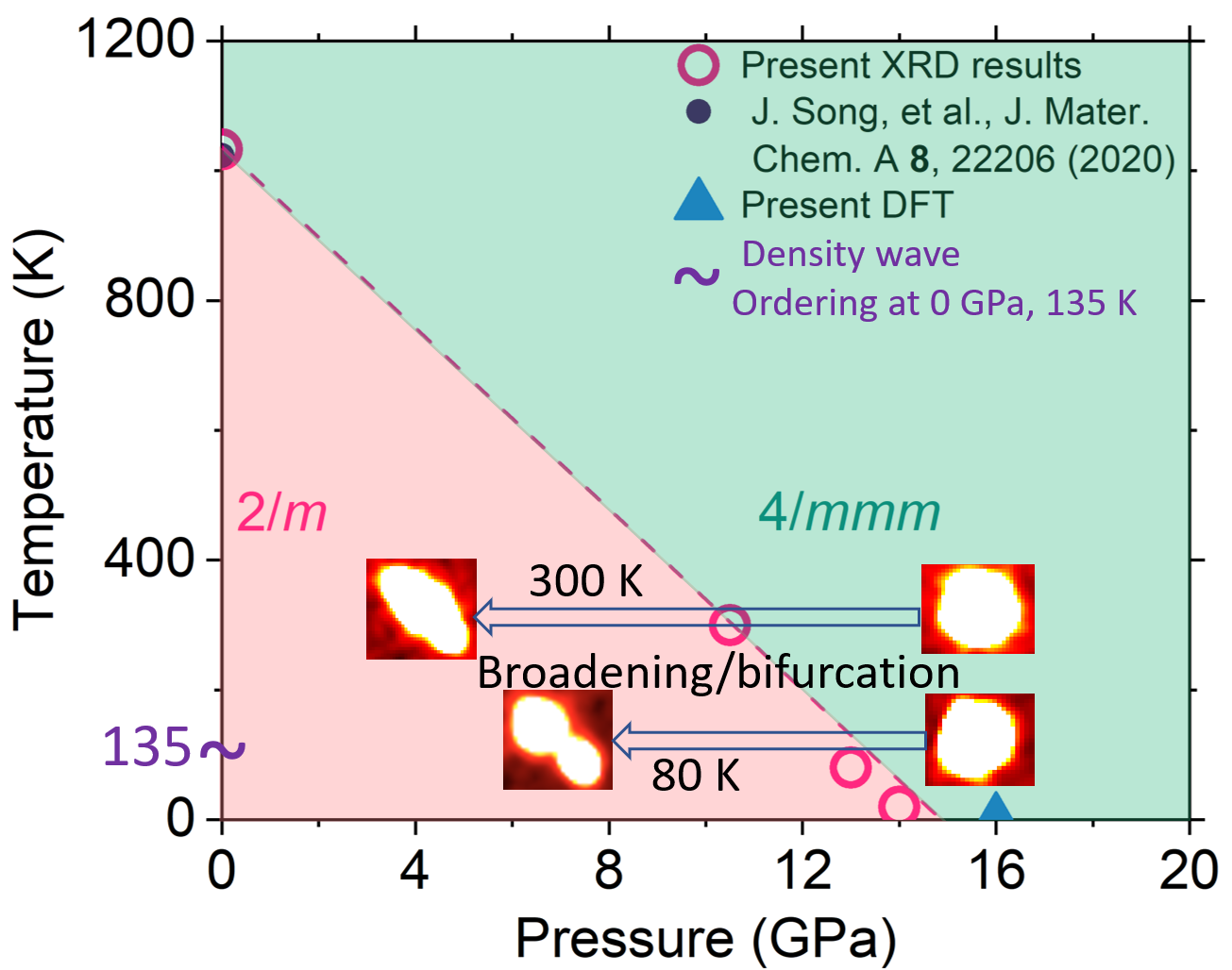}
\caption{\label{fig:phase}%
Pressure--temperature phase diagram showing the tetragonal-to-monoclinic transition; dashed line indicates the phase boundary between $4/mmm$ and $2/m$ symmetries.
Bragg reflections at RT and 80 K illustrate the evolution from $4/mmm$ to $2/m$, with broadening and anisotropy in the monoclinic phase.
Complementary \emph{ab initio} calculations predict $P_c \approx 16$ GPa at 0 K, in agreement with experiments.
}
\end{figure}

Here, $\mathbf{a}_{B}, \mathbf{b}_{B}, \mathbf{c}_{B}$ correspond to the basis vectors of the $B$-centered cell, while $\mathbf{a}_{P}, \mathbf{b}_{P}, \mathbf{c}_{P}$ denote the basis vectors of the primitive cell. See supporting information (SI) \cite{la4ni3o10suppmat2025a} for details. We thus employ the use of the non-standard $F$4/$mmm$ for the tetragonal and $B2_1/a$ for the monoclinic phases to emphasize the importance of maintaining a consistent crystallographic framework in view of connecting both and to avoid the impression of a large monoclinic distortion that arises in the standard setting $P$2$_1$/$c$ with $\beta$  $\approx$ 100$^{\circ}$, thus enabling a more transparent and meaningful description of how the distortion evolves under temperature and pressure. To map the evolution of this structural transition by pressure, single-crystal X-ray diffraction (SXRD) under pressure was performed at the beamline ID15b of the ESRF. Application of hydrostatic pressure suppresses the transition, as elucidated by the phase diagram shown in Figure \ref{fig:phase}, which reveals a pronounced reduction of the transition temperature to 20 K under an applied pressure of 14 GPa.

\begin{figure}[ht]
\centering
\includegraphics[width=140mm]{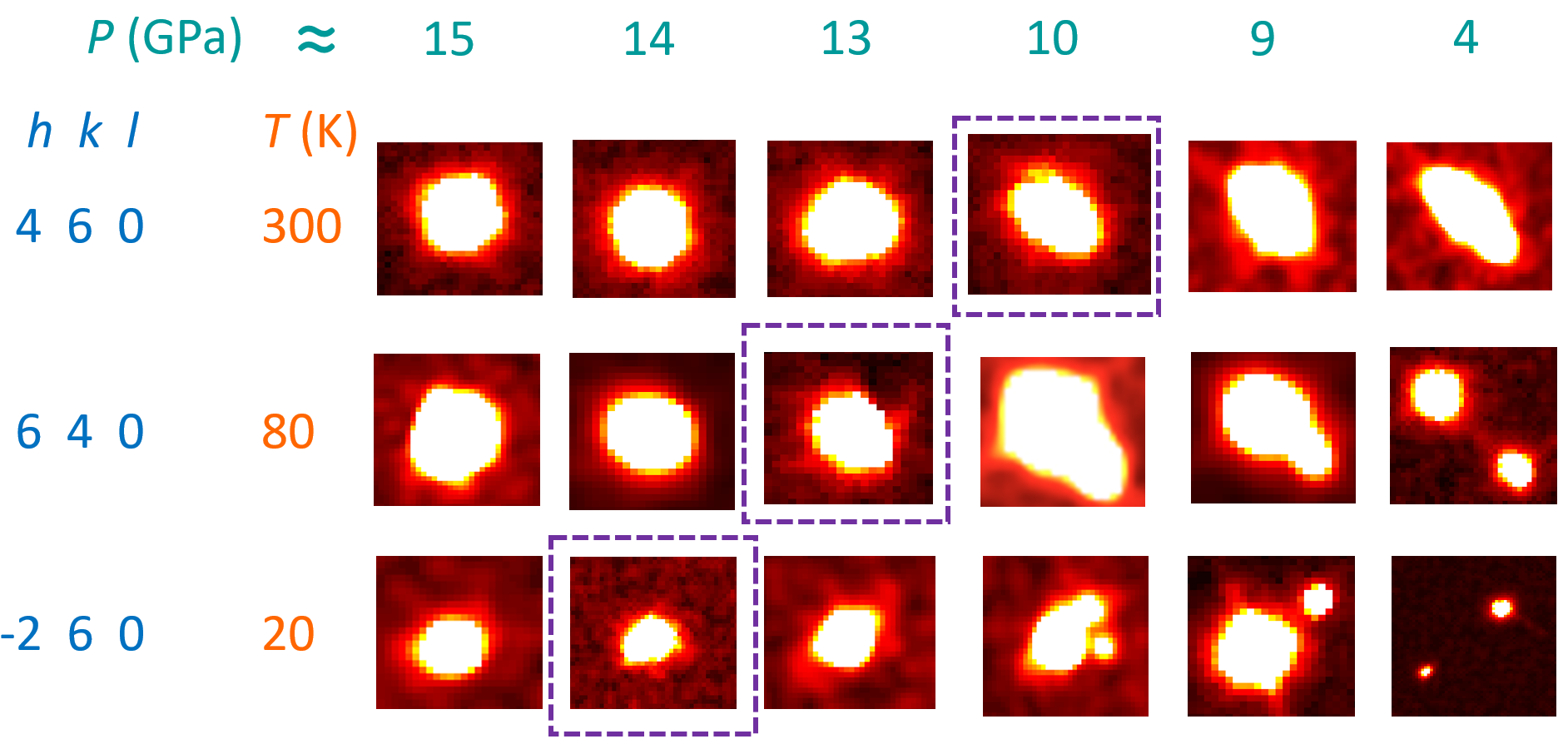}
\caption{\label{fig:unwarp}%
Reconstructed reciprocal-space $hk0$ planes for the non-standard tetragonal $F$4/$mmm$ and monoclinic $B2_1/a$ settings ($a \approx 5.41$~\AA, $b \approx 5.43$~\AA, $c \approx 27$~\AA, $\beta \neq 90^\circ$), measured at 300, 80, and 20~K. Upon reduction of pressure, broadening and splitting of the reflections reveal a $4/mmm \rightarrow 2/m$ structural transition accompanied by pseudo-merohedral twinning. Purple dashed squares indicate the onset of the transition, where the reflections first become anisotropically broadened. Both the anisotropy and the bifurcation exhibit a pronounced temperature dependence.}
\end{figure}

\begin{figure}[ht]
\centering
\includegraphics[width=140mm]{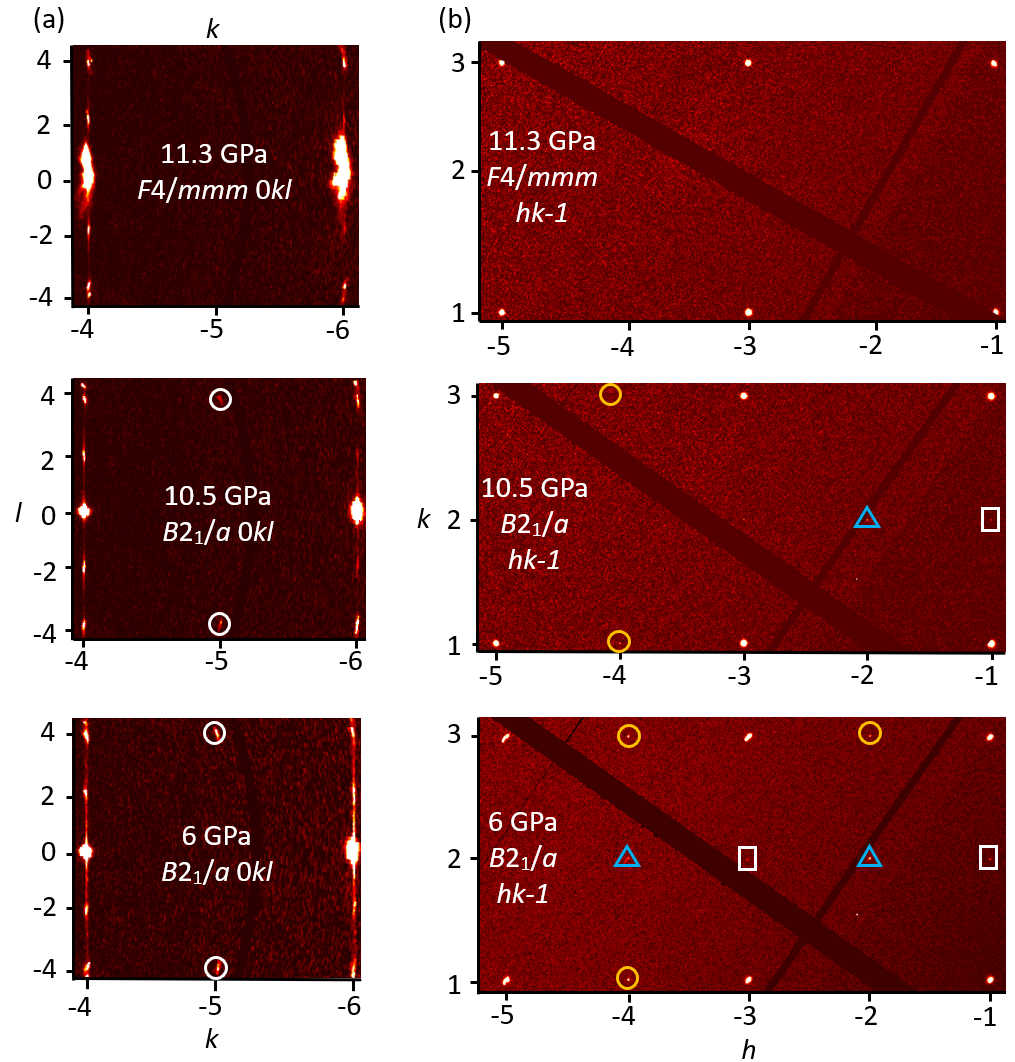}
\caption{\label{fig:unwarp2}%
(a) Excerpts of the $0kl$ plane; reflections $(0\,\overline{5}\,\overline{4})$ and $(0\,\overline{5}\,4)$ (white circles) correspond to a two-fold superstructure. (b) $hk\overline{1}$ plane showing weak reflections below 11.3 GPa (yellow circles) associated with loss of four-fold symmetry; white rectangles denote new reflections that are allowed by the $B$-centering and sky-blue triangles mark domain-wall related scattering.}
\end{figure}

\begin{figure}[ht]
\includegraphics[width=160mm]{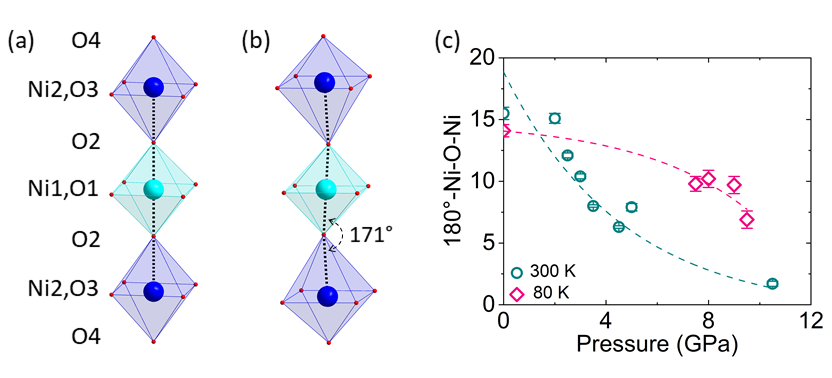}
\caption{\label{fig:crystal_struc} (a)--(b). In the tetragonal phase the bond angle of Ni1-O2-Ni2 is 180$^\circ$ at 15 GPa (RT), while in the monoclinic phase at 6 GPa (RT) the bond angle of Ni1-O2-Ni2 is 171.1$^\circ$. (c) Evolution of the octahedral tilt as a function of pressure at 300 and 80 K. Dashed lines are a guide to the eye. Estimated standard deviations (esds) of the bond angles were calculated from the structural refinement using Jana2006 \cite{petricekv2014a}.}
\end{figure}

\begin{figure}[ht]
\includegraphics[width=100mm]{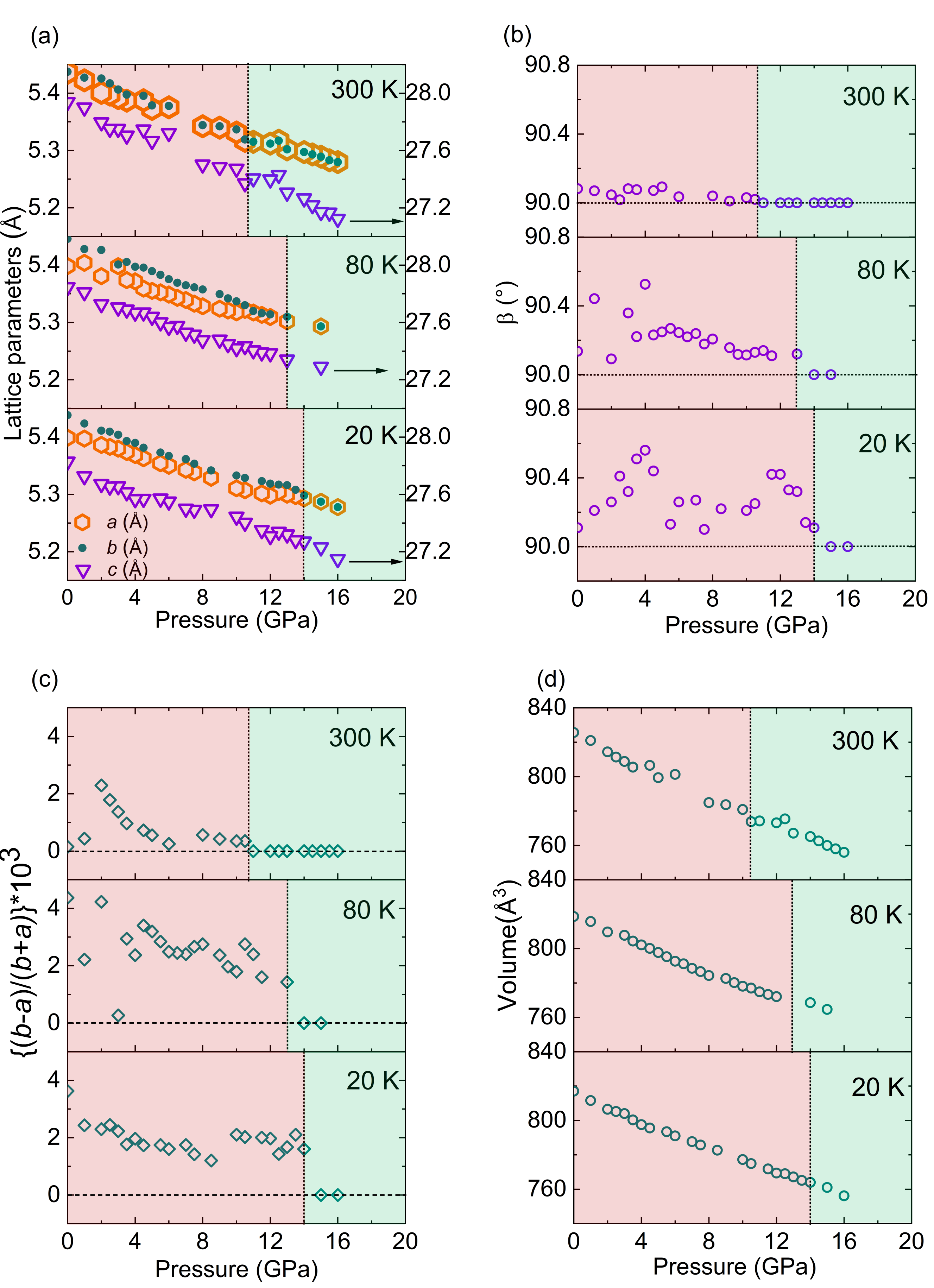}
\caption{\label{fig:lattice parameters}%
(a)--(d) Lattice parameters, distortion and the volume
within the pressure range of 0--16 GPa for temperatures
at 300, 80 and 20 K. In the tetragonal phase (4/$mmm$) (green-shaded
region),
the lattice parameters correspond to the non-standard $F$-centered
setting of $I$4/$mmm$ and for the monoclinic phase (2/$m$) (pink-shaded region) we use the $B$2$_1$/$a$
non-standard space group in order to preserve the same setting
of the lattice which allows for a comparison. Estimated standard deviations (esds) computed from the refinement are smaller than the symbols which can be understood from the numerical values listed in Table S5 in the SI \cite{la4ni3o10suppmat2025a}. Fits of a Birch-Murnaghan equation of state (EOS) to the
volume-pressure data has resulted in V$_0$ = 825(1) \AA${^3}$ and
K$_0$ = 155(4) GPa at 300 K, V$_0$ = 818(1) \AA${^3}$ and K$_0$ =
182(4) GPa at 80 K, and V$_0$ = 815(1) \AA${^3}$ and K$_0$ = 185(2)
GPa at 20 K. See the SI for details \cite{la4ni3o10suppmat2025a}.}
\end{figure}

Establishing the critical pressure and temperature points for the phase diagram not only required rigorous scrutiny of reflections in the $hk0$ plane as shown in Figure \ref{fig:unwarp}, but also tracking the appearance of the 2-fold superstructure and the evolution of the structural parameters as indicated in Figures \ref{fig:unwarp2}, \ref{fig:crystal_struc} and \ref{fig:lattice parameters}. The reflections, nearly circular in the high-pressure tetragonal phase, become progressively broadened and anisotropic as the transition sets in, eventually splitting into distinct components at lower pressures. Purple dashed squares indicate the onset of the broadening associated with the transition towards monoclinic from tetragonal in Figure \ref{fig:unwarp}. This bifurcation of the reflections becomes particularly pronounced at low temperatures 80 and 20 K, showing that the evolution of the distortion in the structure is significantly tuned by both pressure and temperature.

Furthermore, we show that the emergence of the monoclinic ($2/m$) phase in La$_4$Ni$_3$O$_{10}$ cannot be interpreted as a trivial distortion of the tetragonal lattice. As shown in Figure~\ref{fig:unwarp2}(a), reflections such as $(0\,\overline{5}\,\overline{4})$ and $(0\,\overline{5}\,4)$  (white circles) that appear below 11.3 GPa correspond to a 2-fold monoclinic superstructure. When transformed into the standard $I$-centered 'pseudo-tetragonal' lattice ($a = b \approx 3.7$~\AA, $c \approx 27$~\AA), these reflections can alternatively be indexed as $(2.5\ 2.5\, \overline{4})$ and $(2.5\, 2.5\, 4)$, corresponding to positions $\mathbf{q}^1 = (\frac{1}{2}, \frac{1}{2}, 0)$ and $\mathbf{q}^2 = (\frac{1}{2}, -\frac{1}{2}, 0)$ in this pseudo-tetragonal description. These vectors are not independent modulation wavevectors but rather reflect a different indexing convention for the same monoclinic superlattice reflections, highlighting the 2-fold nature of the monoclinic structure relative to the parent tetragonal lattice. Rather than invoking higher-dimensional ($3+d-dimensional$) superspace formalism \cite{van2007incommensurate}, where the intensities of the main Bragg reflections are kept distinct from the superlattice peaks, we reproduce the solution of the monoclinic structure in 3-dimensions by indexing and integrating as a superstructure, echoing earlier works \cite{Song2020a, Zhang2020a, Zhu2024a, zhang2025a, Li2025c}.

Figure \ref{fig:unwarp2}(b) reveals the emergence of new reflections that deviate from the expected centering conditions at pressures below 11.3 GPa, consistent with the recent observations of Li et al. \cite{Li2025c}. Reflections such as $(\overline{3}\, 2\, \overline{1})$ and $(\overline{1}\, 2\, \overline{1})$ highlighted by white rectangles are fully compatible with the $B$-centering. In contrast, those marked by yellow circles like $(\overline{4}\, 3\, \overline{1})$, $(\overline{4}\, 1\, \overline{1})$, and $(\overline{2}\, 3\, \overline{1})$ originate from a twin domain rotated by 90$^\circ$ about the $c$-axis. Sky-blue triangles denote reflections arising from domain-wall scattering, which correlates with the pronounced strengthening of diffuse lines (Fig. \ref{fig:unwarp2} (a)) in the monoclinic phase.

 Figure \ref{fig:crystal_struc} (a--c) shows the deviation of the Ni1-O2-Ni2 bond angle from 180$^\circ$ that corresponds to the tilt of the NiO$_6$ octahedra along \textbf{c} upon entering the 2/$m$ phase, thereby highlighting the effect of the phase transition on an atomic level. Figure \ref{fig:crystal_struc} also illustrates that La$_4$Ni$_3$O$_{10}$ consists of three layers of Ni planes, with each Ni atom surrounded by six oxygen to form an octahedral site. The inner layer is sandwiched between two outer layers. The outer layers of different blocks are stacked with a rock salt layer of LaO. It is worth noting that the octahedral coordination is strongly distorted. Each Ni atom in the inner layer has two apical oxygen (i.e. along the c-axis in the pseudo-tetragonal description of the structure), which are close together, and four oxygen in the basal plane, which are further apart . The octahedra in the outer layers differ. Indeed, the apical O close to the rock salt layer (O4) and the four O in the basal plane (O3) form a square-based pyramid, and the second apical O (O2) shared with the inner layer Ni is far away. At 15 GPa, the octahedra is perfect. As it was already noticed by the previous work, at low pressure, the NiO$_6$ blocks are tilted from the pseudo $c$-axis.

Evolution of the lattice parameters can be inferred from Figure \ref{fig:lattice parameters}(a)--(b) the distortion toward monoclinic symmetry is minimal, particularly at 300 K where it is negligible. Much like the structural transition at ambient pressure outlined in the SI \cite{la4ni3o10suppmat2025a} in Figure S1, it is also noted that in Figure \ref{fig:lattice parameters} (b) there is a jump in the value of $\beta$ at the onset of the 2/$m$ phase, which could be attributed to a weakly first-order phase transition.
The loss of symmetry from 4/$mmm$ to 2/$m$ generates four twin domains whose relation is explained by Figure S3 in the SI \cite{la4ni3o10suppmat2025a}. Essentially, these domains are described by distinct twin laws, and their presence is evident from reflection splitting in the diffraction data as shown in Fig.~\ref{fig:unwarp}. The degree of overlap between the domains referred to as the 'twin obliquity' is partial and indicates pseudo-merohedral twinning as observed from Figure \ref{fig:unwarp}. Detailed information pertaining to the concept of twinning by phase transformation can be obtained from Parsons \cite{parson2003a}. The splitting appears to increase at 80 and 20 K, which contributes to the increase of the lattice distortion as defined by $(b-a)/(b+a)$ at lower temperatures as depicted in Figure \ref{fig:lattice parameters}(c). Lattice parameters and Volume as inferred from \ref{fig:lattice parameters}(d) decreases monotonously with increasing pressure and the equation of state (EOS) fit depicting its compressibility is in good agreement with \cite{Zhu2024a}. Details of the EOS fit are shown in Figure S4 in the SI \cite{la4ni3o10suppmat2025a}. Recent work by Li \textit{et al.}~\cite{Li2025c}, based on measurements at the ID27 beamline at the ESRF on flux-grown crystals, proposes the presence of an intermediate phase with $Bmab$ symmetry under 11 GPa at room temperature. In contrast, our experiments were performed on flux-grown crystals synthesized using the same technique as Li \textit{et al.} at beamline ID15b of ESRF, whose capabilities are analogous to ID27. While Li \textit{et al.} conducted measurements only at room temperature, where $\beta$ is nearly $90^\circ$, we carried out low-temperature XRD measurements under pressure. Under these conditions at temperatures of 80 and 20 K, we clearly observe a deviation of $\beta$ from $90^\circ$, ruling out orthorhombic symmetry. At room temperature and around 10.5--11 GPa, where $\beta$ is nearly $90^\circ$ (see Fig.~\ref{fig:lattice parameters}(b)), the structure may be interpreted as orthorhombic $Bmab$.

However, a full structural refinement at 10.5~GPa, near the 11~GPa pressure point reported by Li \textit{et al.}~\cite{Li2025c}, indicates that the monoclinic $B2_1/a$ model, compared with the orthorhombic $Bmab$ model, results in an improvement of $R_F(\mathrm{obs})$ by approximately 1\% and a substantial reduction in the residual electron density by about 50\% (see Table~S4 in the SI~\cite{la4ni3o10suppmat2025a}). This preference for the monoclinic phase is further reinforced by the Hamilton-R test \cite{hamilton1965a, Internationaltablesvolc}, which demonstrates that the $B2_1/a$ model provides a statistically significant better fit to the diffraction data, with a probability of less than 1\% that this result arises by chance, confirming that the monoclinic structure more accurately represents the crystal under these conditions. Table S7 in the SI~\cite{la4ni3o10suppmat2025a} also shows similar tests at 230 and 80 K under ambient pressure, where the $B2_1/a$ model shows a significantly better fit to the SXRD data than obtained with $Bmab$ symmetry, while the monoclinic lattice distortion is more pronounced. The minor difference between our refinement and that of Li \textit{et al.}~\cite{Li2025c} arises from the choice of criterion of observability. While they adopt a more permissive 2$\sigma$ threshold, we employ a 3$\sigma$ cutoff, resulting in a reduced number of observed reflections in our refinement. This choice has no effect on the weighted $R$ factors and does not lead to any significant change in the refined structural parameters within experimental uncertainty. It should also be noted that the less than ideal goodness-of-fit values ($\mathrm{g.o.f.} \approx 4$, see Table~S5 in the SI \cite{la4ni3o10suppmat2025a}) at higher pressures may arise due to residual strain and de-twinning effects under high-pressure conditions as the crystal evolves toward the tetragonal phase. These results indicate that the transition from tetragonal ($4/mmm$) to monoclinic ($2/m$) proceeds directly without stabilizing an intermediate orthorhombic phase. Moreover, the coexisting metastable orthorhombic $Bmab$ phase reported by Zhang \textit{et.al} \cite{Zhang2020c} on crystals grown by high-pressure floating zone technique appears to be absent in flux-grown crystals.

Tiny deviations of axial angles from $90^\circ$ are not uncommon and have been reported in other systems, such as the CDW compound SrAl$_4$~\cite{ramakrishnan2024a}, incommensurately modulated Rb$_2$ZnCl$_4$~\cite{kotla2025a}, antiferromagnetic Eu$_2$Sb$_3$~\cite{chapuis1980a}, the incommensurate phase of ferroelectric KNbO$_3$~\cite{shoker2025a}, and the superstructure phase of CoSn$_2$~\cite{nandi2025a}. The refined atomic coordinates are provided in Tables S1 and S2 the SI~\cite{la4ni3o10suppmat2025a}. Table~S5 in the SI~\cite{la4ni3o10suppmat2025a} summarizes the crystallographic information. The lattice parameters indicate that the monoclinic distortion is minute, rendering the lattice pseudo-tetragonal.

\begin{figure}[ht]
    \centering
    \includegraphics[width=160mm]{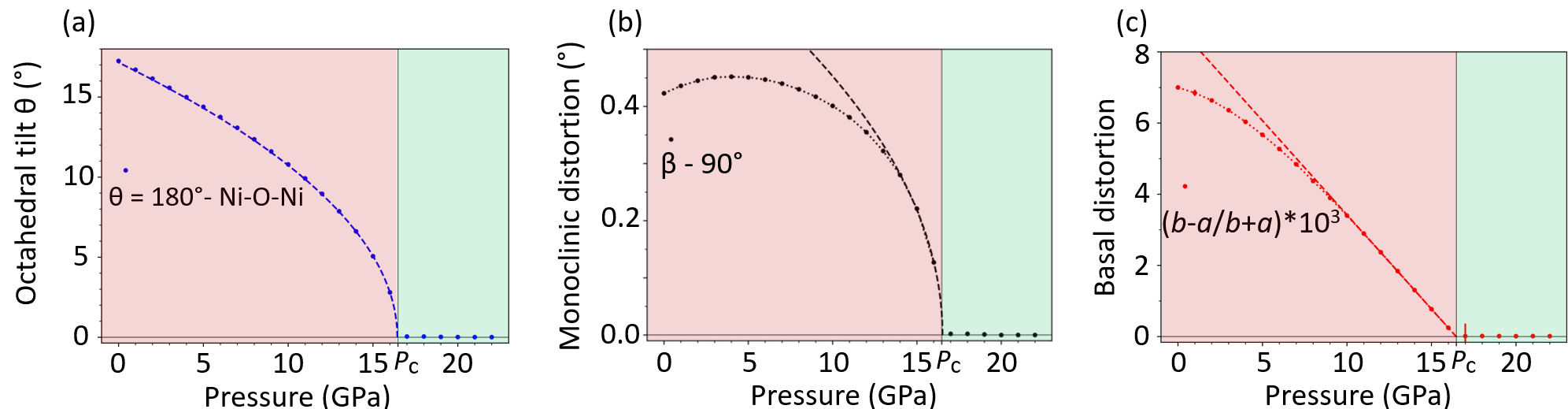}
    \caption{(a)--(c) Density-functional theory GGA PBE crystal relaxation as a function of applied external pressure. (a) Blue dots and lines: NiO octahedra tilt angle $\theta = 180 -$ Ni1-O2-Ni2 in degrees. (b) Black dots and lines: Monoclinic distortion $\beta - 90$ in degrees. (c) Red dots and lines: Basal distortion $(b-a)/(a+b)$ scaled by $10^3$. Critical pressure $P_c \approx 16$ GPa. Both horizontal and vertical error bars are smaller than the dot size when not visible. Dashed lines refer to fits, whereas dotted lines are a guide to the eyes. Green-shaded region marks the tetragonal phase, while the pink-region denotes the moonclinic phase.}
    \label{fig:dft}
\end{figure}

The subtlety of the direct tetragonal-to-monoclinic phase transition realized from XRD can also be understood from Fig.~\ref{fig:dft}, where we present the results of an \textit{ab initio} density-functional theory (DFT) crystal structure relaxation using the PBE \cite{PBE} generalized-gradient approximation (GGA).
The calculations at all pressures were performed within the least symmetric monoclinic spacegroup 14 in the $P2_1/c$ setting, which is a subgroup of both orthorhombic $Bmab$ and tetragonal $I4/mmm$, namely  $P2_1/c \subset Bmab \subset Fmmm \subset I4/mmm$.
Therefore, our crystal relaxation is free to span all these four structures and can capture all transitions between these phases.
The figure illustrates the evolution with pressure of the three order parameters associated to these transitions, namely:
the orthorhombic distortion factor $o$ taken as the basal distortion $o = (b-a)/(a+b)$, associated with the orthorhombic-tetragonal $Fmmm \to I4/mmm$ phase transition;
the NiO octahedra tilt angle $\theta = 180 -$ bond angle of Ni1-O2-Ni2 (see Figure \ref{fig:crystal_struc}c), associated with the $Bmab \to Fmmm$ phase transition;
the 'monoclinicity' (monoclinic distortion) $\beta - 90$ degrees, associated with the monoclinic-orthorhombic $P2_1/c \to Bmab$ phase transition. Here, $\beta$ is the angle between the $a$ and $c$ vectors in the $P2_1/c$  monoclinic cell. After applying the $P2_1/c \to B2_1/a$ transformation matrix, $\beta$ is $\approx$ 90$^{\circ}$, compared to $\approx$ 100$^{\circ}$ in the primitive $P2_1/c$ setting, consistent with the experimental analysis.

 Figure \ref{fig:dft} unequivocally shows that, for \textit{ab initio} DFT PBE, the system undergoes only one phase transition, from the monoclinic $P2_1/c$ directly to the tetragonal $I4/mmm$.
\textit{There is no room for intermediate structures}: both the orthorhombic $Bmab$ found by Ref.~\cite{Li2025c}, and also a hypothetical orthorhombic $Fmmm$ without octahedra distortions, are excluded.
Indeed, all three order parameters nullify at the same critical pressure found to be $P_c = 16.47 \pm 0.04$~GPa.
The reported error reflects the spread between the three independent  $P_c$ values obtained from fits to each order parameter and does not include the error associated with the PBE approximation.

It is well known that the PBE approximation overestimates lattice parameters by at least 1\%, and the error increases for example on the $c$ parameter of layered hexagonal crystals.
At $T = 0$~K and $P = 0$~GPa our DFT PBE relaxation  provided $a=5.43$~\AA, $b=5.51$~\AA, and (in the $B2_1/a$ setting) $c=27.89$~\AA\ with $\beta = 90.42$~deg, and finally $\theta = 17.25$~deg.
With respect to the experiment at T = 20 K, P = 0 GPa, we see that here the PBE overestimation is slightly below 1\% on $a$, slightly above 1\% on $b$, and remarkably only 0.3\% on $c$.
There is no common consensus on the PBE error on lattice angle parameters.
Here we see that PBE and XRD at least agree on the order of magnitude.
It can be expected that the systematic overestimation on lattice parameters leads also to overestimation of critical pressures.
Estimating the error on pressure as the value here required to shrink the $b$ parameter by 1\%, yields $\Delta P = 3$~GPa as the most pessimistic bound; keeping into account also the other lattice parameters leads to a more realistic estimate of $\Delta P = 1$~GPa.
Regardless, while the quantitative result on $P_c$ might be affected by a large PBE error, the qualitative result on the non-existence of intermediate phases is much less questionable.

In Fig.~\ref{fig:dft} we show by dashed lines our best fits on the three last points closest to the critical pressure.
We have used a standard scaling $\sim (P_c - P)^\alpha$ with $\alpha=1/2$ both for the octahedra tilt angle $\theta$ and for the monoclinic $\beta$ angle, while a linear fit has been used for the basal distortion $b-a$.
The square-root scaling is closely respected by the $\theta$ parameter down to $P=0$.
The $\theta$ scaling and magnitude seems in agreement with the experiment (see Fig.~\ref{fig:crystal_struc} c at 80~K).
On the other hand, we observe a quick breakdown of the scaling for the monoclinic $\beta$ parameter.
Letting the critical coefficient free in the fit yields $\alpha=0.4$ but without any improvement on the low pressure side.
Remarkably, the monoclinic angle $\beta$ presents a maximum at $\sim$4 GPa signaling a departure from a simpler mean-field description.
It would be interesting to check the existence of such maximum in XRD data, but the lowest level of monoclinicity $\beta \sim 10^{-1}$~deg, close to the experimental accuracy, will make it difficult.
Finally, we also remark that the basal distortion $o$ is linear down to 9~GPa, and then we observe a departure from linearity again in the region where the system presents its maximum monoclinicity. Overall, the theoretical results do not conform to a simple mean-field description of the phase transition.

On the other hand, our \textit{ab initio} DFT PBE calculation of the crystal structure carried within the $P2_1/c$ structure cannot state anything about the possibility of a density-wave ordering associated with either charge or spin.
We could in principle check the possibility of a commensurate DW by carrying the calculation within the supercell at the corresponding superperiodicity, but this is impossible in the case of an \textit{incommensurate} DW.


Despite our extensive temperature-dependent high-pressure SXRD measurements down to 20 K, no signatures of incommensurate modulation associated with density wave (DW) ordering were detected, contrary to what was observed by Zhang et al. \cite{Zhang2020a} under ambient pressure. To eliminate the possibility that weak satellite reflections were obscured by background scattering from the diamond anvil cell or residual pressure effects, we conducted low-temperature SXRD experiments under ambient-pressure at P24 beamline DESY, Hamburg using the same batch of single-crystals. Although the main Bragg reflections were over-saturated, these measurements below 130 K likewise revealed no incommensurate satellites in any region of reciprocal space as shown in Figure S5 in
the SI \cite{la4ni3o10suppmat2025a}, thereby suggesting the weak nature of any incommensurate modulation. Crystallographic data for the data collected in DESY are provided in Tables S3 and S6 in the SI \cite{la4ni3o10suppmat2025a}.

\begin{figure}[ht]
    \centering
    \includegraphics[width=160mm]{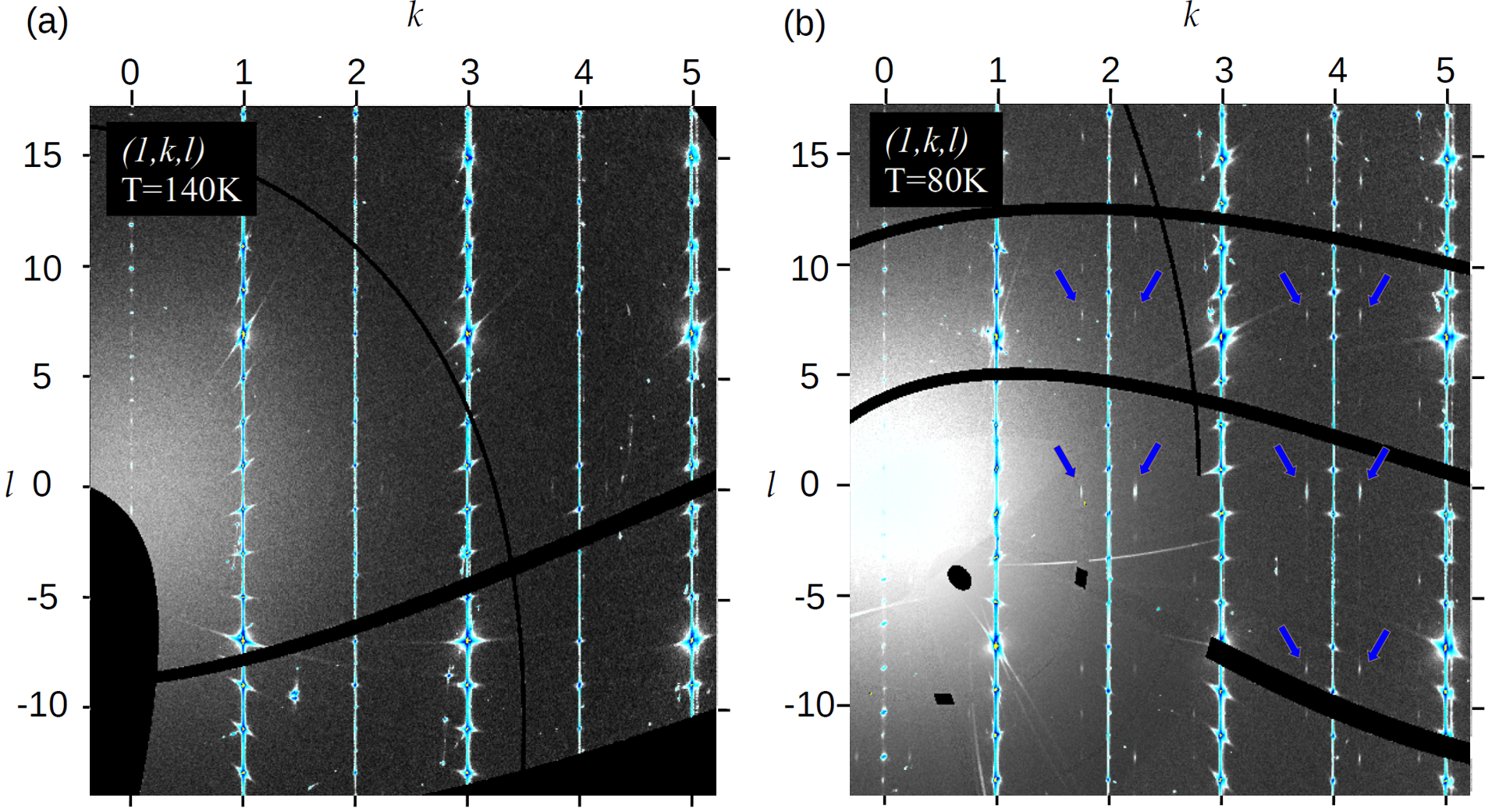}
    \caption{Reconstructed reciprocal layer of the $1kl$ plane in the $B$-centered setting} depicting weak incommensurate satellites (\textbf{q} $\approx$ 0.76 \textbf{b*} + 1 \textbf{c*}) denoted by blue arrows at 80 K under ambient pressures. Our observation on flux-grown crystals is analogous to the results of crystals grown by high-pressure floating-zone technique by Zhang et al. \cite{Zhang2020a}.
    \label{incomm}
\end{figure}

To conclusively verify their presence, we performed preliminary measurements at the ID28 beamline at the ESRF at 140 and 80 K under ambient pressures, where the incommensurate satellites could be clearly observed, presumably due to the higher brilliance enabled by the multi-bend achromat lattice design of the fourth-generation synchrotron as shown in Figure \ref{incomm}. The weak intensity of the satellites with average intensities of approximately $10^{-4}$ relative to the principal Bragg reflections indicates that the modulation is likely weak and the underlying three-dimensional lattice remains essentially undistorted, as commonly observed in other incommensurate systems such as Mo$_2$S$_3$, CuV$_2$S$_4$, EuAl$_4$, SrAl$_4$, Sm$_2$Ru$_3$Ge$_5$, and Gd$_2$Os$_3$Si$_5$ \cite{schuttewj1993b, ramakrishnan2019a, ramakrishnan2022a, kotla2025a, ramakrishnan2024a, bugaris2017charge, sharma2024a}.

The present work shows first observation of incommensurate satellite reflections in flux-grown crystals, whereas prior work by Zhang et al. \cite{Zhang2020a} was performed on crystals grown using the floating-zone method. This demonstrates that the manifestation of the weak satellites is impervious to the crystal growth technique and can therefore be regarded as an intrinsic property of La$_4$Ni$_3$O$_{10}$. Notably, following Zhang et al.'s~\cite{Zhang2020a} work, subsequent investigations have not consistently reproduced the reported low-temperature single-crystal XRD results. The difficulty in detecting the satellites may reflect a combination of their weak intensity and experimental limitations.
We now turn to the spectroscopic signatures, where the observations from X-ray diffraction are fully consistent with Raman spectroscopy. The DW-driven distortions is understood through the emergence of additional phonon modes, thereby confirming that the DW-driven lattice distortion is a robust intrinsic feature of the system.

From the space group $P2_1/c$ (no. 14, standard setting), at room temperature, the Wyckoff positions are identified to be 8 $\textit{4e}$ and 1 $\textit{2a}$. Thus, the number of phonon modes is expected to be: $\Gamma$=24A$_g$+24B$_g$+27A$_u$+27B$_u$.
The Raman-active phonons are 24 A$_g$ and 24 B$_g$, which are active in parallel and crossed polarizations, respectively, when the Poynting vector is perpendicular to the $(bc)$ plane (here we use the standard setting with $b$ as unique axis). The Raman activity is reported in table~\ref{tab:Symmetries}. All these modes have single degeneracy (See also Figures S8--S10 in the SI \cite{la4ni3o10suppmat2025a}).

\begin{figure}[ht]
\center
\includegraphics[width=120mm]{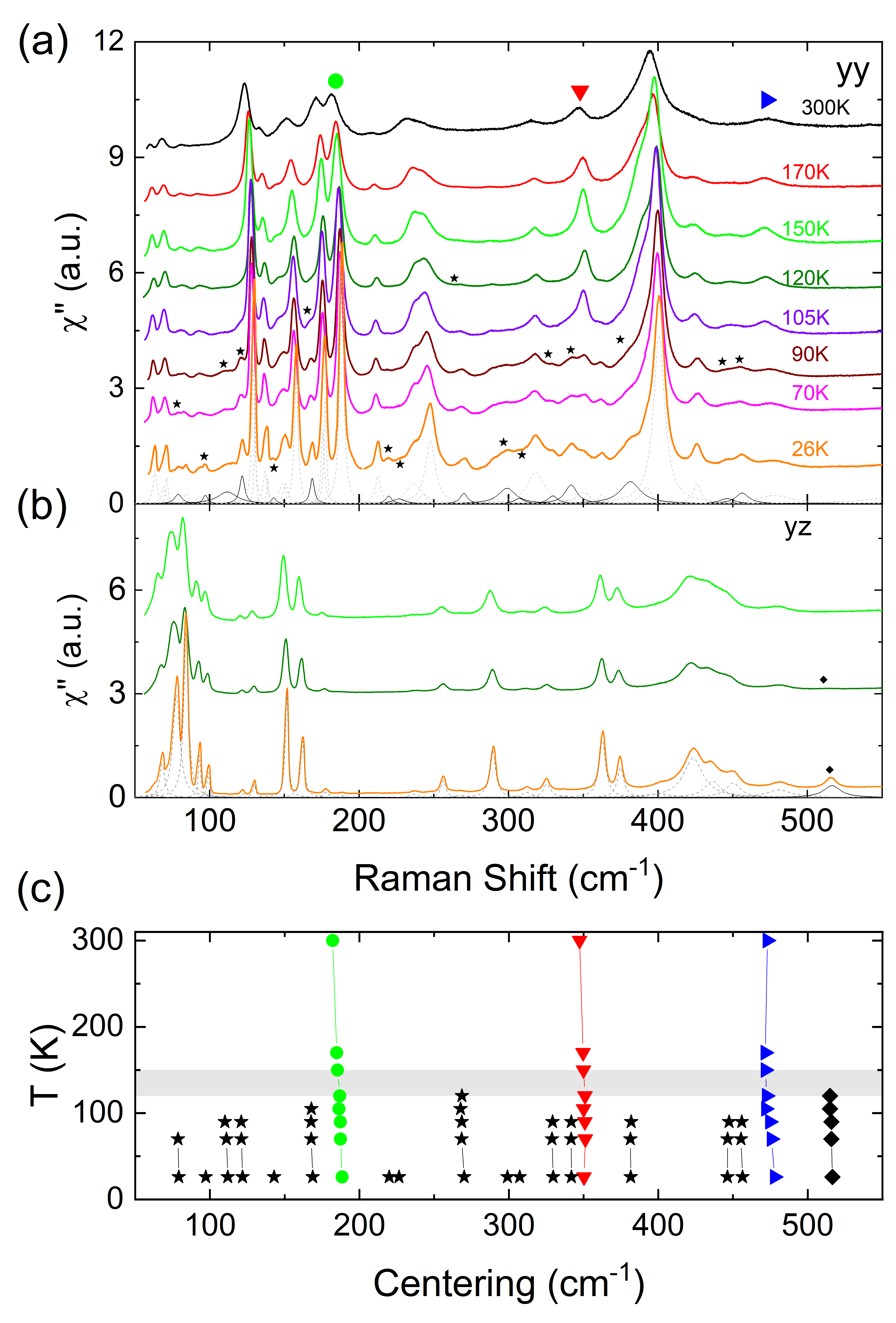}
\caption{(a)-(b) Spectra as a function of temperature in $yy$ (c) and $yz$ (d) configurations of polarizations of light. The modes that emerge across the transition at $\sim$~130~K are highlighted with diamond ($yz$) and stars ($yy$). Their fits at 26~K are shown by plain black lines. The normal-state modes that strongly change with temperature are shown with triangles. The green dot mode is a common-behavior phonon modes for comparison. Their behavior is reported in Fig.S11 of the SI \cite{la4ni3o10suppmat2025a} (c) Phonon energy as a function of temperatures for the new modes and the modes with peculiar behaviors. The grey region indicates the transition temperature range.}
\label{fig:Raman-T}
\end{figure}

As shown Fig.~\ref{fig:Raman-T} (a)-(b), 15 modes in parallel and 19 in crossed polarization configurations
are identified by polarized Raman spectroscopy at 170~K (so above the transition temperature), consistent with the reported space group $P2_1/c$. New modes appear at temperatures between 120~K and 150~K, as show Fig.~\ref{fig:Raman-T}.(a)-(b). At 26~K, a total of 31 modes in parallel configuration and 20 ones in crossed polarization configurations are measured. This evaluation of the number of modes appearing at low temperature has been done in a conservative way (see Figure S6 in the SI \cite{la4ni3o10suppmat2025a}), so these are minimum numbers of modes. When we adopt the spectral fitting approach (see the SI \cite{la4ni3o10suppmat2025a}), a total of 38 modes in parallel and 23 ones in cross configurations of polarization of light are measured at 26~K.

\begin{figure}[ht]
\center
\includegraphics[width=120mm]{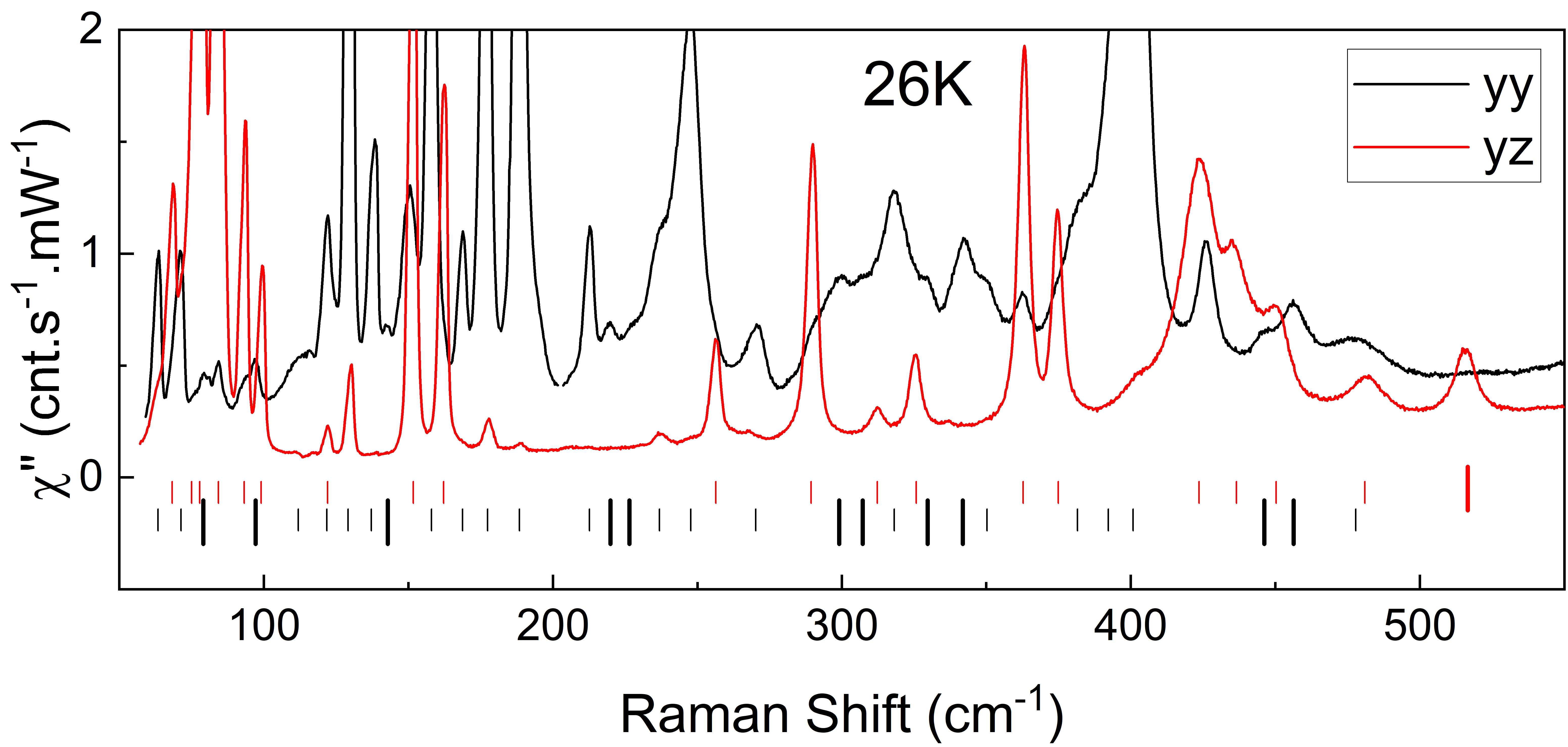}
\caption{Raman spectra of La$_4$Ni$_3$O$_{10}$ with polarizations in the $(bc)$ plane (standard-setting $P2_1/c$) and in both $yy$ and $yz$ configurations, at 26~K. Clearly identified phonons are reported as ticks (red : $yz$, black: $yy$). The longer and thicker ticks identify the modes emerging below 120-140~K.}
\label{fig:Raman-Tpolar}
\end{figure}

The new modes appearing below $\sim$120~K are interpreted as phonons, naturally originating from a transition, either a purely structural distortion or alternatively an electronic one which backfolds the phonon modes originally at finite $\vec{Q}$ into the $\Gamma$ point of the Brillouin zone, reminiscent of the observation of stripe order in hole-doped nickelate compounds \cite{yamamoto_raman_1998, blumberg_charge_1998}. Each mode as presented in Figure \ref{fig:Raman-Tpolar} clearly follows a strict Raman selection rules, being either active in crossed or in parallel polarization configurations.

The number of modes in parallel polarization as well as the total number of modes exceeds the authorized ones in the $P2_1/c$ room temperature space group. Thus we can conclude to a lowering of symmetry below $\sim~ 120$~K. Recent results by Gim $et~al.$ \cite{gim2025orbital} presented the phonon modes up to 120 meV (=970 cm$^{-1}$). They do not report any new mode at low temperature whereas we observe many new ones in this energy range. Other recent publications \cite{suthar2025multiorbital, deswal2025dynamics} reported the appearance of new modes at low temperature. Interestingly, Suthar et al. observed a total of 48 modes at low temperature (24 in each polarization configurations), so it is still consistent with the room temperature space group. Our results, while being quite similar in many aspects (behavior of peculiar phonons, energies of phonon modes as described in the SI \cite{la4ni3o10suppmat2025a}), show more than 48 modes at low temperature, thus ruling out the $P2_1/c$ as the low temperature space group. In Table S8 in the SI \cite{la4ni3o10suppmat2025a}, we report a full list of our detected phonon modes. Generally, the samples' quality and exact composition seems to have important impact on the Raman responses, pointing to this characterization technique as a sensible one.

From our results in Fig.~\ref{fig:Raman-T} (c), the transition temperature to this lower symmetry space group is at $\sim$~130~K. Besides, certain modes detected at room temperature (large colored symbols in Fig.~\ref{fig:Raman-T}.(a)) show anomalous temperature dependence upon cooling, also in correlation with the emergence of these new modes (see Fig.S11 in SI \cite{la4ni3o10suppmat2025a}). We naturally interpret this transition temperature as the onset of the Density Wave, which is reported to be at about 135~K by Zhang \textit{et al.}\cite{Zhang2020a}. \\

To conclude, we report that the monoclinic phase consistently manifests as a twinned two-fold superstructure below 1030 K which is easily observed by XRD.  The monoclinic distortion is exceedingly small, thereby maintaining an overall pseudo-tetragonal crystallographic framework. Pressurizing the material from ambient conditions to 14 GPa leads to a gradual suppression of the distortion  over a broad temperature range from 1030 K to 20 K, reflecting its delicate energetic balance. Guided by \textit{ab initio} calculations and pressure-temperature dependent XRD measurements, we show that the tetragonal-to-monoclinic phase transition is subtle and direct, without an intermediate orthorhombic $Bmab$ phase. By adopting a consistent crystallographic description, we are able to elucidate how the symmetries of the two phases evolve and are interconnected, thereby resolving the long-standing ambiguities among the tetragonal, orthorhombic and monoclinic structures. Crucially from XRD, we present the first observation of incommensurate satellite reflections in flux-grown crystals associated with DW ordering, thereby establishing that the associated lattice modulation is an intrinsic property of the material and not contingent on the method of synthesis. The results corroborate the Raman-active modes indicative of DW-driven lattice distortions. Determination of the symmetry of the incommensurately modulated phase requires a comprehensive investigation employing the superspace approach to disentangle the atomic modulation and uncover the degrees of freedom underlying the density-wave ordering.

\section*{\label{sec:la4ni3o10_experiment}Methods}

\subsection*{\label{sec:la4ni3o10_crystal_growth}%
Crystal growth}

There are two methods to synthesize single crystals of La$_4$Ni$_3$O$_{10}$. One is using the floating zone technique with an image furnace under oxygen pressure \cite{Zhang2020a} and the other one is by the flux method \cite{Li2024a}. We used the latter technique resulting in flattened cuboid black crystals of La$_4$Ni$_3$O$_{10}$, with long edges of approximately 50-100 microns.
La$_2$O$_3$ and Ni were ground together in stoichiometric ratio, then thoroughly
mixed with K$_2$CO$_3$ flux in the molar ratio Ni:K$_2$CO$_3$ 1:35. The mixture was
placed in an alumina crucible with a lid to prevent K$_2$CO$_3$ evaporation, then heated in a box furnace at 1050$^\circ$C for 48 hours, cooled slowly to
975$^{\circ}$C at 0.4$^{\circ}$C/h, and cooled with the furnace inertia to room temperature.
The remaining flux was then washed away with distilled water.
Powder X-ray diffraction data were collected with a Bruker D8-Endeavor with Cu K$\alpha$ radiation. The pattern of the pulverized reaction product shows
La$_4$Ni$_3$O$_{10}$ as the majority phase, along with NiO and La$_2$NiO$_4$ as side products of the reaction.

\subsection*{\label{sec:EPMA}%
Scanning Electron Microscope and Electron probe micro-analysis}
Morphology of grains were investigated with a FESEM Zeiss Ultra+ scanning electron microscope. Energy dispersive X-ray analysis on selected crystals was carried out using a Bruker SDD detector and Spirit software.
EPMA analysis were performed on a Jeol – 8800A electron microprobe equipped with five wavelength-dispersive spectrometers. Three of them were used to determine the chemical composition of grains: PETJ for Lanthanum (L$\alpha$), LiFH for Nickel (K$\alpha$), and LDE1H (W/Si) for oxygen (K$\alpha$). Analysis were made at 15 keV accelerating voltage, with a 30 nA probe current with 2 $\mu$m beam diameter. Similar conditions (counting times, current, samples and standards together in the chamber) were respected. KTiOPO$_4$ for Oxygen (40.4\%, LaB$_6$ for Lanthanum and pure Nickel were used as standards. Concentrations were calculated using a $\Phi$($\rho$z) procedure. Both techniques confirm the La:Ni 4:3 atomic ratio.

\subsection*{\label{sec:la4ni3o10_PXRD}%
Temperature dependent powder XRD}

High temperature synchrotron powder XRD data were measured at BM01 (Swiss-Norwegian Beamlines, SNBL) at ESRF in Grenoble, France, at a wavelength of $\lambda = 0.71913$ \AA{} and using a 2D PILATUS 2M detector \cite{Dyadkin:ie5157}. The carefully ground powder sample was filled into a 0.1 mm diameter quartz capillary and heated with a resistive capillary heater \cite{Marshall:ok5082}. Data were recorded every 2K from 290 to 1223 K in heating and cooling and Rietveld refinements were performed with the Fullprof software \cite{RODRIGUEZCARVAJAL199355}.

\subsection*{\label{sec:la4ni3o10_SXRD}%
Temperature and Pressure dependent single crystal XRD}
High-pressure experiments were performed at ID15b \cite{Garbarino2024-ID15B} ESRF Grenoble France, using  membrane-driven diamond-anvil cells (DACs) with 500 $\mu$m diamond culets employing
a radiation of a wavelength of $\lambda = 0.4099$ \AA{}.
Diffracted intensity was collected on  a EIGER 9M area detector
during continuous rotation of the crystal about
the $\omega$ axis, in frames of $0.5$ deg rotation
and 1 second exposure time.
Stainless steel gasket were used for the pressure chambers. The pressure transmitting medium used was Helium, loaded at 1.2 kbar, to ensure high hydrostatic pressure conditions up to the highest pressure reached in this study. The pressure was measured using the R1-line emission of a ruby ball placed close to the sample using the International Practical Pressure Scale IPPS-Ruby2020 equation of state~\cite{Shen2020}. The ruby signal is measured before and after each measurement in order to control the pressure drift during acquisitions. The recorded pressure is set at the average of these two pressure values and the uncertainty is set as the half of the difference between these two values. It is typically found smaller than the symbol size used for the figures in this paper. The homogeneity of the pressure in the DAC was followed from both the width and the splitting between the R1 and R2 ruby lines~\cite{Takemura2001, Dewaele2007}.

Under ambient pressure temperature-dependent single-crystal x-ray diffraction (SXRD) was measured
 at station EH1 of beamline P24 of PETRA-III extension
at DESY in Hamburg, Germany, employing radiation of
a wavelength of $\lambda = 0.5002$ \AA{}.
Diffracted X rays were detected by a Pilatus 1M CdTe detector.
The temperature of the sample was regulated with a
CRYOCOOL open-flow cryostat, employing helium
as cryogenic gas. Crystal of 50 microns
was selected for the SXRD experiment at beamline P24.
Diffracted intensity was collected on the detector
during continuous rotation of the crystal about
the $\phi$ axis, in frames of $0.1$ deg rotation
and 0.1 second exposure time.
Each run of data collection comprises a 10 times
repeated measurement of $3600$ frames,
corresponding to a total rotation
of the crystal by 360 deg, repeated 10 times.
These data were binned to 360 frames
of 1 deg of rotation and 10 seconds
exposure time employing the software
\texttt{addrunscbf} and \texttt{combcbf} \cite{paulmann2022a}.
Due to the anisotropic shape of the reflection the EVAL15 suite
\cite{schreursamm2010a} could not be used successfully. Instead,
data processing has been done with the EVAL14
\cite{duisenberga2003a} to integrate the data.
In EVAL14 the concept of ray tracing is not used,
as EVAL14 defines a box, where inside box there is a reflection and
outside the box no reflection. The border of the box is used
for evaluating the background level.
Integration now is simply adding up
intensities of pixels inside the box.
SADABS \cite{sheldrick2008} has been
used for the absorption correction for
all the data sets. JANA2006 \cite{petricekv2014a} was
used for the structure refinements at 230 and 80 K.
Preliminary single crystal XRD measurements
were also performed on ID28 in ESRF using the Pilatus 1M area
Si-detector without any attenuation
of the beam at 140 K and 80 K close to the wavelengths of DESY, where incommensurate satellites
are observed at 80 K.

\subsection*{\label{sec:la4ni3o10_Raman}%
Raman scattering}
Raman measurements were performed with a 532~nm solid-state laser with an incident laser power between 0.5-2 mW. A Trivista 777 spectrometer equipped with ultra-low noise, cryogenically-cooled PyLon CCDs was used in single-stage configurations for which we have access to Raman signal down to 70 cm$^{-1}$. A cryo-free cryostat \cite{zeman2025design} was used to perform measurements down to 2~K (26~K with laser heating included).    \\

\begin{table}[ht]
\centering
\caption{The Raman observable symmetries for a given scattering orientation of light with respect to the global symmetry of the lattice of La$_4$Ni$_3$O$_{10}$. X,Y and Z are taken in the standard setting of the space group $P2_1/c$. The Poynting vector is perpendicular to X, while the polarization Y and Z are in the (bc) plane but not aligned with a particular axis. Y is defined to be perpendicular to Z.}
\begin{ruledtabular}
\begin{tabular}{cc}
Porto notation & $P2_1/c$                              \\
X(YY)$\mathrm{\overline{X}}$         & 24 A$_{g}$       \\
X(YZ)$\mathrm{\overline{X}}$         & 24 B$_g$       \\
\end{tabular}
\label{tab:Symmetries}
\end{ruledtabular}
\end{table}

\subsection*{\textit{Ab initio} density-functional theory calculation}
The density-functional theory (DFT) crystal relaxation has been performed using the \texttt{ABINIT} code in the the PBE \cite{PBE} generalized-gradient approximation (GGA) with pseudopotentials from the PsuedoDojo table.
As explained in the main text, the calculation has been carried out in the lowest symmetry monoclinic $P2_1/c$ elementary and largest cell able to encompass all crystal structures up to the $I4/mmm$.
Within this cell the total energy was found converged within 1~mHa using a plane-waves cutoff of 100~Ryd and a Brillouin zone $k$-point sampling of 2$\times$4$\times$4 shifted by 1/2$\times$1/2$\times$1/2 (as explained in the crystallographic part, within the $P2_1/c$ setting the Ruddlesden-Popper (RP) stacking axis is oriented along $a$ and is doubled).
The relaxation was stopped when the forces and stresses reduced below a threshold of only 5$\cdot 10^{-6}$ Ha/Bohr: this was an absolutely critical parameter of the calculation to get the right phase diagram.
With just only one order of magnitude more, 5$\cdot 10^{-5}$ Ha/Bohr, the relaxation could have stopped in the tetragonal (or even in one of the orthorhombic) at low pressure, when the minimum of the total energy was on the $P2_1/c$, and vice-versa at highest pressure, depending on the starting crystal structure.
It was not even a question of relative minima: just only tiny forces/stresses between energetically very close structures.
Finally, the errors on lattice parameters have been estimated by calculating the stiffness tensor from the data out of the last four relaxation steps, and using its inverse, the compliance tensor, applied to the residual stress to evaluate the residual strain which is then taken as the lattice error estimate.
We did the same to estimate the error on the octahedra tilt angle by relying on elastic constants associated to the internal atomic positions.

\section*{Data availability}
Data is available upon reasonable request from the authors.

\begin{acknowledgments}
High pressure X-ray diffraction experiments was performed in the
beamline ID15b of the ESRF under the proposal HC5916.
Beamtime was allocated for proposal R-20250763 for which
we acknowledge DESY (Hamburg, Germany),
a member of the Helmholtz Association HGF, for
the provision of experimental facilities and
We thank Martin Tolkiehn, Preeti Porkiyal and
Heiko Schultz for their assistance with data
collection at beamline P24 of PETRA-III at DESY.
Computer time has been provided by GRICAD, project \texttt{mbqft}.
S. R. and P. R. thank P. Monceau, J. E. Lorenzo and M. Gibert for fruitful discussions.
V.~O.\ also thanks J.~Even, C.~Katan, A. Cano and Q. Meier for useful discussions on RP symmetries and their phase transitions.
\end{acknowledgments}

\section*{Funding statement}
We thank the support from the Agence Nationale de la Recherche under the project
SUPERNICKEL (Grant No.ANR-21-CE30-0041-04).
M.-A.M. and Y. G. thank the European Research Council (ERC)
under the European Union's Horizon 2020 research and innovation program (Grant Agreement No. 865826).

\section*{Author contributions}

E.P. and A.H.-A. synthesized the single crystals. V.O. carried out the DFT calculations. X-ray diffraction experiments were performed by S.R., E.P., G.G., M.D., O.P., A.P., D.V., D.C., L.N., C.P., J.B., A.B., S.vS., P.T. and P.R. EPMA was conducted by M.Q. and S.P. SXRD data analysis was carried out by S.R., while PXRD analysis was done by E.P. Raman measurements and analysis were done by Y.G. and M.A-M. The manuscript was written by S.R., Y.G., V.O., E.P., M.A-M. and P.R. with inputs from all authors.
The project was initiated by P.T. and P.R. and the work was supervised by P.R.. Fundings were acquired by A.P., P.T., M.A-M. and P.R..

\section*{Corresponding authors}
Correspondence to Sitaram Ramakrishnan (email address: sitaram.ramakrishnan@neel.cnrs.fr) and Pierre Rodi\`ere (email address: pierre.rodiere@neel.cnrs.fr).

\section*{Ethics declaration}
\subsection*{Competing interests}
The authors declare no competing interests.


\begin{thebibliography}{78}%
\makeatletter
\providecommand \@ifxundefined [1]{%
 \@ifx{#1\undefined}
}%
\providecommand \@ifnum [1]{%
 \ifnum #1\expandafter \@firstoftwo
 \else \expandafter \@secondoftwo
 \fi
}%
\providecommand \@ifx [1]{%
 \ifx #1\expandafter \@firstoftwo
 \else \expandafter \@secondoftwo
 \fi
}%
\providecommand \natexlab [1]{#1}%
\providecommand \enquote  [1]{``#1''}%
\providecommand \bibnamefont  [1]{#1}%
\providecommand \bibfnamefont [1]{#1}%
\providecommand \citenamefont [1]{#1}%
\providecommand \href@noop [0]{\@secondoftwo}%
\providecommand \href [0]{\begingroup \@sanitize@url \@href}%
\providecommand \@href[1]{\@@startlink{#1}\@@href}%
\providecommand \@@href[1]{\endgroup#1\@@endlink}%
\providecommand \@sanitize@url [0]{\catcode `\\12\catcode `\$12\catcode
  `\&12\catcode `\#12\catcode `\^12\catcode `\_12\catcode `\%12\relax}%
\providecommand \@@startlink[1]{}%
\providecommand \@@endlink[0]{}%
\providecommand \url  [0]{\begingroup\@sanitize@url \@url }%
\providecommand \@url [1]{\endgroup\@href {#1}{\urlprefix }}%
\providecommand \urlprefix  [0]{URL }%
\providecommand \Eprint [0]{\href }%
\providecommand \doibase [0]{https://doi.org/}%
\providecommand \selectlanguage [0]{\@gobble}%
\providecommand \bibinfo  [0]{\@secondoftwo}%
\providecommand \bibfield  [0]{\@secondoftwo}%
\providecommand \translation [1]{[#1]}%
\providecommand \BibitemOpen [0]{}%
\providecommand \bibitemStop [0]{}%
\providecommand \bibitemNoStop [0]{.\EOS\space}%
\providecommand \EOS [0]{\spacefactor3000\relax}%
\providecommand \BibitemShut  [1]{\csname bibitem#1\endcsname}%
\let\auto@bib@innerbib\@empty
\bibitem [{\citenamefont {{Chu, C. W. and Hor, P. H. and Meng, R. L. and Gao,
  L. and Huang, Z. J. and Wang, and Y. Q.}}(1987)}]{Chu1987a}%
  \BibitemOpen
  \bibfield  {author} {\bibinfo {author} {\bibnamefont {{Chu, C. W. and Hor, P.
  H. and Meng, R. L. and Gao, L. and Huang, Z. J. and Wang, and Y. Q.}}},\
  }\bibfield  {title} {\bibinfo {title} {Evidence for superconductivity above
  40 {K} in the {La-Ba-Cu-O} compound system},\ }\href@noop {} {\bibfield
  {journal} {\bibinfo  {journal} {Phys. Rev. Lett.}\ }\textbf {\bibinfo
  {volume} {58}},\ \bibinfo {pages} {405} (\bibinfo {year} {1987})}\BibitemShut
  {NoStop}%
\bibitem [{\citenamefont {{H. Takagi, H. Eisaki, S. Uchida, A. Maeda, S.
  Tajima, K. Uchinokura and S. Tanaka}}(1988)}]{Takagi1988a}%
  \BibitemOpen
  \bibfield  {author} {\bibinfo {author} {\bibnamefont {{H. Takagi, H. Eisaki,
  S. Uchida, A. Maeda, S. Tajima, K. Uchinokura and S. Tanaka}}},\ }\bibfield
  {title} {\bibinfo {title} {{Transport and optical studies of single crystals
  of the 80-K Bi–Sr–Ca–Cu–O superconductor}},\ }\href@noop {}
  {\bibfield  {journal} {\bibinfo  {journal} {Nature}\ }\textbf {\bibinfo
  {volume} {332}},\ \bibinfo {pages} {236–238} (\bibinfo {year}
  {1988})}\BibitemShut {NoStop}%
\bibitem [{\citenamefont {{B. Keimer, S. A. Kivelson, M. R. Norman, S. Uchida
  and J. Zaanen}}(2015)}]{Keimer2015a}%
  \BibitemOpen
  \bibfield  {author} {\bibinfo {author} {\bibnamefont {{B. Keimer, S. A.
  Kivelson, M. R. Norman, S. Uchida and J. Zaanen}}},\ }\bibfield  {title}
  {\bibinfo {title} {{From quantum matter to high-temperature superconductivity
  in copper oxides}},\ }\href@noop {} {\bibfield  {journal} {\bibinfo
  {journal} {Nature}\ }\textbf {\bibinfo {volume} {518}},\ \bibinfo {pages}
  {179–186} (\bibinfo {year} {2015})}\BibitemShut {NoStop}%
\bibitem [{\citenamefont {Tranquada}\ \emph {et~al.}(1994)\citenamefont
  {Tranquada}, \citenamefont {Buttrey}, \citenamefont {Sachan},\ and\
  \citenamefont {Lorenzo}}]{tranquada1994a}%
  \BibitemOpen
  \bibfield  {author} {\bibinfo {author} {\bibfnamefont {J.~M.}\ \bibnamefont
  {Tranquada}}, \bibinfo {author} {\bibfnamefont {D.~J.}\ \bibnamefont
  {Buttrey}}, \bibinfo {author} {\bibfnamefont {V.}~\bibnamefont {Sachan}},\
  and\ \bibinfo {author} {\bibfnamefont {J.~E.}\ \bibnamefont {Lorenzo}},\
  }\bibfield  {title} {\bibinfo {title} {Simultaneous ordering of holes and
  spins in {${\mathrm{La}}_{2}$Ni${\mathrm{O}}_{4.125}$}},\ }\href@noop {}
  {\bibfield  {journal} {\bibinfo  {journal} {Phys. Rev. Lett.}\ }\textbf
  {\bibinfo {volume} {73}},\ \bibinfo {pages} {1003} (\bibinfo {year}
  {1994})}\BibitemShut {NoStop}%
\bibitem [{\citenamefont {Zachar}\ \emph {et~al.}(1998)\citenamefont {Zachar},
  \citenamefont {Kivelson},\ and\ \citenamefont {Emery}}]{Kivelson1998a}%
  \BibitemOpen
  \bibfield  {author} {\bibinfo {author} {\bibfnamefont {O.}~\bibnamefont
  {Zachar}}, \bibinfo {author} {\bibfnamefont {S.~A.}\ \bibnamefont
  {Kivelson}},\ and\ \bibinfo {author} {\bibfnamefont {V.~J.}\ \bibnamefont
  {Emery}},\ }\bibfield  {title} {\bibinfo {title} {Landau theory of stripe
  phases in cuprates and nickelates},\ }\href@noop {} {\bibfield  {journal}
  {\bibinfo  {journal} {Phys. Rev. B}\ }\textbf {\bibinfo {volume} {57}},\
  \bibinfo {pages} {1422} (\bibinfo {year} {1998})}\BibitemShut {NoStop}%
\bibitem [{\citenamefont {Zhang}\ \emph {et~al.}(2016)\citenamefont {Zhang},
  \citenamefont {Chen}, \citenamefont {Phelan}, \citenamefont {Zheng},
  \citenamefont {Norman},\ and\ \citenamefont {Mitchell}}]{Zhang2016d}%
  \BibitemOpen
  \bibfield  {author} {\bibinfo {author} {\bibfnamefont {J.}~\bibnamefont
  {Zhang}}, \bibinfo {author} {\bibfnamefont {Y.-S.}\ \bibnamefont {Chen}},
  \bibinfo {author} {\bibfnamefont {D.}~\bibnamefont {Phelan}}, \bibinfo
  {author} {\bibfnamefont {H.}~\bibnamefont {Zheng}}, \bibinfo {author}
  {\bibfnamefont {M.~R.}\ \bibnamefont {Norman}},\ and\ \bibinfo {author}
  {\bibfnamefont {J.~F.}\ \bibnamefont {Mitchell}},\ }\bibfield  {title}
  {\bibinfo {title} {Stacked charge stripes in the quasi-2d trilayer nickelate
  {La$_4$Ni$_3$O$_8$}},\ }\href {https://doi.org/10.1073/pnas.1606637113}
  {\bibfield  {journal} {\bibinfo  {journal} {Proceedings of the National
  Academy of Sciences}\ }\textbf {\bibinfo {volume} {113}},\ \bibinfo {pages}
  {8945–8950} (\bibinfo {year} {2016})}\BibitemShut {NoStop}%
\bibitem [{\citenamefont {Wang}\ \emph
  {et~al.}(2024{\natexlab{a}})\citenamefont {Wang}, \citenamefont {Lee},\ and\
  \citenamefont {Goodge}}]{wang2024f}%
  \BibitemOpen
  \bibfield  {author} {\bibinfo {author} {\bibfnamefont {B.~Y.}\ \bibnamefont
  {Wang}}, \bibinfo {author} {\bibfnamefont {K.}~\bibnamefont {Lee}},\ and\
  \bibinfo {author} {\bibfnamefont {B.~H.}\ \bibnamefont {Goodge}},\ }\bibfield
   {title} {\bibinfo {title} {Experimental progress in superconducting
  nickelates},\ }\href
  {https://doi.org/https://doi.org/10.1146/annurev-conmatphys-032922-093307}
  {\bibfield  {journal} {\bibinfo  {journal} {Annual Review of Condensed Matter
  Physics}\ }\textbf {\bibinfo {volume} {15}},\ \bibinfo {pages} {305}
  (\bibinfo {year} {2024}{\natexlab{a}})}\BibitemShut {NoStop}%
\bibitem [{\citenamefont {Wang}\ \emph
  {et~al.}(2025{\natexlab{a}})\citenamefont {Wang}, \citenamefont {Jiang},
  \citenamefont {Ying}, \citenamefont {Wu}, \citenamefont {Cheng},
  \citenamefont {Hu},\ and\ \citenamefont {Chen}}]{wang2025h}%
  \BibitemOpen
  \bibfield  {author} {\bibinfo {author} {\bibfnamefont {Y.}~\bibnamefont
  {Wang}}, \bibinfo {author} {\bibfnamefont {K.}~\bibnamefont {Jiang}},
  \bibinfo {author} {\bibfnamefont {J.}~\bibnamefont {Ying}}, \bibinfo {author}
  {\bibfnamefont {T.}~\bibnamefont {Wu}}, \bibinfo {author} {\bibfnamefont
  {J.}~\bibnamefont {Cheng}}, \bibinfo {author} {\bibfnamefont
  {J.}~\bibnamefont {Hu}},\ and\ \bibinfo {author} {\bibfnamefont
  {X.}~\bibnamefont {Chen}},\ }\bibfield  {title} {\bibinfo {title} {Recent
  progress in nickelate superconductors},\ }\href@noop {} {\bibfield  {journal}
  {\bibinfo  {journal} {National Science Review}\ }\textbf {\bibinfo {volume}
  {12}},\ \bibinfo {pages} {nwaf373} (\bibinfo {year}
  {2025}{\natexlab{a}})}\BibitemShut {NoStop}%
\bibitem [{\citenamefont {{D. Li, K. Lee, B. Y. Wang, M. Osada, S. Crossley, H.
  R. Lee, Y. Cui, Y. Hikita and Harold Y. Hwang}}(2019)}]{Li2019a}%
  \BibitemOpen
  \bibfield  {author} {\bibinfo {author} {\bibnamefont {{D. Li, K. Lee, B. Y.
  Wang, M. Osada, S. Crossley, H. R. Lee, Y. Cui, Y. Hikita and Harold Y.
  Hwang}}},\ }\bibfield  {title} {\bibinfo {title} {{Superconductivity in an
  infinite-layer nickelate}},\ }\href@noop {} {\bibfield  {journal} {\bibinfo
  {journal} {Nature}\ }\textbf {\bibinfo {volume} {572}},\ \bibinfo {pages}
  {624–627} (\bibinfo {year} {2019})}\BibitemShut {NoStop}%
\bibitem [{\citenamefont {Li}\ \emph {et~al.}(2020{\natexlab{a}})\citenamefont
  {Li}, \citenamefont {Wang}, \citenamefont {Lee}, \citenamefont {Harvey},
  \citenamefont {Osada}, \citenamefont {Goodge}, \citenamefont {Kourkoutis},\
  and\ \citenamefont {Hwang}}]{Li2020a}%
  \BibitemOpen
  \bibfield  {author} {\bibinfo {author} {\bibfnamefont {D.}~\bibnamefont
  {Li}}, \bibinfo {author} {\bibfnamefont {B.~Y.}\ \bibnamefont {Wang}},
  \bibinfo {author} {\bibfnamefont {K.}~\bibnamefont {Lee}}, \bibinfo {author}
  {\bibfnamefont {S.~P.}\ \bibnamefont {Harvey}}, \bibinfo {author}
  {\bibfnamefont {M.}~\bibnamefont {Osada}}, \bibinfo {author} {\bibfnamefont
  {B.~H.}\ \bibnamefont {Goodge}}, \bibinfo {author} {\bibfnamefont {L.~F.}\
  \bibnamefont {Kourkoutis}},\ and\ \bibinfo {author} {\bibfnamefont {H.~Y.}\
  \bibnamefont {Hwang}},\ }\bibfield  {title} {\bibinfo {title}
  {Superconducting dome in {Nd$_{1-x}$Sr$_{x}$NiO$_2$} infinite layer films},\
  }\href@noop {} {\bibfield  {journal} {\bibinfo  {journal} {Phys. Rev. Lett.}\
  }\textbf {\bibinfo {volume} {125}},\ \bibinfo {pages} {027001} (\bibinfo
  {year} {2020}{\natexlab{a}})}\BibitemShut {NoStop}%
\bibitem [{\citenamefont {Zeng}\ \emph {et~al.}(2020)\citenamefont {Zeng},
  \citenamefont {Tang}, \citenamefont {Yin}, \citenamefont {Li}, \citenamefont
  {Li}, \citenamefont {Huang}, \citenamefont {Hu}, \citenamefont {Liu},
  \citenamefont {Omar}, \citenamefont {Jani}, \citenamefont {Lim},
  \citenamefont {Han}, \citenamefont {Wan}, \citenamefont {Yang}, \citenamefont
  {Pennycook}, \citenamefont {Wee},\ and\ \citenamefont {Ariando}}]{Zeng2020m}%
  \BibitemOpen
  \bibfield  {author} {\bibinfo {author} {\bibfnamefont {S.}~\bibnamefont
  {Zeng}}, \bibinfo {author} {\bibfnamefont {C.~S.}\ \bibnamefont {Tang}},
  \bibinfo {author} {\bibfnamefont {X.}~\bibnamefont {Yin}}, \bibinfo {author}
  {\bibfnamefont {C.}~\bibnamefont {Li}}, \bibinfo {author} {\bibfnamefont
  {M.}~\bibnamefont {Li}}, \bibinfo {author} {\bibfnamefont {Z.}~\bibnamefont
  {Huang}}, \bibinfo {author} {\bibfnamefont {J.}~\bibnamefont {Hu}}, \bibinfo
  {author} {\bibfnamefont {W.}~\bibnamefont {Liu}}, \bibinfo {author}
  {\bibfnamefont {G.~J.}\ \bibnamefont {Omar}}, \bibinfo {author}
  {\bibfnamefont {H.}~\bibnamefont {Jani}}, \bibinfo {author} {\bibfnamefont
  {Z.~S.}\ \bibnamefont {Lim}}, \bibinfo {author} {\bibfnamefont
  {K.}~\bibnamefont {Han}}, \bibinfo {author} {\bibfnamefont {D.}~\bibnamefont
  {Wan}}, \bibinfo {author} {\bibfnamefont {P.}~\bibnamefont {Yang}}, \bibinfo
  {author} {\bibfnamefont {S.~J.}\ \bibnamefont {Pennycook}}, \bibinfo {author}
  {\bibfnamefont {A.~T.~S.}\ \bibnamefont {Wee}},\ and\ \bibinfo {author}
  {\bibfnamefont {A.}~\bibnamefont {Ariando}},\ }\bibfield  {title} {\bibinfo
  {title} {Phase diagram and superconducting dome of infinite-layer
  {${\mathrm{Nd}}_{1\ensuremath{-}x}{\mathrm{Sr}}_{x}{\mathrm{NiO}}_{2}$} thin
  films},\ }\href@noop {} {\bibfield  {journal} {\bibinfo  {journal} {Phys.
  Rev. Lett.}\ }\textbf {\bibinfo {volume} {125}},\ \bibinfo {pages} {147003}
  (\bibinfo {year} {2020})}\BibitemShut {NoStop}%
\bibitem [{\citenamefont {Li}\ \emph {et~al.}(2020{\natexlab{b}})\citenamefont
  {Li}, \citenamefont {He}, \citenamefont {Si}, \citenamefont {Zhu},
  \citenamefont {Zhang},\ and\ \citenamefont {Wen}}]{Li2020jk}%
  \BibitemOpen
  \bibfield  {author} {\bibinfo {author} {\bibfnamefont {Q.}~\bibnamefont
  {Li}}, \bibinfo {author} {\bibfnamefont {C.}~\bibnamefont {He}}, \bibinfo
  {author} {\bibfnamefont {J.}~\bibnamefont {Si}}, \bibinfo {author}
  {\bibfnamefont {X.}~\bibnamefont {Zhu}}, \bibinfo {author} {\bibfnamefont
  {Y.}~\bibnamefont {Zhang}},\ and\ \bibinfo {author} {\bibfnamefont {H.-H.}\
  \bibnamefont {Wen}},\ }\bibfield  {title} {\bibinfo {title} {Absence of
  superconductivity in bulk {Nd$_{1-x}$Sr$_x$NiO$_2$}},\ }\href@noop {}
  {\bibfield  {journal} {\bibinfo  {journal} {Communications Materials}\
  }\textbf {\bibinfo {volume} {1}},\ \bibinfo {pages} {16} (\bibinfo {year}
  {2020}{\natexlab{b}})}\BibitemShut {NoStop}%
\bibitem [{\citenamefont {Suyolcu}\ \emph {et~al.}(2025)\citenamefont
  {Suyolcu}, \citenamefont {Puphal},\ and\ \citenamefont
  {Hepting}}]{Suyolcu2025}%
  \BibitemOpen
  \bibfield  {author} {\bibinfo {author} {\bibfnamefont {Y.~E.}\ \bibnamefont
  {Suyolcu}}, \bibinfo {author} {\bibfnamefont {P.}~\bibnamefont {Puphal}},\
  and\ \bibinfo {author} {\bibfnamefont {M.}~\bibnamefont {Hepting}},\
  }\bibfield  {title} {\bibinfo {title} {Three generations of infinite-layer
  nickelate crystals},\ }\href {https://doi.org/10.1557/s43579-025-00689-x}
  {\bibfield  {journal} {\bibinfo  {journal} {MRS Communications}\ }\textbf
  {\bibinfo {volume} {15}},\ \bibinfo {pages} {169} (\bibinfo {year}
  {2025})}\BibitemShut {NoStop}%
\bibitem [{\citenamefont {{H. Sun, M. Huo, X. Hu, J. Li, Z. Liu, Y. Han, L.
  Tang, Z. Mao, P. Yang, B. Wang, J. Cheng, D. Yao, G.-M. Zhang and M.
  Wang}}(2023)}]{Sun2023a}%
  \BibitemOpen
  \bibfield  {author} {\bibinfo {author} {\bibnamefont {{H. Sun, M. Huo, X. Hu,
  J. Li, Z. Liu, Y. Han, L. Tang, Z. Mao, P. Yang, B. Wang, J. Cheng, D. Yao,
  G.-M. Zhang and M. Wang}}},\ }\bibfield  {title} {\bibinfo {title}
  {{Signatures of superconductivity near 80 K in a nickelate under high
  pressure}},\ }\href@noop {} {\bibfield  {journal} {\bibinfo  {journal}
  {Nature}\ }\textbf {\bibinfo {volume} {621}},\ \bibinfo {pages} {493–498}
  (\bibinfo {year} {2023})}\BibitemShut {NoStop}%
\bibitem [{\citenamefont {Hou}\ \emph {et~al.}(2023)\citenamefont {Hou},
  \citenamefont {Yang}, \citenamefont {Liu}, \citenamefont {Li}, \citenamefont
  {Shan}, \citenamefont {Ma}, \citenamefont {Wang}, \citenamefont {Wang},
  \citenamefont {Guo}, \citenamefont {Sun}, \citenamefont {Uwatoko},
  \citenamefont {Wang}, \citenamefont {Zhang}, \citenamefont {Wang},\ and\
  \citenamefont {Cheng}}]{hou2023a}%
  \BibitemOpen
  \bibfield  {author} {\bibinfo {author} {\bibfnamefont {J.}~\bibnamefont
  {Hou}}, \bibinfo {author} {\bibfnamefont {P.-T.}\ \bibnamefont {Yang}},
  \bibinfo {author} {\bibfnamefont {Z.-Y.}\ \bibnamefont {Liu}}, \bibinfo
  {author} {\bibfnamefont {J.-Y.}\ \bibnamefont {Li}}, \bibinfo {author}
  {\bibfnamefont {P.-F.}\ \bibnamefont {Shan}}, \bibinfo {author}
  {\bibfnamefont {L.}~\bibnamefont {Ma}}, \bibinfo {author} {\bibfnamefont
  {G.}~\bibnamefont {Wang}}, \bibinfo {author} {\bibfnamefont {N.-N.}\
  \bibnamefont {Wang}}, \bibinfo {author} {\bibfnamefont {H.-Z.}\ \bibnamefont
  {Guo}}, \bibinfo {author} {\bibfnamefont {J.-P.}\ \bibnamefont {Sun}},
  \bibinfo {author} {\bibfnamefont {Y.}~\bibnamefont {Uwatoko}}, \bibinfo
  {author} {\bibfnamefont {M.}~\bibnamefont {Wang}}, \bibinfo {author}
  {\bibfnamefont {G.-M.}\ \bibnamefont {Zhang}}, \bibinfo {author}
  {\bibfnamefont {B.-S.}\ \bibnamefont {Wang}},\ and\ \bibinfo {author}
  {\bibfnamefont {J.-G.}\ \bibnamefont {Cheng}},\ }\bibfield  {title} {\bibinfo
  {title} {Emergence of high-temperature superconducting phase in pressurized
  {La$_3$Ni$_2$O$_7$} crystals},\ }\href@noop {} {\bibfield  {journal}
  {\bibinfo  {journal} {Chinese Physics Letters}\ }\textbf {\bibinfo {volume}
  {40}},\ \bibinfo {pages} {117302} (\bibinfo {year} {2023})}\BibitemShut
  {NoStop}%
\bibitem [{\citenamefont {Zhang}\ \emph
  {et~al.}(2024{\natexlab{a}})\citenamefont {Zhang}, \citenamefont {Su},
  \citenamefont {Huang}, \citenamefont {Shan}, \citenamefont {Sun},
  \citenamefont {Huo}, \citenamefont {Ye}, \citenamefont {Zhang}, \citenamefont
  {Yang}, \citenamefont {Xu}, \citenamefont {Su}, \citenamefont {Li},
  \citenamefont {Smidman}, \citenamefont {Wang}, \citenamefont {Jiao},\ and\
  \citenamefont {Yuan}}]{Zhang2024b}%
  \BibitemOpen
  \bibfield  {author} {\bibinfo {author} {\bibfnamefont {Y.}~\bibnamefont
  {Zhang}}, \bibinfo {author} {\bibfnamefont {D.}~\bibnamefont {Su}}, \bibinfo
  {author} {\bibfnamefont {Y.}~\bibnamefont {Huang}}, \bibinfo {author}
  {\bibfnamefont {Z.}~\bibnamefont {Shan}}, \bibinfo {author} {\bibfnamefont
  {H.}~\bibnamefont {Sun}}, \bibinfo {author} {\bibfnamefont {M.}~\bibnamefont
  {Huo}}, \bibinfo {author} {\bibfnamefont {K.}~\bibnamefont {Ye}}, \bibinfo
  {author} {\bibfnamefont {J.}~\bibnamefont {Zhang}}, \bibinfo {author}
  {\bibfnamefont {Z.}~\bibnamefont {Yang}}, \bibinfo {author} {\bibfnamefont
  {Y.}~\bibnamefont {Xu}}, \bibinfo {author} {\bibfnamefont {Y.}~\bibnamefont
  {Su}}, \bibinfo {author} {\bibfnamefont {R.}~\bibnamefont {Li}}, \bibinfo
  {author} {\bibfnamefont {M.}~\bibnamefont {Smidman}}, \bibinfo {author}
  {\bibfnamefont {M.}~\bibnamefont {Wang}}, \bibinfo {author} {\bibfnamefont
  {L.}~\bibnamefont {Jiao}},\ and\ \bibinfo {author} {\bibfnamefont
  {H.}~\bibnamefont {Yuan}},\ }\bibfield  {title} {\bibinfo {title}
  {High-temperature superconductivity with zero resistance and strange-metal
  behaviour in {La$_3$Ni$_2$O$_{7-\delta}$}},\ }\href@noop {} {\bibfield
  {journal} {\bibinfo  {journal} {Nature Physics}\ }\textbf {\bibinfo {volume}
  {20}},\ \bibinfo {pages} {1269} (\bibinfo {year}
  {2024}{\natexlab{a}})}\BibitemShut {NoStop}%
\bibitem [{\citenamefont {Wang}\ \emph
  {et~al.}(2025{\natexlab{b}})\citenamefont {Wang}, \citenamefont {Wang},
  \citenamefont {Lu}, \citenamefont {Calder}, \citenamefont {Yan},
  \citenamefont {Shi}, \citenamefont {Hou}, \citenamefont {Ma}, \citenamefont
  {Zhang}, \citenamefont {Sun}, \citenamefont {Wang}, \citenamefont {Meng},
  \citenamefont {Liu},\ and\ \citenamefont {Cheng}}]{Wang2025zx}%
  \BibitemOpen
  \bibfield  {author} {\bibinfo {author} {\bibfnamefont {G.}~\bibnamefont
  {Wang}}, \bibinfo {author} {\bibfnamefont {N.}~\bibnamefont {Wang}}, \bibinfo
  {author} {\bibfnamefont {T.}~\bibnamefont {Lu}}, \bibinfo {author}
  {\bibfnamefont {S.}~\bibnamefont {Calder}}, \bibinfo {author} {\bibfnamefont
  {J.}~\bibnamefont {Yan}}, \bibinfo {author} {\bibfnamefont {L.}~\bibnamefont
  {Shi}}, \bibinfo {author} {\bibfnamefont {J.}~\bibnamefont {Hou}}, \bibinfo
  {author} {\bibfnamefont {L.}~\bibnamefont {Ma}}, \bibinfo {author}
  {\bibfnamefont {L.}~\bibnamefont {Zhang}}, \bibinfo {author} {\bibfnamefont
  {J.}~\bibnamefont {Sun}}, \bibinfo {author} {\bibfnamefont {B.}~\bibnamefont
  {Wang}}, \bibinfo {author} {\bibfnamefont {S.}~\bibnamefont {Meng}}, \bibinfo
  {author} {\bibfnamefont {M.}~\bibnamefont {Liu}},\ and\ \bibinfo {author}
  {\bibfnamefont {J.}~\bibnamefont {Cheng}},\ }\bibfield  {title} {\bibinfo
  {title} {Chemical versus physical pressure effects on the structure
  transition of bilayer nickelates},\ }\href@noop {} {\bibfield  {journal}
  {\bibinfo  {journal} {npj Quantum Materials}\ }\textbf {\bibinfo {volume}
  {10}},\ \bibinfo {pages} {1} (\bibinfo {year}
  {2025}{\natexlab{b}})}\BibitemShut {NoStop}%
\bibitem [{\citenamefont {Wang}\ \emph
  {et~al.}(2024{\natexlab{b}})\citenamefont {Wang}, \citenamefont {Wang},
  \citenamefont {Shen}, \citenamefont {Hou}, \citenamefont {Ma}, \citenamefont
  {Shi}, \citenamefont {Ren}, \citenamefont {Gu}, \citenamefont {Ma},
  \citenamefont {Yang}, \citenamefont {Liu}, \citenamefont {Guo}, \citenamefont
  {Sun}, \citenamefont {Zhang}, \citenamefont {Calder}, \citenamefont {Yan},
  \citenamefont {Wang}, \citenamefont {Uwatoko},\ and\ \citenamefont
  {Cheng}}]{wang2024a}%
  \BibitemOpen
  \bibfield  {author} {\bibinfo {author} {\bibfnamefont {G.}~\bibnamefont
  {Wang}}, \bibinfo {author} {\bibfnamefont {N.~N.}\ \bibnamefont {Wang}},
  \bibinfo {author} {\bibfnamefont {X.~L.}\ \bibnamefont {Shen}}, \bibinfo
  {author} {\bibfnamefont {J.}~\bibnamefont {Hou}}, \bibinfo {author}
  {\bibfnamefont {L.}~\bibnamefont {Ma}}, \bibinfo {author} {\bibfnamefont
  {L.~F.}\ \bibnamefont {Shi}}, \bibinfo {author} {\bibfnamefont {Z.~A.}\
  \bibnamefont {Ren}}, \bibinfo {author} {\bibfnamefont {Y.~D.}\ \bibnamefont
  {Gu}}, \bibinfo {author} {\bibfnamefont {H.~M.}\ \bibnamefont {Ma}}, \bibinfo
  {author} {\bibfnamefont {P.~T.}\ \bibnamefont {Yang}}, \bibinfo {author}
  {\bibfnamefont {Z.~Y.}\ \bibnamefont {Liu}}, \bibinfo {author} {\bibfnamefont
  {H.~Z.}\ \bibnamefont {Guo}}, \bibinfo {author} {\bibfnamefont {J.~P.}\
  \bibnamefont {Sun}}, \bibinfo {author} {\bibfnamefont {G.~M.}\ \bibnamefont
  {Zhang}}, \bibinfo {author} {\bibfnamefont {S.}~\bibnamefont {Calder}},
  \bibinfo {author} {\bibfnamefont {J.-Q.}\ \bibnamefont {Yan}}, \bibinfo
  {author} {\bibfnamefont {B.~S.}\ \bibnamefont {Wang}}, \bibinfo {author}
  {\bibfnamefont {Y.}~\bibnamefont {Uwatoko}},\ and\ \bibinfo {author}
  {\bibfnamefont {J.-G.}\ \bibnamefont {Cheng}},\ }\bibfield  {title} {\bibinfo
  {title} {Pressure-induced superconductivity in polycrystalline
  {La$_3$Ni$_2$O$_{7-\delta}$}},\ }\href@noop {} {\bibfield  {journal}
  {\bibinfo  {journal} {Phys. Rev. X}\ }\textbf {\bibinfo {volume} {14}},\
  \bibinfo {pages} {011040} (\bibinfo {year} {2024}{\natexlab{b}})}\BibitemShut
  {NoStop}%
\bibitem [{\citenamefont {Wang}\ \emph
  {et~al.}(2024{\natexlab{c}})\citenamefont {Wang}, \citenamefont {Li},
  \citenamefont {Xie}, \citenamefont {Liu}, \citenamefont {Sun}, \citenamefont
  {Huang}, \citenamefont {Gao}, \citenamefont {Nakagawa}, \citenamefont {Fu},
  \citenamefont {Dong}, \citenamefont {Cao}, \citenamefont {Yu}, \citenamefont
  {Kawaguchi}, \citenamefont {Kadobayashi}, \citenamefont {Wang}, \citenamefont
  {Jin}, \citenamefont {Mao},\ and\ \citenamefont {Liu}}]{wang2024au}%
  \BibitemOpen
  \bibfield  {author} {\bibinfo {author} {\bibfnamefont {L.}~\bibnamefont
  {Wang}}, \bibinfo {author} {\bibfnamefont {Y.}~\bibnamefont {Li}}, \bibinfo
  {author} {\bibfnamefont {S.-Y.}\ \bibnamefont {Xie}}, \bibinfo {author}
  {\bibfnamefont {F.}~\bibnamefont {Liu}}, \bibinfo {author} {\bibfnamefont
  {H.}~\bibnamefont {Sun}}, \bibinfo {author} {\bibfnamefont {C.}~\bibnamefont
  {Huang}}, \bibinfo {author} {\bibfnamefont {Y.}~\bibnamefont {Gao}}, \bibinfo
  {author} {\bibfnamefont {T.}~\bibnamefont {Nakagawa}}, \bibinfo {author}
  {\bibfnamefont {B.}~\bibnamefont {Fu}}, \bibinfo {author} {\bibfnamefont
  {B.}~\bibnamefont {Dong}}, \bibinfo {author} {\bibfnamefont {Z.}~\bibnamefont
  {Cao}}, \bibinfo {author} {\bibfnamefont {R.}~\bibnamefont {Yu}}, \bibinfo
  {author} {\bibfnamefont {S.~I.}\ \bibnamefont {Kawaguchi}}, \bibinfo {author}
  {\bibfnamefont {H.}~\bibnamefont {Kadobayashi}}, \bibinfo {author}
  {\bibfnamefont {M.}~\bibnamefont {Wang}}, \bibinfo {author} {\bibfnamefont
  {C.}~\bibnamefont {Jin}}, \bibinfo {author} {\bibfnamefont {H.-k.}\
  \bibnamefont {Mao}},\ and\ \bibinfo {author} {\bibfnamefont {H.}~\bibnamefont
  {Liu}},\ }\bibfield  {title} {\bibinfo {title} {Structure responsible for the
  superconducting state in {La$_3$Ni$_2$O$_7$} at high-pressure and
  low-temperature conditions},\ }\href@noop {} {\bibfield  {journal} {\bibinfo
  {journal} {Journal of the American Chemical Society}\ }\textbf {\bibinfo
  {volume} {146}},\ \bibinfo {pages} {7506} (\bibinfo {year}
  {2024}{\natexlab{c}})}\BibitemShut {NoStop}%
\bibitem [{\citenamefont {Zhu}\ \emph {et~al.}(2024)\citenamefont {Zhu},
  \citenamefont {Peng}, \citenamefont {Zhang}, \citenamefont {Pan},
  \citenamefont {Chen}, \citenamefont {Chen}, \citenamefont {Ren},
  \citenamefont {Liu}, \citenamefont {Hao}, \citenamefont {Li}, \citenamefont
  {Xing}, \citenamefont {Lan}, \citenamefont {Han}, \citenamefont {Wang},
  \citenamefont {Jia}, \citenamefont {Wo}, \citenamefont {Gu}, \citenamefont
  {Gu}, \citenamefont {Ji}, \citenamefont {Wang}, \citenamefont {Gou},
  \citenamefont {Shen}, \citenamefont {Ying}, \citenamefont {Chen},
  \citenamefont {Yang}, \citenamefont {Cao}, \citenamefont {Zheng},
  \citenamefont {Zeng}, \citenamefont {Guo},\ and\ \citenamefont
  {Zhao}}]{Zhu2024a}%
  \BibitemOpen
  \bibfield  {author} {\bibinfo {author} {\bibfnamefont {Y.}~\bibnamefont
  {Zhu}}, \bibinfo {author} {\bibfnamefont {D.}~\bibnamefont {Peng}}, \bibinfo
  {author} {\bibfnamefont {E.}~\bibnamefont {Zhang}}, \bibinfo {author}
  {\bibfnamefont {B.}~\bibnamefont {Pan}}, \bibinfo {author} {\bibfnamefont
  {X.}~\bibnamefont {Chen}}, \bibinfo {author} {\bibfnamefont {L.}~\bibnamefont
  {Chen}}, \bibinfo {author} {\bibfnamefont {H.}~\bibnamefont {Ren}}, \bibinfo
  {author} {\bibfnamefont {F.}~\bibnamefont {Liu}}, \bibinfo {author}
  {\bibfnamefont {Y.}~\bibnamefont {Hao}}, \bibinfo {author} {\bibfnamefont
  {N.}~\bibnamefont {Li}}, \bibinfo {author} {\bibfnamefont {Z.}~\bibnamefont
  {Xing}}, \bibinfo {author} {\bibfnamefont {F.}~\bibnamefont {Lan}}, \bibinfo
  {author} {\bibfnamefont {J.}~\bibnamefont {Han}}, \bibinfo {author}
  {\bibfnamefont {J.}~\bibnamefont {Wang}}, \bibinfo {author} {\bibfnamefont
  {D.}~\bibnamefont {Jia}}, \bibinfo {author} {\bibfnamefont {H.}~\bibnamefont
  {Wo}}, \bibinfo {author} {\bibfnamefont {Y.}~\bibnamefont {Gu}}, \bibinfo
  {author} {\bibfnamefont {Y.}~\bibnamefont {Gu}}, \bibinfo {author}
  {\bibfnamefont {L.}~\bibnamefont {Ji}}, \bibinfo {author} {\bibfnamefont
  {W.}~\bibnamefont {Wang}}, \bibinfo {author} {\bibfnamefont {H.}~\bibnamefont
  {Gou}}, \bibinfo {author} {\bibfnamefont {Y.}~\bibnamefont {Shen}}, \bibinfo
  {author} {\bibfnamefont {T.}~\bibnamefont {Ying}}, \bibinfo {author}
  {\bibfnamefont {X.}~\bibnamefont {Chen}}, \bibinfo {author} {\bibfnamefont
  {W.}~\bibnamefont {Yang}}, \bibinfo {author} {\bibfnamefont {H.}~\bibnamefont
  {Cao}}, \bibinfo {author} {\bibfnamefont {C.}~\bibnamefont {Zheng}}, \bibinfo
  {author} {\bibfnamefont {Q.}~\bibnamefont {Zeng}}, \bibinfo {author}
  {\bibfnamefont {J.-g.}\ \bibnamefont {Guo}},\ and\ \bibinfo {author}
  {\bibfnamefont {J.}~\bibnamefont {Zhao}},\ }\bibfield  {title} {\bibinfo
  {title} {Superconductivity in pressurized trilayer
  {La$_4$Ni$_3$O$_{10-\delta}$} single crystals},\ }\href@noop {} {\bibfield
  {journal} {\bibinfo  {journal} {Nature}\ }\textbf {\bibinfo {volume} {631}}
  (\bibinfo {year} {2024})}\BibitemShut {NoStop}%
\bibitem [{\citenamefont {Zhang}\ \emph
  {et~al.}(2025{\natexlab{a}})\citenamefont {Zhang}, \citenamefont {Pei},
  \citenamefont {Peng}, \citenamefont {Du}, \citenamefont {Hu}, \citenamefont
  {Cao}, \citenamefont {Wang}, \citenamefont {Wu}, \citenamefont {Li},
  \citenamefont {Liu}, \citenamefont {Wen}, \citenamefont {Song}, \citenamefont
  {Zhao}, \citenamefont {Li}, \citenamefont {Cao}, \citenamefont {Zhu},
  \citenamefont {Zhang}, \citenamefont {Yu}, \citenamefont {Cheng},
  \citenamefont {Zhang}, \citenamefont {Li}, \citenamefont {Zhao},
  \citenamefont {Chen}, \citenamefont {Jin}, \citenamefont {Guo}, \citenamefont
  {Wu}, \citenamefont {Yang}, \citenamefont {Zeng}, \citenamefont {Yan},
  \citenamefont {Yang},\ and\ \citenamefont {Qi}}]{zhang2025a}%
  \BibitemOpen
  \bibfield  {author} {\bibinfo {author} {\bibfnamefont {M.}~\bibnamefont
  {Zhang}}, \bibinfo {author} {\bibfnamefont {C.}~\bibnamefont {Pei}}, \bibinfo
  {author} {\bibfnamefont {D.}~\bibnamefont {Peng}}, \bibinfo {author}
  {\bibfnamefont {X.}~\bibnamefont {Du}}, \bibinfo {author} {\bibfnamefont
  {W.}~\bibnamefont {Hu}}, \bibinfo {author} {\bibfnamefont {Y.}~\bibnamefont
  {Cao}}, \bibinfo {author} {\bibfnamefont {Q.}~\bibnamefont {Wang}}, \bibinfo
  {author} {\bibfnamefont {J.}~\bibnamefont {Wu}}, \bibinfo {author}
  {\bibfnamefont {Y.}~\bibnamefont {Li}}, \bibinfo {author} {\bibfnamefont
  {H.}~\bibnamefont {Liu}}, \bibinfo {author} {\bibfnamefont {C.}~\bibnamefont
  {Wen}}, \bibinfo {author} {\bibfnamefont {J.}~\bibnamefont {Song}}, \bibinfo
  {author} {\bibfnamefont {Y.}~\bibnamefont {Zhao}}, \bibinfo {author}
  {\bibfnamefont {C.}~\bibnamefont {Li}}, \bibinfo {author} {\bibfnamefont
  {W.}~\bibnamefont {Cao}}, \bibinfo {author} {\bibfnamefont {S.}~\bibnamefont
  {Zhu}}, \bibinfo {author} {\bibfnamefont {Q.}~\bibnamefont {Zhang}}, \bibinfo
  {author} {\bibfnamefont {N.}~\bibnamefont {Yu}}, \bibinfo {author}
  {\bibfnamefont {P.}~\bibnamefont {Cheng}}, \bibinfo {author} {\bibfnamefont
  {L.}~\bibnamefont {Zhang}}, \bibinfo {author} {\bibfnamefont
  {Z.}~\bibnamefont {Li}}, \bibinfo {author} {\bibfnamefont {J.}~\bibnamefont
  {Zhao}}, \bibinfo {author} {\bibfnamefont {Y.}~\bibnamefont {Chen}}, \bibinfo
  {author} {\bibfnamefont {C.}~\bibnamefont {Jin}}, \bibinfo {author}
  {\bibfnamefont {H.}~\bibnamefont {Guo}}, \bibinfo {author} {\bibfnamefont
  {C.}~\bibnamefont {Wu}}, \bibinfo {author} {\bibfnamefont {F.}~\bibnamefont
  {Yang}}, \bibinfo {author} {\bibfnamefont {Q.}~\bibnamefont {Zeng}}, \bibinfo
  {author} {\bibfnamefont {S.}~\bibnamefont {Yan}}, \bibinfo {author}
  {\bibfnamefont {L.}~\bibnamefont {Yang}},\ and\ \bibinfo {author}
  {\bibfnamefont {Y.}~\bibnamefont {Qi}},\ }\bibfield  {title} {\bibinfo
  {title} {Superconductivity in trilayer nickelate {La$_4$Ni$_3$O$_{10}$} under
  pressure},\ }\href@noop {} {\bibfield  {journal} {\bibinfo  {journal} {Phys.
  Rev. X}\ }\textbf {\bibinfo {volume} {15}},\ \bibinfo {pages} {021005}
  (\bibinfo {year} {2025}{\natexlab{a}})}\BibitemShut {NoStop}%
\bibitem [{\citenamefont {Zhang}\ \emph
  {et~al.}(2025{\natexlab{b}})\citenamefont {Zhang}, \citenamefont {Peng},
  \citenamefont {Zhu}, \citenamefont {Chen}, \citenamefont {Cui}, \citenamefont
  {Wang}, \citenamefont {Wang}, \citenamefont {Zeng},\ and\ \citenamefont
  {Zhao}}]{zhang2025b}%
  \BibitemOpen
  \bibfield  {author} {\bibinfo {author} {\bibfnamefont {E.}~\bibnamefont
  {Zhang}}, \bibinfo {author} {\bibfnamefont {D.}~\bibnamefont {Peng}},
  \bibinfo {author} {\bibfnamefont {Y.}~\bibnamefont {Zhu}}, \bibinfo {author}
  {\bibfnamefont {L.}~\bibnamefont {Chen}}, \bibinfo {author} {\bibfnamefont
  {B.}~\bibnamefont {Cui}}, \bibinfo {author} {\bibfnamefont {X.}~\bibnamefont
  {Wang}}, \bibinfo {author} {\bibfnamefont {W.}~\bibnamefont {Wang}}, \bibinfo
  {author} {\bibfnamefont {Q.}~\bibnamefont {Zeng}},\ and\ \bibinfo {author}
  {\bibfnamefont {J.}~\bibnamefont {Zhao}},\ }\bibfield  {title} {\bibinfo
  {title} {Bulk superconductivity in pressurized trilayer nickelate
  {Pr$_4$Ni$_3$O$_{10}$} single crystals},\ }\href@noop {} {\bibfield
  {journal} {\bibinfo  {journal} {Phys. Rev. X}\ }\textbf {\bibinfo {volume}
  {15}},\ \bibinfo {pages} {021008} (\bibinfo {year}
  {2025}{\natexlab{b}})}\BibitemShut {NoStop}%
\bibitem [{\citenamefont {Shi}\ \emph {et~al.}(2025)\citenamefont {Shi},
  \citenamefont {Li}, \citenamefont {Wang}, \citenamefont {Peng}, \citenamefont
  {Yang}, \citenamefont {Li}, \citenamefont {Fan}, \citenamefont {Jiang},
  \citenamefont {He}, \citenamefont {Zeng}, \citenamefont {Song}, \citenamefont
  {Ge}, \citenamefont {Xiang}, \citenamefont {Wang}, \citenamefont {Ying},
  \citenamefont {Wu},\ and\ \citenamefont {Chen}}]{Shi2025z}%
  \BibitemOpen
  \bibfield  {author} {\bibinfo {author} {\bibfnamefont {M.}~\bibnamefont
  {Shi}}, \bibinfo {author} {\bibfnamefont {Y.}~\bibnamefont {Li}}, \bibinfo
  {author} {\bibfnamefont {Y.}~\bibnamefont {Wang}}, \bibinfo {author}
  {\bibfnamefont {D.}~\bibnamefont {Peng}}, \bibinfo {author} {\bibfnamefont
  {S.}~\bibnamefont {Yang}}, \bibinfo {author} {\bibfnamefont {H.}~\bibnamefont
  {Li}}, \bibinfo {author} {\bibfnamefont {K.}~\bibnamefont {Fan}}, \bibinfo
  {author} {\bibfnamefont {K.}~\bibnamefont {Jiang}}, \bibinfo {author}
  {\bibfnamefont {J.}~\bibnamefont {He}}, \bibinfo {author} {\bibfnamefont
  {Q.}~\bibnamefont {Zeng}}, \bibinfo {author} {\bibfnamefont {D.}~\bibnamefont
  {Song}}, \bibinfo {author} {\bibfnamefont {B.}~\bibnamefont {Ge}}, \bibinfo
  {author} {\bibfnamefont {Z.}~\bibnamefont {Xiang}}, \bibinfo {author}
  {\bibfnamefont {Z.}~\bibnamefont {Wang}}, \bibinfo {author} {\bibfnamefont
  {J.}~\bibnamefont {Ying}}, \bibinfo {author} {\bibfnamefont {T.}~\bibnamefont
  {Wu}},\ and\ \bibinfo {author} {\bibfnamefont {X.}~\bibnamefont {Chen}},\
  }\bibfield  {title} {\bibinfo {title} {Absence of superconductivity and
  density-wave transition in ambient-pressure tetragonal la4ni3o10},\
  }\href@noop {} {\bibfield  {journal} {\bibinfo  {journal} {Nature
  Communications}\ }\textbf {\bibinfo {volume} {16}},\ \bibinfo {pages} {2887}
  (\bibinfo {year} {2025})}\BibitemShut {NoStop}%
\bibitem [{\citenamefont {Li}\ \emph {et~al.}(2025{\natexlab{a}})\citenamefont
  {Li}, \citenamefont {Hao}, \citenamefont {Guo}, \citenamefont {Jia},
  \citenamefont {Dou}, \citenamefont {Peng}, \citenamefont {Zhang},
  \citenamefont {Xiao}, \citenamefont {Zhang}, \citenamefont {Zhou},
  \citenamefont {Huang}, \citenamefont {Wang}, \citenamefont {Guo},
  \citenamefont {Mezouar}, \citenamefont {Elias F. S.~Rodrigues}, \citenamefont
  {Luo}, \citenamefont {Yang}, \citenamefont {Zeng}, \citenamefont {Zhou},
  \citenamefont {Gou}, \citenamefont {Zheng}, \citenamefont {Liu},\ and\
  \citenamefont {Zhang}}]{Li2025c}%
  \BibitemOpen
  \bibfield  {author} {\bibinfo {author} {\bibfnamefont {F.}~\bibnamefont
  {Li}}, \bibinfo {author} {\bibfnamefont {Y.}~\bibnamefont {Hao}}, \bibinfo
  {author} {\bibfnamefont {N.}~\bibnamefont {Guo}}, \bibinfo {author}
  {\bibfnamefont {D.}~\bibnamefont {Jia}}, \bibinfo {author} {\bibfnamefont
  {J.}~\bibnamefont {Dou}}, \bibinfo {author} {\bibfnamefont {D.}~\bibnamefont
  {Peng}}, \bibinfo {author} {\bibfnamefont {G.}~\bibnamefont {Zhang}},
  \bibinfo {author} {\bibfnamefont {L.}~\bibnamefont {Xiao}}, \bibinfo {author}
  {\bibfnamefont {J.}~\bibnamefont {Zhang}}, \bibinfo {author} {\bibfnamefont
  {W.}~\bibnamefont {Zhou}}, \bibinfo {author} {\bibfnamefont {Y.}~\bibnamefont
  {Huang}}, \bibinfo {author} {\bibfnamefont {X.}~\bibnamefont {Wang}},
  \bibinfo {author} {\bibfnamefont {Z.}~\bibnamefont {Guo}}, \bibinfo {author}
  {\bibfnamefont {M.}~\bibnamefont {Mezouar}}, \bibinfo {author} {\bibfnamefont
  {J.}~\bibnamefont {Elias F. S.~Rodrigues}}, \bibinfo {author} {\bibfnamefont
  {J.}~\bibnamefont {Luo}}, \bibinfo {author} {\bibfnamefont {J.}~\bibnamefont
  {Yang}}, \bibinfo {author} {\bibfnamefont {Q.}~\bibnamefont {Zeng}}, \bibinfo
  {author} {\bibfnamefont {R.}~\bibnamefont {Zhou}}, \bibinfo {author}
  {\bibfnamefont {H.}~\bibnamefont {Gou}}, \bibinfo {author} {\bibfnamefont
  {Q.}~\bibnamefont {Zheng}}, \bibinfo {author} {\bibfnamefont
  {G.}~\bibnamefont {Liu}},\ and\ \bibinfo {author} {\bibfnamefont
  {J.}~\bibnamefont {Zhang}},\ }\bibfield  {title} {\bibinfo {title}
  {Single-crystal structure determination of superconducting
  {La$_4$Ni$_3$O$_{10-\delta}$} under high pressure},\ }\href@noop {}
  {\bibfield  {journal} {\bibinfo  {journal} {Advanced Materials}\ }\textbf
  {\bibinfo {volume} {n/a}},\ \bibinfo {pages} {e07365} (\bibinfo {year}
  {2025}{\natexlab{a}})}\BibitemShut {NoStop}%
\bibitem [{\citenamefont {Nagell}\ \emph {et~al.}(2017)\citenamefont {Nagell},
  \citenamefont {Sławiński}, \citenamefont {Vajeeston}, \citenamefont
  {Fjellvåg},\ and\ \citenamefont {Sjåstad}}]{nagell2017a}%
  \BibitemOpen
  \bibfield  {author} {\bibinfo {author} {\bibfnamefont {M.~U.}\ \bibnamefont
  {Nagell}}, \bibinfo {author} {\bibfnamefont {W.~A.}\ \bibnamefont
  {Sławiński}}, \bibinfo {author} {\bibfnamefont {P.}~\bibnamefont
  {Vajeeston}}, \bibinfo {author} {\bibfnamefont {H.}~\bibnamefont
  {Fjellvåg}},\ and\ \bibinfo {author} {\bibfnamefont {A.~O.}\ \bibnamefont
  {Sjåstad}},\ }\bibfield  {title} {\bibinfo {title} {Temperature induced
  transitions in {La}$_4${(Co$_{1-x}$Ni$_x$)}$_3${O}$_{10+\delta}$; oxygen
  stoichiometry and mobility},\ }\href@noop {} {\bibfield  {journal} {\bibinfo
  {journal} {Solid State Ionics}\ }\textbf {\bibinfo {volume} {305}},\ \bibinfo
  {pages} {7} (\bibinfo {year} {2017})}\BibitemShut {NoStop}%
\bibitem [{\citenamefont {Song}\ \emph {et~al.}(2020)\citenamefont {Song},
  \citenamefont {Ning}, \citenamefont {Boukamp}, \citenamefont {Bassat},\ and\
  \citenamefont {Bouwmeester}}]{Song2020a}%
  \BibitemOpen
  \bibfield  {author} {\bibinfo {author} {\bibfnamefont {J.}~\bibnamefont
  {Song}}, \bibinfo {author} {\bibfnamefont {D.}~\bibnamefont {Ning}}, \bibinfo
  {author} {\bibfnamefont {B.}~\bibnamefont {Boukamp}}, \bibinfo {author}
  {\bibfnamefont {J.-M.}\ \bibnamefont {Bassat}},\ and\ \bibinfo {author}
  {\bibfnamefont {H.~J.~M.}\ \bibnamefont {Bouwmeester}},\ }\bibfield  {title}
  {\bibinfo {title} {Structure{,} electrical conductivity and oxygen transport
  properties of ruddlesden–popper phases {Ln$_{n+1}$Ni$_n$O3$_{n+1}$} {(Ln =
  La{,} Pr and Nd; n = 1{,} 2 and 3)}},\ }\href@noop {} {\bibfield  {journal}
  {\bibinfo  {journal} {J. Mater. Chem. A}\ }\textbf {\bibinfo {volume} {8}},\
  \bibinfo {pages} {22206} (\bibinfo {year} {2020})}\BibitemShut {NoStop}%
\bibitem [{\citenamefont {Zhang}\ \emph
  {et~al.}(2020{\natexlab{a}})\citenamefont {Zhang}, \citenamefont {Phelan},
  \citenamefont {Botana}, \citenamefont {Chen}, \citenamefont {Zheng},
  \citenamefont {Krogstad}, \citenamefont {Wang}, \citenamefont {Qiu},
  \citenamefont {Rodriguez-Rivera}, \citenamefont {Osborn}, \citenamefont
  {Rosenkranz}, \citenamefont {Norman},\ and\ \citenamefont
  {Mitchell}}]{Zhang2020a}%
  \BibitemOpen
  \bibfield  {author} {\bibinfo {author} {\bibfnamefont {J.}~\bibnamefont
  {Zhang}}, \bibinfo {author} {\bibfnamefont {D.}~\bibnamefont {Phelan}},
  \bibinfo {author} {\bibfnamefont {A.~S.}\ \bibnamefont {Botana}}, \bibinfo
  {author} {\bibfnamefont {Y.-S.}\ \bibnamefont {Chen}}, \bibinfo {author}
  {\bibfnamefont {H.}~\bibnamefont {Zheng}}, \bibinfo {author} {\bibfnamefont
  {M.}~\bibnamefont {Krogstad}}, \bibinfo {author} {\bibfnamefont {S.~G.}\
  \bibnamefont {Wang}}, \bibinfo {author} {\bibfnamefont {Y.}~\bibnamefont
  {Qiu}}, \bibinfo {author} {\bibfnamefont {J.~A.}\ \bibnamefont
  {Rodriguez-Rivera}}, \bibinfo {author} {\bibfnamefont {R.}~\bibnamefont
  {Osborn}}, \bibinfo {author} {\bibfnamefont {S.}~\bibnamefont {Rosenkranz}},
  \bibinfo {author} {\bibfnamefont {M.~R.}\ \bibnamefont {Norman}},\ and\
  \bibinfo {author} {\bibfnamefont {J.~F.}\ \bibnamefont {Mitchell}},\
  }\bibfield  {title} {\bibinfo {title} {Intertwined density waves in a
  metallic nickelate},\ }\href@noop {} {\bibfield  {journal} {\bibinfo
  {journal} {Nature Communications}\ }\textbf {\bibinfo {volume} {11}},\
  \bibinfo {pages} {6003} (\bibinfo {year} {2020}{\natexlab{a}})}\BibitemShut
  {NoStop}%
\bibitem [{\citenamefont {Xu}\ \emph {et~al.}(2025{\natexlab{a}})\citenamefont
  {Xu}, \citenamefont {Chen}, \citenamefont {Huo}, \citenamefont {Hu},
  \citenamefont {Wang}, \citenamefont {Wu}, \citenamefont {Li}, \citenamefont
  {Wu}, \citenamefont {Wang}, \citenamefont {Yao}, \citenamefont {Dong},\ and\
  \citenamefont {Wang}}]{xu2025f}%
  \BibitemOpen
  \bibfield  {author} {\bibinfo {author} {\bibfnamefont {S.}~\bibnamefont
  {Xu}}, \bibinfo {author} {\bibfnamefont {C.-Q.}\ \bibnamefont {Chen}},
  \bibinfo {author} {\bibfnamefont {M.}~\bibnamefont {Huo}}, \bibinfo {author}
  {\bibfnamefont {D.}~\bibnamefont {Hu}}, \bibinfo {author} {\bibfnamefont
  {H.}~\bibnamefont {Wang}}, \bibinfo {author} {\bibfnamefont {Q.}~\bibnamefont
  {Wu}}, \bibinfo {author} {\bibfnamefont {R.}~\bibnamefont {Li}}, \bibinfo
  {author} {\bibfnamefont {D.}~\bibnamefont {Wu}}, \bibinfo {author}
  {\bibfnamefont {M.}~\bibnamefont {Wang}}, \bibinfo {author} {\bibfnamefont
  {D.-X.}\ \bibnamefont {Yao}}, \bibinfo {author} {\bibfnamefont
  {T.}~\bibnamefont {Dong}},\ and\ \bibinfo {author} {\bibfnamefont
  {N.}~\bibnamefont {Wang}},\ }\bibfield  {title} {\bibinfo {title} {Origin of
  the density wave instability in trilayer nickelate
  $\mathrm{L}{\mathrm{a}}_{4}\mathrm{N}{\mathrm{i}}_{3}{\mathrm{o}}_{10}$
  revealed by optical and ultrafast spectroscopy},\ }\href@noop {} {\bibfield
  {journal} {\bibinfo  {journal} {Phys. Rev. B}\ }\textbf {\bibinfo {volume}
  {111}},\ \bibinfo {pages} {075140} (\bibinfo {year}
  {2025}{\natexlab{a}})}\BibitemShut {NoStop}%
\bibitem [{\citenamefont {Li}\ \emph {et~al.}(2025{\natexlab{b}})\citenamefont
  {Li}, \citenamefont {Gong}, \citenamefont {Zhu}, \citenamefont {Chen},
  \citenamefont {Zhang}, \citenamefont {Zhang}, \citenamefont {Li},
  \citenamefont {Yin}, \citenamefont {Wang}, \citenamefont {Zhao},
  \citenamefont {Feng}, \citenamefont {Du},\ and\ \citenamefont
  {Yan}}]{Li2025e}%
  \BibitemOpen
  \bibfield  {author} {\bibinfo {author} {\bibfnamefont {M.}~\bibnamefont
  {Li}}, \bibinfo {author} {\bibfnamefont {J.}~\bibnamefont {Gong}}, \bibinfo
  {author} {\bibfnamefont {Y.}~\bibnamefont {Zhu}}, \bibinfo {author}
  {\bibfnamefont {Z.}~\bibnamefont {Chen}}, \bibinfo {author} {\bibfnamefont
  {J.}~\bibnamefont {Zhang}}, \bibinfo {author} {\bibfnamefont
  {E.}~\bibnamefont {Zhang}}, \bibinfo {author} {\bibfnamefont
  {Y.}~\bibnamefont {Li}}, \bibinfo {author} {\bibfnamefont {R.}~\bibnamefont
  {Yin}}, \bibinfo {author} {\bibfnamefont {S.}~\bibnamefont {Wang}}, \bibinfo
  {author} {\bibfnamefont {J.}~\bibnamefont {Zhao}}, \bibinfo {author}
  {\bibfnamefont {D.-L.}\ \bibnamefont {Feng}}, \bibinfo {author}
  {\bibfnamefont {Z.}~\bibnamefont {Du}},\ and\ \bibinfo {author}
  {\bibfnamefont {Y.-J.}\ \bibnamefont {Yan}},\ }\bibfield  {title} {\bibinfo
  {title} {Direct visualization of an incommensurate unidirectional charge
  density wave in
  {$\mathrm{L}{\mathrm{a}}_{4}\mathrm{N}{\mathrm{i}}_{3}{\mathrm{O}}_{10}$}},\
  }\href@noop {} {\bibfield  {journal} {\bibinfo  {journal} {Phys. Rev. B}\
  }\textbf {\bibinfo {volume} {112}},\ \bibinfo {pages} {045132} (\bibinfo
  {year} {2025}{\natexlab{b}})}\BibitemShut {NoStop}%
\bibitem [{\citenamefont {Khasanov}\ \emph {et~al.}(2026)\citenamefont
  {Khasanov}, \citenamefont {Sazgari}, \citenamefont {Hicken}, \citenamefont
  {Plokhikh}, \citenamefont {Medarde}, \citenamefont {Pomjakushina},
  \citenamefont {Keller}, \citenamefont {Pomjakushin}, \citenamefont
  {Bartkowiak}, \citenamefont {Kr\'olak}, \citenamefont {Winiarski},
  \citenamefont {Steppke}, \citenamefont {Krieger}, \citenamefont {Luetkens},
  \citenamefont {Klimczuk}, \citenamefont {Schneider}, \citenamefont
  {Gawryluk},\ and\ \citenamefont {Guguchia}}]{khasanov2026a}%
  \BibitemOpen
  \bibfield  {author} {\bibinfo {author} {\bibfnamefont {R.}~\bibnamefont
  {Khasanov}}, \bibinfo {author} {\bibfnamefont {V.}~\bibnamefont {Sazgari}},
  \bibinfo {author} {\bibfnamefont {T.~J.}\ \bibnamefont {Hicken}}, \bibinfo
  {author} {\bibfnamefont {I.}~\bibnamefont {Plokhikh}}, \bibinfo {author}
  {\bibfnamefont {M.}~\bibnamefont {Medarde}}, \bibinfo {author} {\bibfnamefont
  {E.}~\bibnamefont {Pomjakushina}}, \bibinfo {author} {\bibfnamefont
  {L.}~\bibnamefont {Keller}}, \bibinfo {author} {\bibfnamefont
  {V.}~\bibnamefont {Pomjakushin}}, \bibinfo {author} {\bibfnamefont
  {M.}~\bibnamefont {Bartkowiak}}, \bibinfo {author} {\bibfnamefont
  {S.}~\bibnamefont {Kr\'olak}}, \bibinfo {author} {\bibfnamefont {M.~J.}\
  \bibnamefont {Winiarski}}, \bibinfo {author} {\bibfnamefont {A.}~\bibnamefont
  {Steppke}}, \bibinfo {author} {\bibfnamefont {J.~A.}\ \bibnamefont
  {Krieger}}, \bibinfo {author} {\bibfnamefont {H.}~\bibnamefont {Luetkens}},
  \bibinfo {author} {\bibfnamefont {T.}~\bibnamefont {Klimczuk}}, \bibinfo
  {author} {\bibfnamefont {C.~W.}\ \bibnamefont {Schneider}}, \bibinfo {author}
  {\bibfnamefont {D.~J.}\ \bibnamefont {Gawryluk}},\ and\ \bibinfo {author}
  {\bibfnamefont {Z.}~\bibnamefont {Guguchia}},\ }\bibfield  {title} {\bibinfo
  {title} {Pressure and oxygen-isotope substitution on density-wave transitions
  in ${\mathrm{la}}_{4}{\mathrm{ni}}_{3}{\mathrm{o}}_{10}$},\ }\href@noop {}
  {\bibfield  {journal} {\bibinfo  {journal} {Phys. Rev. Res.}\ }\textbf
  {\bibinfo {volume} {8}},\ \bibinfo {pages} {013249} (\bibinfo {year}
  {2026})}\BibitemShut {NoStop}%
\bibitem [{\citenamefont {Wu}\ \emph {et~al.}(2001)\citenamefont {Wu},
  \citenamefont {Neumeier},\ and\ \citenamefont {Hundley}}]{Wu2001a}%
  \BibitemOpen
  \bibfield  {author} {\bibinfo {author} {\bibfnamefont {G.}~\bibnamefont
  {Wu}}, \bibinfo {author} {\bibfnamefont {J.~J.}\ \bibnamefont {Neumeier}},\
  and\ \bibinfo {author} {\bibfnamefont {M.~F.}\ \bibnamefont {Hundley}},\
  }\bibfield  {title} {\bibinfo {title} {Magnetic susceptibility, heat
  capacity, and pressure dependence of the electrical resistivity of
  {${\mathrm{La}}_{3}{\mathrm{Ni}}_{2}{\mathrm{O}}_{7}$ and
  ${\mathrm{La}}_{4}{\mathrm{Ni}}_{3}{\mathrm{O}}_{10}$}},\ }\href@noop {}
  {\bibfield  {journal} {\bibinfo  {journal} {Phys. Rev. B}\ }\textbf {\bibinfo
  {volume} {63}},\ \bibinfo {pages} {245120} (\bibinfo {year}
  {2001})}\BibitemShut {NoStop}%
\bibitem [{\citenamefont {Sakakibara}\ \emph {et~al.}(2024)\citenamefont
  {Sakakibara}, \citenamefont {Ochi}, \citenamefont {Nagata}, \citenamefont
  {Ueki}, \citenamefont {Sakurai}, \citenamefont {Matsumoto}, \citenamefont
  {Terashima}, \citenamefont {Hirose}, \citenamefont {Ohta}, \citenamefont
  {Kato}, \citenamefont {Takano},\ and\ \citenamefont {Kuroki}}]{Sakakibara24}%
  \BibitemOpen
  \bibfield  {author} {\bibinfo {author} {\bibfnamefont {H.}~\bibnamefont
  {Sakakibara}}, \bibinfo {author} {\bibfnamefont {M.}~\bibnamefont {Ochi}},
  \bibinfo {author} {\bibfnamefont {H.}~\bibnamefont {Nagata}}, \bibinfo
  {author} {\bibfnamefont {Y.}~\bibnamefont {Ueki}}, \bibinfo {author}
  {\bibfnamefont {H.}~\bibnamefont {Sakurai}}, \bibinfo {author} {\bibfnamefont
  {R.}~\bibnamefont {Matsumoto}}, \bibinfo {author} {\bibfnamefont
  {K.}~\bibnamefont {Terashima}}, \bibinfo {author} {\bibfnamefont
  {K.}~\bibnamefont {Hirose}}, \bibinfo {author} {\bibfnamefont
  {H.}~\bibnamefont {Ohta}}, \bibinfo {author} {\bibfnamefont {M.}~\bibnamefont
  {Kato}}, \bibinfo {author} {\bibfnamefont {Y.}~\bibnamefont {Takano}},\ and\
  \bibinfo {author} {\bibfnamefont {K.}~\bibnamefont {Kuroki}},\ }\bibfield
  {title} {\bibinfo {title} {Theoretical analysis on the possibility of
  superconductivity in the trilayer ruddlesden-popper nickelate
  {${\mathrm{La}}_{4}{\mathrm{Ni}}_{3}{\mathrm{O}}_{10}$ under pressure and its
  experimental examination: Comparison with
  ${\mathrm{La}}_{3}{\mathrm{Ni}}_{2}{\mathrm{O}}_{7}$}},\ }\href
  {https://doi.org/10.1103/PhysRevB.109.144511} {\bibfield  {journal} {\bibinfo
   {journal} {Phys. Rev. B}\ }\textbf {\bibinfo {volume} {109}},\ \bibinfo
  {pages} {144511} (\bibinfo {year} {2024})}\BibitemShut {NoStop}%
\bibitem [{\citenamefont {Huang}\ and\ \citenamefont {Zhou}(2024)}]{Huang24}%
  \BibitemOpen
  \bibfield  {author} {\bibinfo {author} {\bibfnamefont {J.}~\bibnamefont
  {Huang}}\ and\ \bibinfo {author} {\bibfnamefont {T.}~\bibnamefont {Zhou}},\
  }\bibfield  {title} {\bibinfo {title} {Interlayer pairing-induced partially
  gapped fermi surface in trilayer
  {${\mathrm{La}}_{4}{\mathrm{Ni}}_{3}{\mathrm{O}}_{10}$ superconductors}},\
  }\href {https://doi.org/10.1103/PhysRevB.110.L060506} {\bibfield  {journal}
  {\bibinfo  {journal} {Phys. Rev. B}\ }\textbf {\bibinfo {volume} {110}},\
  \bibinfo {pages} {L060506} (\bibinfo {year} {2024})}\BibitemShut {NoStop}%
\bibitem [{\citenamefont {Ouyang}\ \emph {et~al.}(2025)\citenamefont {Ouyang},
  \citenamefont {He},\ and\ \citenamefont {Lu}}]{Ouyang25}%
  \BibitemOpen
  \bibfield  {author} {\bibinfo {author} {\bibfnamefont {Z.}~\bibnamefont
  {Ouyang}}, \bibinfo {author} {\bibfnamefont {R.-Q.}\ \bibnamefont {He}},\
  and\ \bibinfo {author} {\bibfnamefont {Z.-Y.}\ \bibnamefont {Lu}},\
  }\bibfield  {title} {\bibinfo {title} {Phase diagrams and two key factors to
  superconductivity of ruddlesden-popper nickelates},\ }\href
  {https://doi.org/10.1103/1412-nfzm} {\bibfield  {journal} {\bibinfo
  {journal} {Phys. Rev. B}\ }\textbf {\bibinfo {volume} {112}},\ \bibinfo
  {pages} {045127} (\bibinfo {year} {2025})}\BibitemShut {NoStop}%
\bibitem [{\citenamefont {Chen}\ \emph {et~al.}(2024)\citenamefont {Chen},
  \citenamefont {Luo}, \citenamefont {Wang}, \citenamefont {W\'u},\ and\
  \citenamefont {Yao}}]{Chen2024a}%
  \BibitemOpen
  \bibfield  {author} {\bibinfo {author} {\bibfnamefont {C.-Q.}\ \bibnamefont
  {Chen}}, \bibinfo {author} {\bibfnamefont {Z.}~\bibnamefont {Luo}}, \bibinfo
  {author} {\bibfnamefont {M.}~\bibnamefont {Wang}}, \bibinfo {author}
  {\bibfnamefont {W.}~\bibnamefont {W\'u}},\ and\ \bibinfo {author}
  {\bibfnamefont {D.-X.}\ \bibnamefont {Yao}},\ }\bibfield  {title} {\bibinfo
  {title} {Trilayer multiorbital models of
  {${\mathrm{La}}_{4}{\mathrm{Ni}}_{3}{\mathrm{O}}_{10}$}},\ }\href@noop {}
  {\bibfield  {journal} {\bibinfo  {journal} {Phys. Rev. B}\ }\textbf {\bibinfo
  {volume} {110}},\ \bibinfo {pages} {014503} (\bibinfo {year}
  {2024})}\BibitemShut {NoStop}%
\bibitem [{\citenamefont {Zhang}\ \emph
  {et~al.}(2024{\natexlab{b}})\citenamefont {Zhang}, \citenamefont {Lin},
  \citenamefont {Moreo}, \citenamefont {Maier},\ and\ \citenamefont
  {Dagotto}}]{Zhang2024h}%
  \BibitemOpen
  \bibfield  {author} {\bibinfo {author} {\bibfnamefont {Y.}~\bibnamefont
  {Zhang}}, \bibinfo {author} {\bibfnamefont {L.-F.}\ \bibnamefont {Lin}},
  \bibinfo {author} {\bibfnamefont {A.}~\bibnamefont {Moreo}}, \bibinfo
  {author} {\bibfnamefont {T.~A.}\ \bibnamefont {Maier}},\ and\ \bibinfo
  {author} {\bibfnamefont {E.}~\bibnamefont {Dagotto}},\ }\bibfield  {title}
  {\bibinfo {title} {Prediction of ${s}^{\ifmmode\pm\else\textpm\fi{}}$-wave
  superconductivity enhanced by electronic doping in trilayer nickelates
  {${\mathrm{La}}_{4}{\mathrm{Ni}}_{3}{\mathrm{O}}_{10}$ under Pressure}},\
  }\href@noop {} {\bibfield  {journal} {\bibinfo  {journal} {Phys. Rev. Lett.}\
  }\textbf {\bibinfo {volume} {133}},\ \bibinfo {pages} {136001} (\bibinfo
  {year} {2024}{\natexlab{b}})}\BibitemShut {NoStop}%
\bibitem [{\citenamefont {Zhang}\ \emph
  {et~al.}(2025{\natexlab{c}})\citenamefont {Zhang}, \citenamefont {Sun},
  \citenamefont {Liu}, \citenamefont {Liu}, \citenamefont {Chen},\ and\
  \citenamefont {Yang}}]{Zhang2025j}%
  \BibitemOpen
  \bibfield  {author} {\bibinfo {author} {\bibfnamefont {M.}~\bibnamefont
  {Zhang}}, \bibinfo {author} {\bibfnamefont {H.}~\bibnamefont {Sun}}, \bibinfo
  {author} {\bibfnamefont {Y.-B.}\ \bibnamefont {Liu}}, \bibinfo {author}
  {\bibfnamefont {Q.}~\bibnamefont {Liu}}, \bibinfo {author} {\bibfnamefont
  {W.-Q.}\ \bibnamefont {Chen}},\ and\ \bibinfo {author} {\bibfnamefont
  {F.}~\bibnamefont {Yang}},\ }\bibfield  {title} {\bibinfo {title}
  {Spin-density wave and superconductivity in
  {${\mathrm{La}}_{4}{\mathrm{Ni}}_{3}{\mathrm{O}}_{10}$} under ambient
  pressure},\ }\href@noop {} {\bibfield  {journal} {\bibinfo  {journal} {Phys.
  Rev. B}\ }\textbf {\bibinfo {volume} {111}},\ \bibinfo {pages} {144502}
  (\bibinfo {year} {2025}{\natexlab{c}})}\BibitemShut {NoStop}%
\bibitem [{\citenamefont {Xu}\ \emph {et~al.}(2025{\natexlab{b}})\citenamefont
  {Xu}, \citenamefont {Wang}, \citenamefont {Huo}, \citenamefont {Hu},
  \citenamefont {Wu}, \citenamefont {Yue}, \citenamefont {Wu}, \citenamefont
  {Wang}, \citenamefont {Dong},\ and\ \citenamefont {Wang}}]{Xu2025a}%
  \BibitemOpen
  \bibfield  {author} {\bibinfo {author} {\bibfnamefont {S.}~\bibnamefont
  {Xu}}, \bibinfo {author} {\bibfnamefont {H.}~\bibnamefont {Wang}}, \bibinfo
  {author} {\bibfnamefont {M.}~\bibnamefont {Huo}}, \bibinfo {author}
  {\bibfnamefont {D.}~\bibnamefont {Hu}}, \bibinfo {author} {\bibfnamefont
  {Q.}~\bibnamefont {Wu}}, \bibinfo {author} {\bibfnamefont {L.}~\bibnamefont
  {Yue}}, \bibinfo {author} {\bibfnamefont {D.}~\bibnamefont {Wu}}, \bibinfo
  {author} {\bibfnamefont {M.}~\bibnamefont {Wang}}, \bibinfo {author}
  {\bibfnamefont {T.}~\bibnamefont {Dong}},\ and\ \bibinfo {author}
  {\bibfnamefont {N.}~\bibnamefont {Wang}},\ }\bibfield  {title} {\bibinfo
  {title} {Collapse of density wave and emergence of superconductivity in
  pressurized-{La$_4$Ni$_3$O$_{10}$} evidenced by ultrafast spectroscopy},\
  }\href@noop {} {\bibfield  {journal} {\bibinfo  {journal} {Nature
  Communications}\ }\textbf {\bibinfo {volume} {16}},\ \bibinfo {pages} {7039}
  (\bibinfo {year} {2025}{\natexlab{b}})}\BibitemShut {NoStop}%
\bibitem [{\citenamefont {Kumar}\ \emph {et~al.}(2020)\citenamefont {Kumar},
  \citenamefont {Øystein Fjellvåg}, \citenamefont {Sjåstad},\ and\
  \citenamefont {Fjellvåg}}]{kumar2020a}%
  \BibitemOpen
  \bibfield  {author} {\bibinfo {author} {\bibfnamefont {S.}~\bibnamefont
  {Kumar}}, \bibinfo {author} {\bibnamefont {Øystein Fjellvåg}}, \bibinfo
  {author} {\bibfnamefont {A.~O.}\ \bibnamefont {Sjåstad}},\ and\ \bibinfo
  {author} {\bibfnamefont {H.}~\bibnamefont {Fjellvåg}},\ }\bibfield  {title}
  {\bibinfo {title} {Physical properties of ruddlesden-popper (n=3) nickelate:
  {La$_4$Ni$_3$O$_{10}$}},\ }\href@noop {} {\bibfield  {journal} {\bibinfo
  {journal} {Journal of Magnetism and Magnetic Materials}\ }\textbf {\bibinfo
  {volume} {496}},\ \bibinfo {pages} {165915} (\bibinfo {year}
  {2020})}\BibitemShut {NoStop}%
\bibitem [{\citenamefont {Zhang}\ \emph
  {et~al.}(2020{\natexlab{b}})\citenamefont {Zhang}, \citenamefont {Zheng},
  \citenamefont {Chen}, \citenamefont {Ren}, \citenamefont {Yonemura},
  \citenamefont {Huq},\ and\ \citenamefont {Mitchell}}]{Zhang2020c}%
  \BibitemOpen
  \bibfield  {author} {\bibinfo {author} {\bibfnamefont {J.}~\bibnamefont
  {Zhang}}, \bibinfo {author} {\bibfnamefont {H.}~\bibnamefont {Zheng}},
  \bibinfo {author} {\bibfnamefont {Y.-S.}\ \bibnamefont {Chen}}, \bibinfo
  {author} {\bibfnamefont {Y.}~\bibnamefont {Ren}}, \bibinfo {author}
  {\bibfnamefont {M.}~\bibnamefont {Yonemura}}, \bibinfo {author}
  {\bibfnamefont {A.}~\bibnamefont {Huq}},\ and\ \bibinfo {author}
  {\bibfnamefont {J.~F.}\ \bibnamefont {Mitchell}},\ }\bibfield  {title}
  {\bibinfo {title} {High oxygen pressure floating zone growth and crystal
  structure of the metallic nickelates
  {${R}_{4}{\mathrm{Ni}}_{3}{\mathrm{O}}_{10}$
  ($R=\mathrm{La},\mathrm{Pr}$)}},\ }\href@noop {} {\bibfield  {journal}
  {\bibinfo  {journal} {Phys. Rev. Mater.}\ }\textbf {\bibinfo {volume} {4}},\
  \bibinfo {pages} {083402} (\bibinfo {year} {2020}{\natexlab{b}})}\BibitemShut
  {NoStop}%
\bibitem [{\citenamefont {Rout}\ \emph {et~al.}(2020)\citenamefont {Rout},
  \citenamefont {Mudi}, \citenamefont {Hoffmann}, \citenamefont {Spachmann},
  \citenamefont {Klingeler},\ and\ \citenamefont {Singh}}]{rout2020a}%
  \BibitemOpen
  \bibfield  {author} {\bibinfo {author} {\bibfnamefont {D.}~\bibnamefont
  {Rout}}, \bibinfo {author} {\bibfnamefont {S.~R.}\ \bibnamefont {Mudi}},
  \bibinfo {author} {\bibfnamefont {M.}~\bibnamefont {Hoffmann}}, \bibinfo
  {author} {\bibfnamefont {S.}~\bibnamefont {Spachmann}}, \bibinfo {author}
  {\bibfnamefont {R.}~\bibnamefont {Klingeler}},\ and\ \bibinfo {author}
  {\bibfnamefont {S.}~\bibnamefont {Singh}},\ }\bibfield  {title} {\bibinfo
  {title} {Structural and physical properties of trilayer nickelates
  {${R}_{4}{\mathrm{Ni}}_{3}{\mathrm{O}}_{10}$ ($R=\mathrm{La}$, Pr, and
  Nd)}},\ }\href@noop {} {\bibfield  {journal} {\bibinfo  {journal} {Phys. Rev.
  B}\ }\textbf {\bibinfo {volume} {102}},\ \bibinfo {pages} {195144} (\bibinfo
  {year} {2020})}\BibitemShut {NoStop}%
\bibitem [{\citenamefont {Li}\ \emph {et~al.}(2017)\citenamefont {Li},
  \citenamefont {Zhou}, \citenamefont {Nummy}, \citenamefont {Zhang},
  \citenamefont {Pardo}, \citenamefont {Pickett}, \citenamefont {Mitchell},\
  and\ \citenamefont {Dessau}}]{Li2017m}%
  \BibitemOpen
  \bibfield  {author} {\bibinfo {author} {\bibfnamefont {H.}~\bibnamefont
  {Li}}, \bibinfo {author} {\bibfnamefont {X.}~\bibnamefont {Zhou}}, \bibinfo
  {author} {\bibfnamefont {T.}~\bibnamefont {Nummy}}, \bibinfo {author}
  {\bibfnamefont {J.}~\bibnamefont {Zhang}}, \bibinfo {author} {\bibfnamefont
  {V.}~\bibnamefont {Pardo}}, \bibinfo {author} {\bibfnamefont {W.~E.}\
  \bibnamefont {Pickett}}, \bibinfo {author} {\bibfnamefont {J.~F.}\
  \bibnamefont {Mitchell}},\ and\ \bibinfo {author} {\bibfnamefont {D.~S.}\
  \bibnamefont {Dessau}},\ }\bibfield  {title} {\bibinfo {title} {Fermiology
  and electron dynamics of trilayer nickelate {La$_4$Ni$_3$O$_{10}$}},\
  }\href@noop {} {\bibfield  {journal} {\bibinfo  {journal} {Nature
  Communications}\ }\textbf {\bibinfo {volume} {8}} (\bibinfo {year}
  {2017})}\BibitemShut {NoStop}%
\bibitem [{la4()}]{la4ni3o10suppmat2025a}%
  \BibitemOpen
  \href@noop {} {}\bibinfo {note} {See Supplemental Material at [URL will be
  inserted by publisher] for details on the diffraction, Raman experiments and
  values of the structural parameters.}\BibitemShut {Stop}%
\bibitem [{\citenamefont {Puggioni}\ and\ \citenamefont
  {Rondinelli}(2018)}]{Puggioni2018a}%
  \BibitemOpen
  \bibfield  {author} {\bibinfo {author} {\bibfnamefont {D.}~\bibnamefont
  {Puggioni}}\ and\ \bibinfo {author} {\bibfnamefont {J.~M.}\ \bibnamefont
  {Rondinelli}},\ }\bibfield  {title} {\bibinfo {title} {Crystal structure
  stability and electronic properties of the layered nickelate
  {${\mathrm{La}}_{4}{\mathrm{Ni}}_{3}{\mathrm{O}}_{10}$}},\ }\href@noop {}
  {\bibfield  {journal} {\bibinfo  {journal} {Phys. Rev. B}\ }\textbf {\bibinfo
  {volume} {97}},\ \bibinfo {pages} {115116} (\bibinfo {year}
  {2018})}\BibitemShut {NoStop}%
\bibitem [{\citenamefont {van Smaalen}(2007)}]{van2007incommensurate}%
  \BibitemOpen
  \bibfield  {author} {\bibinfo {author} {\bibfnamefont {S.}~\bibnamefont {van
  Smaalen}},\ }\href@noop {} {\emph {\bibinfo {title} {Incommensurate
  Crystallography}}},\ International Union of Crystallography Monographs on
  Crystallography\ (\bibinfo  {publisher} {OUP Oxford},\ \bibinfo {year}
  {2007})\BibitemShut {NoStop}%
\bibitem [{\citenamefont {Petricek}\ \emph {et~al.}(2014)\citenamefont
  {Petricek}, \citenamefont {Dusek},\ and\ \citenamefont
  {Palatinus}}]{petricekv2014a}%
  \BibitemOpen
  \bibfield  {author} {\bibinfo {author} {\bibfnamefont {V.}~\bibnamefont
  {Petricek}}, \bibinfo {author} {\bibfnamefont {M.}~\bibnamefont {Dusek}},\
  and\ \bibinfo {author} {\bibfnamefont {L.}~\bibnamefont {Palatinus}},\
  }\bibfield  {title} {\bibinfo {title} {Crystallographic computing system
  {JANA2006:} general features},\ }\href
  {https://doi.org/10.1515/zkri-2014-1737} {\bibfield  {journal} {\bibinfo
  {journal} {Z. Kristallogr.}\ }\textbf {\bibinfo {volume} {229}},\ \bibinfo
  {pages} {345} (\bibinfo {year} {2014})}\BibitemShut {NoStop}%
\bibitem [{\citenamefont {Parsons}(2003)}]{parson2003a}%
  \BibitemOpen
  \bibfield  {author} {\bibinfo {author} {\bibfnamefont {S.}~\bibnamefont
  {Parsons}},\ }\bibfield  {title} {\bibinfo {title} {{Introduction to
  twinning}},\ }\href@noop {} {\bibfield  {journal} {\bibinfo  {journal} {Acta
  Crystallographica Section D}\ }\textbf {\bibinfo {volume} {59}},\ \bibinfo
  {pages} {1995} (\bibinfo {year} {2003})}\BibitemShut {NoStop}%
\bibitem [{\citenamefont {Hamilton}(1965)}]{hamilton1965a}%
  \BibitemOpen
  \bibfield  {author} {\bibinfo {author} {\bibfnamefont {W.~C.}\ \bibnamefont
  {Hamilton}},\ }\bibfield  {title} {\bibinfo {title} {Significance tests on
  the crystallographic {R} factor},\ }\href@noop {} {\bibfield  {journal}
  {\bibinfo  {journal} {Acta Crystallographica}\ }\textbf {\bibinfo {volume}
  {18}},\ \bibinfo {pages} {502} (\bibinfo {year} {1965})}\BibitemShut
  {NoStop}%
\bibitem [{\citenamefont {Prince}\ and\ \citenamefont
  {Spiegelman}(2006)}]{Internationaltablesvolc}%
  \BibitemOpen
  \bibfield  {author} {\bibinfo {author} {\bibfnamefont {E.}~\bibnamefont
  {Prince}}\ and\ \bibinfo {author} {\bibfnamefont {C.~H.}\ \bibnamefont
  {Spiegelman}},\ }\href@noop {} {\emph {\bibinfo {title} {International Tables
  for Crystallography Vol. C}}}\ (\bibinfo {year} {2006})\ pp.\ \bibinfo
  {pages} {702--706}\BibitemShut {NoStop}%
\bibitem [{\citenamefont {Ramakrishnan}\ \emph {et~al.}(2024)\citenamefont
  {Ramakrishnan}, \citenamefont {Kotla}, \citenamefont {Pi}, \citenamefont
  {Maity}, \citenamefont {Chen}, \citenamefont {Bao}, \citenamefont {Guo},
  \citenamefont {Kado}, \citenamefont {Agarwal}, \citenamefont {Eisele},
  \citenamefont {Nohara}, \citenamefont {Noohinejad}, \citenamefont {Weng},
  \citenamefont {Ramakrishnan}, \citenamefont {Thamizhavel},\ and\
  \citenamefont {van Smaalen}}]{ramakrishnan2024a}%
  \BibitemOpen
  \bibfield  {author} {\bibinfo {author} {\bibfnamefont {S.}~\bibnamefont
  {Ramakrishnan}}, \bibinfo {author} {\bibfnamefont {S.~R.}\ \bibnamefont
  {Kotla}}, \bibinfo {author} {\bibfnamefont {H.}~\bibnamefont {Pi}}, \bibinfo
  {author} {\bibfnamefont {B.~B.}\ \bibnamefont {Maity}}, \bibinfo {author}
  {\bibfnamefont {J.}~\bibnamefont {Chen}}, \bibinfo {author} {\bibfnamefont
  {J.-K.}\ \bibnamefont {Bao}}, \bibinfo {author} {\bibfnamefont
  {Z.}~\bibnamefont {Guo}}, \bibinfo {author} {\bibfnamefont {M.}~\bibnamefont
  {Kado}}, \bibinfo {author} {\bibfnamefont {H.}~\bibnamefont {Agarwal}},
  \bibinfo {author} {\bibfnamefont {C.}~\bibnamefont {Eisele}}, \bibinfo
  {author} {\bibfnamefont {M.}~\bibnamefont {Nohara}}, \bibinfo {author}
  {\bibfnamefont {L.}~\bibnamefont {Noohinejad}}, \bibinfo {author}
  {\bibfnamefont {H.}~\bibnamefont {Weng}}, \bibinfo {author} {\bibfnamefont
  {S.}~\bibnamefont {Ramakrishnan}}, \bibinfo {author} {\bibfnamefont
  {A.}~\bibnamefont {Thamizhavel}},\ and\ \bibinfo {author} {\bibfnamefont
  {S.}~\bibnamefont {van Smaalen}},\ }\bibfield  {title} {\bibinfo {title}
  {Noncentrosymmetric, transverse structural modulation in
  {$\text{Sr}{\text{Al}}_{4}$, and elucidation of its origin in the
  $\text{Ba}{\text{Al}}_{4}$} family of compounds},\ }\href@noop {} {\bibfield
  {journal} {\bibinfo  {journal} {Phys. Rev. Res.}\ }\textbf {\bibinfo {volume}
  {6}},\ \bibinfo {pages} {023277} (\bibinfo {year} {2024})}\BibitemShut
  {NoStop}%
\bibitem [{\citenamefont {Kotla}\ \emph {et~al.}(2025)\citenamefont {Kotla},
  \citenamefont {Ramakrishnan}, \citenamefont {Schaller}, \citenamefont
  {Rekis}, \citenamefont {Eisele}, \citenamefont {Bao}, \citenamefont
  {Noohinejad}, \citenamefont {{de Laitre}}, \citenamefont {{de Boissieu}},\
  and\ \citenamefont {{van Smaalen}}}]{kotla2025a}%
  \BibitemOpen
  \bibfield  {author} {\bibinfo {author} {\bibfnamefont {S.~R.}\ \bibnamefont
  {Kotla}}, \bibinfo {author} {\bibfnamefont {S.}~\bibnamefont {Ramakrishnan}},
  \bibinfo {author} {\bibfnamefont {A.~M.}\ \bibnamefont {Schaller}}, \bibinfo
  {author} {\bibfnamefont {T.}~\bibnamefont {Rekis}}, \bibinfo {author}
  {\bibfnamefont {C.}~\bibnamefont {Eisele}}, \bibinfo {author} {\bibfnamefont
  {J.-K.}\ \bibnamefont {Bao}}, \bibinfo {author} {\bibfnamefont
  {L.}~\bibnamefont {Noohinejad}}, \bibinfo {author} {\bibfnamefont
  {G.}~\bibnamefont {{de Laitre}}}, \bibinfo {author} {\bibfnamefont
  {M.}~\bibnamefont {{de Boissieu}}},\ and\ \bibinfo {author} {\bibfnamefont
  {S.}~\bibnamefont {{van Smaalen}}},\ }\bibfield  {title} {\bibinfo {title}
  {Deciphering the commensurately modulated monoclinic phase of
  {Rb$_2$ZnCl$_4$} at low temperatures},\ }\href@noop {} {\bibfield  {journal}
  {\bibinfo  {journal} {Journal of Solid State Chemistry}\ }\textbf {\bibinfo
  {volume} {345}},\ \bibinfo {pages} {125226} (\bibinfo {year}
  {2025})}\BibitemShut {NoStop}%
\bibitem [{\citenamefont {Chapuis}\ \emph {et~al.}(1980)\citenamefont
  {Chapuis}, \citenamefont {Hulliger},\ and\ \citenamefont
  {Schmelczer}}]{chapuis1980a}%
  \BibitemOpen
  \bibfield  {author} {\bibinfo {author} {\bibfnamefont {G.}~\bibnamefont
  {Chapuis}}, \bibinfo {author} {\bibfnamefont {F.}~\bibnamefont {Hulliger}},\
  and\ \bibinfo {author} {\bibfnamefont {R.}~\bibnamefont {Schmelczer}},\
  }\bibfield  {title} {\bibinfo {title} {The crystal structure and some
  properties of {Eu$_2$Sb$_3$}},\ }\href@noop {} {\bibfield  {journal}
  {\bibinfo  {journal} {Journal of Solid State Chemistry}\ }\textbf {\bibinfo
  {volume} {31}},\ \bibinfo {pages} {59} (\bibinfo {year} {1980})}\BibitemShut
  {NoStop}%
\bibitem [{\citenamefont {Shoker}\ \emph {et~al.}(2026)\citenamefont {Shoker},
  \citenamefont {Ramakrishnan}, \citenamefont {Croes}, \citenamefont {Cregut},
  \citenamefont {Beyer}, \citenamefont {Dorkenoo}, \citenamefont {Rodi\`ere},
  \citenamefont {Wehinger}, \citenamefont {Garbarino}, \citenamefont {Mezouar},
  \citenamefont {Verseils}, \citenamefont {Fertey}, \citenamefont
  {Cherifi-Hertel}, \citenamefont {Bouvier},\ and\ \citenamefont
  {Guennou}}]{shoker2025a}%
  \BibitemOpen
  \bibfield  {author} {\bibinfo {author} {\bibfnamefont {M.~B.}\ \bibnamefont
  {Shoker}}, \bibinfo {author} {\bibfnamefont {S.}~\bibnamefont
  {Ramakrishnan}}, \bibinfo {author} {\bibfnamefont {B.}~\bibnamefont {Croes}},
  \bibinfo {author} {\bibfnamefont {O.}~\bibnamefont {Cregut}}, \bibinfo
  {author} {\bibfnamefont {N.}~\bibnamefont {Beyer}}, \bibinfo {author}
  {\bibfnamefont {K.~D.}\ \bibnamefont {Dorkenoo}}, \bibinfo {author}
  {\bibfnamefont {P.}~\bibnamefont {Rodi\`ere}}, \bibinfo {author}
  {\bibfnamefont {B.}~\bibnamefont {Wehinger}}, \bibinfo {author}
  {\bibfnamefont {G.}~\bibnamefont {Garbarino}}, \bibinfo {author}
  {\bibfnamefont {M.}~\bibnamefont {Mezouar}}, \bibinfo {author} {\bibfnamefont
  {M.}~\bibnamefont {Verseils}}, \bibinfo {author} {\bibfnamefont
  {P.}~\bibnamefont {Fertey}}, \bibinfo {author} {\bibfnamefont
  {S.}~\bibnamefont {Cherifi-Hertel}}, \bibinfo {author} {\bibfnamefont
  {P.}~\bibnamefont {Bouvier}},\ and\ \bibinfo {author} {\bibfnamefont
  {M.}~\bibnamefont {Guennou}},\ }\bibfield  {title} {\bibinfo {title}
  {Re-emergence of a polar instability at high pressure in
  {${\mathrm{KNbO}}_{3}$}},\ }\href@noop {} {\bibfield  {journal} {\bibinfo
  {journal} {Phys. Rev. Lett.}\ }\textbf {\bibinfo {volume} {136}},\ \bibinfo
  {pages} {056101} (\bibinfo {year} {2026})}\BibitemShut {NoStop}%
\bibitem [{\citenamefont {Nandi}\ \emph {et~al.}(2025)\citenamefont {Nandi},
  \citenamefont {Maity}, \citenamefont {Dan}, \citenamefont {Ali},
  \citenamefont {Patra}, \citenamefont {Mondal}, \citenamefont {Garbarino},
  \citenamefont {Rodière}, \citenamefont {Ramakrishnan}, \citenamefont
  {Singh},\ and\ \citenamefont {Thamizhavel}}]{nandi2025a}%
  \BibitemOpen
  \bibfield  {author} {\bibinfo {author} {\bibfnamefont {S.}~\bibnamefont
  {Nandi}}, \bibinfo {author} {\bibfnamefont {B.}~\bibnamefont {Maity}},
  \bibinfo {author} {\bibfnamefont {S.}~\bibnamefont {Dan}}, \bibinfo {author}
  {\bibfnamefont {K.}~\bibnamefont {Ali}}, \bibinfo {author} {\bibfnamefont
  {B.}~\bibnamefont {Patra}}, \bibinfo {author} {\bibfnamefont
  {A.}~\bibnamefont {Mondal}}, \bibinfo {author} {\bibfnamefont
  {G.}~\bibnamefont {Garbarino}}, \bibinfo {author} {\bibfnamefont
  {P.}~\bibnamefont {Rodière}}, \bibinfo {author} {\bibfnamefont
  {S.}~\bibnamefont {Ramakrishnan}}, \bibinfo {author} {\bibfnamefont
  {B.}~\bibnamefont {Singh}},\ and\ \bibinfo {author} {\bibfnamefont
  {A.}~\bibnamefont {Thamizhavel}},\ }\href@noop {} {\bibinfo {title} {Evidence
  of electronic instability driven structural distortion in the nodal line
  semimetal {CoSn$_2$}}} (\bibinfo {year} {2025}),\ \Eprint
  {https://arxiv.org/abs/2509.23221} {arXiv:2509.23221 [cond-mat.mtrl-sci]}
  \BibitemShut {NoStop}%
\bibitem [{\citenamefont {Perdew}\ \emph {et~al.}(1996)\citenamefont {Perdew},
  \citenamefont {Burke},\ and\ \citenamefont {Ernzerhof}}]{PBE}%
  \BibitemOpen
  \bibfield  {author} {\bibinfo {author} {\bibfnamefont {J.~P.}\ \bibnamefont
  {Perdew}}, \bibinfo {author} {\bibfnamefont {K.}~\bibnamefont {Burke}},\ and\
  \bibinfo {author} {\bibfnamefont {M.}~\bibnamefont {Ernzerhof}},\ }\bibfield
  {title} {\bibinfo {title} {Generalized gradient approximation made simple},\
  }\href {https://doi.org/10.1103/PhysRevLett.77.3865} {\bibfield  {journal}
  {\bibinfo  {journal} {Phys. Rev. Lett.}\ }\textbf {\bibinfo {volume} {77}},\
  \bibinfo {pages} {3865} (\bibinfo {year} {1996})}\BibitemShut {NoStop}%
\bibitem [{\citenamefont {Schutte}\ \emph {et~al.}(1993)\citenamefont
  {Schutte}, \citenamefont {Disselborg},\ and\ \citenamefont
  {de~Boer}}]{schuttewj1993b}%
  \BibitemOpen
  \bibfield  {author} {\bibinfo {author} {\bibfnamefont {W.~J.}\ \bibnamefont
  {Schutte}}, \bibinfo {author} {\bibfnamefont {F.}~\bibnamefont
  {Disselborg}},\ and\ \bibinfo {author} {\bibfnamefont {J.~L.}\ \bibnamefont
  {de~Boer}},\ }\bibfield  {title} {\bibinfo {title} {Determination of the
  two-dimensional incommensurately modulated structure of {Mo$_2$S$_3$}},\
  }\href@noop {} {\bibfield  {journal} {\bibinfo  {journal} {Acta Crystallogr.
  B}\ }\textbf {\bibinfo {volume} {49}},\ \bibinfo {pages} {787} (\bibinfo
  {year} {1993})}\BibitemShut {NoStop}%
\bibitem [{\citenamefont {Ramakrishnan}\ \emph {et~al.}(2019)\citenamefont
  {Ramakrishnan}, \citenamefont {Sch\"onleber}, \citenamefont {H\"ubschle},
  \citenamefont {Eisele}, \citenamefont {Schaller}, \citenamefont {Rekis},
  \citenamefont {Bui}, \citenamefont {Feulner}, \citenamefont {van Smaalen},
  \citenamefont {Bag}, \citenamefont {Ramakrishnan}, \citenamefont {Tolkiehn},\
  and\ \citenamefont {Paulmann}}]{ramakrishnan2019a}%
  \BibitemOpen
  \bibfield  {author} {\bibinfo {author} {\bibfnamefont {S.}~\bibnamefont
  {Ramakrishnan}}, \bibinfo {author} {\bibfnamefont {A.}~\bibnamefont
  {Sch\"onleber}}, \bibinfo {author} {\bibfnamefont {C.~B.}\ \bibnamefont
  {H\"ubschle}}, \bibinfo {author} {\bibfnamefont {C.}~\bibnamefont {Eisele}},
  \bibinfo {author} {\bibfnamefont {A.~M.}\ \bibnamefont {Schaller}}, \bibinfo
  {author} {\bibfnamefont {T.}~\bibnamefont {Rekis}}, \bibinfo {author}
  {\bibfnamefont {N.~H.~A.}\ \bibnamefont {Bui}}, \bibinfo {author}
  {\bibfnamefont {F.}~\bibnamefont {Feulner}}, \bibinfo {author} {\bibfnamefont
  {S.}~\bibnamefont {van Smaalen}}, \bibinfo {author} {\bibfnamefont
  {B.}~\bibnamefont {Bag}}, \bibinfo {author} {\bibfnamefont {S.}~\bibnamefont
  {Ramakrishnan}}, \bibinfo {author} {\bibfnamefont {M.}~\bibnamefont
  {Tolkiehn}},\ and\ \bibinfo {author} {\bibfnamefont {C.}~\bibnamefont
  {Paulmann}},\ }\bibfield  {title} {\bibinfo {title} {Charge density wave and
  lock-in transitions of {CuV$_{2}$S$_{4}$}},\ }\href@noop {} {\bibfield
  {journal} {\bibinfo  {journal} {Phys. Rev. B}\ }\textbf {\bibinfo {volume}
  {99}},\ \bibinfo {pages} {195140} (\bibinfo {year} {2019})}\BibitemShut
  {NoStop}%
\bibitem [{\citenamefont {Ramakrishnan}\ \emph {et~al.}(2022)\citenamefont
  {Ramakrishnan}, \citenamefont {Kotla}, \citenamefont {Rekis}, \citenamefont
  {Bao}, \citenamefont {Eisele}, \citenamefont {Noohinejad}, \citenamefont
  {Tolkiehn}, \citenamefont {Paulmann}, \citenamefont {Singh}, \citenamefont
  {Verma}, \citenamefont {Bag}, \citenamefont {Kulkarni}, \citenamefont
  {Thamizhavel}, \citenamefont {Singh}, \citenamefont {Ramakrishnan},\ and\
  \citenamefont {van Smaalen}}]{ramakrishnan2022a}%
  \BibitemOpen
  \bibfield  {author} {\bibinfo {author} {\bibfnamefont {S.}~\bibnamefont
  {Ramakrishnan}}, \bibinfo {author} {\bibfnamefont {S.~R.}\ \bibnamefont
  {Kotla}}, \bibinfo {author} {\bibfnamefont {T.}~\bibnamefont {Rekis}},
  \bibinfo {author} {\bibfnamefont {J.-K.}\ \bibnamefont {Bao}}, \bibinfo
  {author} {\bibfnamefont {C.}~\bibnamefont {Eisele}}, \bibinfo {author}
  {\bibfnamefont {L.}~\bibnamefont {Noohinejad}}, \bibinfo {author}
  {\bibfnamefont {M.}~\bibnamefont {Tolkiehn}}, \bibinfo {author}
  {\bibfnamefont {C.}~\bibnamefont {Paulmann}}, \bibinfo {author}
  {\bibfnamefont {B.}~\bibnamefont {Singh}}, \bibinfo {author} {\bibfnamefont
  {R.}~\bibnamefont {Verma}}, \bibinfo {author} {\bibfnamefont
  {B.}~\bibnamefont {Bag}}, \bibinfo {author} {\bibfnamefont {R.}~\bibnamefont
  {Kulkarni}}, \bibinfo {author} {\bibfnamefont {A.}~\bibnamefont
  {Thamizhavel}}, \bibinfo {author} {\bibfnamefont {B.}~\bibnamefont {Singh}},
  \bibinfo {author} {\bibfnamefont {S.}~\bibnamefont {Ramakrishnan}},\ and\
  \bibinfo {author} {\bibfnamefont {S.}~\bibnamefont {van Smaalen}},\
  }\bibfield  {title} {\bibinfo {title} {Orthorhombic charge density wave on
  the tetragonal lattice of {EuAl$_{4}$}},\ }\href@noop {} {\bibfield
  {journal} {\bibinfo  {journal} {IUCrJ}\ }\textbf {\bibinfo {volume} {9}},\
  \bibinfo {pages} {378} (\bibinfo {year} {2022})}\BibitemShut {NoStop}%
\bibitem [{\citenamefont {Bugaris}\ \emph {et~al.}(2017)\citenamefont
  {Bugaris}, \citenamefont {Malliakas}, \citenamefont {Han}, \citenamefont
  {Calta}, \citenamefont {Sturza}, \citenamefont {Krogstad}, \citenamefont
  {Osborn}, \citenamefont {Rosenkranz}, \citenamefont {Ruff}, \citenamefont
  {Trimarchi}, \citenamefont {Bud’ko}, \citenamefont {Balasubramanian},
  \citenamefont {Chung},\ and\ \citenamefont {Kanatzidis}}]{bugaris2017charge}%
  \BibitemOpen
  \bibfield  {author} {\bibinfo {author} {\bibfnamefont {D.~E.}\ \bibnamefont
  {Bugaris}}, \bibinfo {author} {\bibfnamefont {C.~D.}\ \bibnamefont
  {Malliakas}}, \bibinfo {author} {\bibfnamefont {F.}~\bibnamefont {Han}},
  \bibinfo {author} {\bibfnamefont {N.~P.}\ \bibnamefont {Calta}}, \bibinfo
  {author} {\bibfnamefont {M.}~\bibnamefont {Sturza}}, \bibinfo {author}
  {\bibfnamefont {M.~J.}\ \bibnamefont {Krogstad}}, \bibinfo {author}
  {\bibfnamefont {R.}~\bibnamefont {Osborn}}, \bibinfo {author} {\bibfnamefont
  {S.}~\bibnamefont {Rosenkranz}}, \bibinfo {author} {\bibfnamefont {J.~P.~C.}\
  \bibnamefont {Ruff}}, \bibinfo {author} {\bibfnamefont {G.}~\bibnamefont
  {Trimarchi}}, \bibinfo {author} {\bibfnamefont {S.~L.}\ \bibnamefont
  {Bud’ko}}, \bibinfo {author} {\bibfnamefont {M.}~\bibnamefont
  {Balasubramanian}}, \bibinfo {author} {\bibfnamefont {D.~Y.}\ \bibnamefont
  {Chung}},\ and\ \bibinfo {author} {\bibfnamefont {M.~G.}\ \bibnamefont
  {Kanatzidis}},\ }\bibfield  {title} {\bibinfo {title} {Charge density wave in
  the new polymorphs of {RE$_2$Ru$_3$Ge$_5$} ({RE=} {Pr,} {Sm,} {Dy})},\
  }\href@noop {} {\bibfield  {journal} {\bibinfo  {journal} {Journal of the
  American Chemical Society}\ }\textbf {\bibinfo {volume} {139}},\ \bibinfo
  {pages} {4130} (\bibinfo {year} {2017})}\BibitemShut {NoStop}%
\bibitem [{\citenamefont {Sharma}\ \emph {et~al.}(2024)\citenamefont {Sharma},
  \citenamefont {Ramakrishnan}, \citenamefont {SS}, \citenamefont {Kotla},
  \citenamefont {Maiti}, \citenamefont {Eisele}, \citenamefont {Agarwal},
  \citenamefont {Noohinejad}, \citenamefont {Tolkiehn}, \citenamefont {Bansal},
  \citenamefont {van Smaalen},\ and\ \citenamefont {Arumugam}}]{sharma2024a}%
  \BibitemOpen
  \bibfield  {author} {\bibinfo {author} {\bibfnamefont {V.}~\bibnamefont
  {Sharma}}, \bibinfo {author} {\bibfnamefont {S.}~\bibnamefont
  {Ramakrishnan}}, \bibinfo {author} {\bibfnamefont {J.}~\bibnamefont {SS}},
  \bibinfo {author} {\bibfnamefont {S.~R.}\ \bibnamefont {Kotla}}, \bibinfo
  {author} {\bibfnamefont {B.}~\bibnamefont {Maiti}}, \bibinfo {author}
  {\bibfnamefont {C.}~\bibnamefont {Eisele}}, \bibinfo {author} {\bibfnamefont
  {H.}~\bibnamefont {Agarwal}}, \bibinfo {author} {\bibfnamefont
  {L.}~\bibnamefont {Noohinejad}}, \bibinfo {author} {\bibfnamefont
  {M.}~\bibnamefont {Tolkiehn}}, \bibinfo {author} {\bibfnamefont
  {D.}~\bibnamefont {Bansal}}, \bibinfo {author} {\bibfnamefont
  {S.}~\bibnamefont {van Smaalen}},\ and\ \bibinfo {author} {\bibfnamefont
  {T.}~\bibnamefont {Arumugam}},\ }\bibfield  {title} {\bibinfo {title} {Room
  temperature charge density wave in a tetragonal polymorph of
  {Gd$_2$Os$_3$Si$_5$} and study of its origin in the {RE$_2$T$_3$X$_5$} {(RE =
  Rare Earth, T = Transition Metal, X = Si, Ge) Series}},\ }\href@noop {}
  {\bibfield  {journal} {\bibinfo  {journal} {Chemistry of Materials}\ }\textbf
  {\bibinfo {volume} {36}},\ \bibinfo {pages} {6888} (\bibinfo {year}
  {2024})}\BibitemShut {NoStop}%
\bibitem [{\citenamefont {Yamamoto}\ \emph {et~al.}(1998)\citenamefont
  {Yamamoto}, \citenamefont {Katsufuji}, \citenamefont {Tanabe},\ and\
  \citenamefont {Tokura}}]{yamamoto_raman_1998}%
  \BibitemOpen
  \bibfield  {author} {\bibinfo {author} {\bibfnamefont {K.}~\bibnamefont
  {Yamamoto}}, \bibinfo {author} {\bibfnamefont {T.}~\bibnamefont {Katsufuji}},
  \bibinfo {author} {\bibfnamefont {T.}~\bibnamefont {Tanabe}},\ and\ \bibinfo
  {author} {\bibfnamefont {Y.}~\bibnamefont {Tokura}},\ }\bibfield  {title}
  {\bibinfo {title} {Raman scattering of the charge-spin stripes in
  {${\mathrm{La}}_{1.67}{\mathrm{Sr}}_{0.33}{\mathrm{NiO}}_{4}$}},\ }\href@noop
  {} {\bibfield  {journal} {\bibinfo  {journal} {Phys. Rev. Lett.}\ }\textbf
  {\bibinfo {volume} {80}},\ \bibinfo {pages} {1493} (\bibinfo {year}
  {1998})}\BibitemShut {NoStop}%
\bibitem [{\citenamefont {Blumberg}\ \emph {et~al.}(1998)\citenamefont
  {Blumberg}, \citenamefont {Klein},\ and\ \citenamefont
  {Cheong}}]{blumberg_charge_1998}%
  \BibitemOpen
  \bibfield  {author} {\bibinfo {author} {\bibfnamefont {G.}~\bibnamefont
  {Blumberg}}, \bibinfo {author} {\bibfnamefont {M.~V.}\ \bibnamefont
  {Klein}},\ and\ \bibinfo {author} {\bibfnamefont {S.-W.}\ \bibnamefont
  {Cheong}},\ }\bibfield  {title} {\bibinfo {title} {Charge and spin dynamics
  of an ordered stripe phase in
  {${\mathrm{La}}_{1\frac{2}{3}}{\mathrm{Sr}}_{\frac{1}{3}}{\mathrm{NiO}}_{4}$}
  investigated by raman spectroscopy},\ }\href@noop {} {\bibfield  {journal}
  {\bibinfo  {journal} {Phys. Rev. Lett.}\ }\textbf {\bibinfo {volume} {80}},\
  \bibinfo {pages} {564} (\bibinfo {year} {1998})}\BibitemShut {NoStop}%
\bibitem [{\citenamefont {Gim}\ \emph {et~al.}(2025)\citenamefont {Gim},
  \citenamefont {Park},\ and\ \citenamefont {Kim}}]{gim2025orbital}%
  \BibitemOpen
  \bibfield  {author} {\bibinfo {author} {\bibfnamefont {D.-H.}\ \bibnamefont
  {Gim}}, \bibinfo {author} {\bibfnamefont {C.~H.}\ \bibnamefont {Park}},\ and\
  \bibinfo {author} {\bibfnamefont {K.~H.}\ \bibnamefont {Kim}},\ }\bibfield
  {title} {\bibinfo {title} {Orbital-selective quasiparticle depletion across
  the density wave transition in trilayer nickelate {La$_4$Ni$_3$O$_{10}$}},\
  }\href@noop {} {\bibfield  {journal} {\bibinfo  {journal} {Physical Review
  Letters}\ }\textbf {\bibinfo {volume} {135}},\ \bibinfo {pages} {136505}
  (\bibinfo {year} {2025})}\BibitemShut {NoStop}%
\bibitem [{\citenamefont {Suthar}\ \emph {et~al.}(2025)\citenamefont {Suthar},
  \citenamefont {Sundaramurthy}, \citenamefont {Bejas}, \citenamefont {Le},
  \citenamefont {Puphal}, \citenamefont {Sosa-Lizama}, \citenamefont {Schulz},
  \citenamefont {Nuss}, \citenamefont {Isobe}, \citenamefont {van Aken} \emph
  {et~al.}}]{suthar2025multiorbital}%
  \BibitemOpen
  \bibfield  {author} {\bibinfo {author} {\bibfnamefont {A.}~\bibnamefont
  {Suthar}}, \bibinfo {author} {\bibfnamefont {V.}~\bibnamefont
  {Sundaramurthy}}, \bibinfo {author} {\bibfnamefont {M.}~\bibnamefont
  {Bejas}}, \bibinfo {author} {\bibfnamefont {C.}~\bibnamefont {Le}}, \bibinfo
  {author} {\bibfnamefont {P.}~\bibnamefont {Puphal}}, \bibinfo {author}
  {\bibfnamefont {P.}~\bibnamefont {Sosa-Lizama}}, \bibinfo {author}
  {\bibfnamefont {A.}~\bibnamefont {Schulz}}, \bibinfo {author} {\bibfnamefont
  {J.}~\bibnamefont {Nuss}}, \bibinfo {author} {\bibfnamefont {M.}~\bibnamefont
  {Isobe}}, \bibinfo {author} {\bibfnamefont {P.~A.}\ \bibnamefont {van Aken}},
  \emph {et~al.},\ }\bibfield  {title} {\bibinfo {title} {Multiorbital
  character of the density wave instability in {La$_4$Ni$_3$O$_{10}$}},\
  }\href@noop {} {\bibfield  {journal} {\bibinfo  {journal} {arXiv preprint
  arXiv:2508.06440}\ } (\bibinfo {year} {2025})}\BibitemShut {NoStop}%
\bibitem [{\citenamefont {Deswal}\ \emph {et~al.}(2025)\citenamefont {Deswal},
  \citenamefont {Kumar}, \citenamefont {Rout}, \citenamefont {Singh},\ and\
  \citenamefont {Kumar}}]{deswal2025dynamics}%
  \BibitemOpen
  \bibfield  {author} {\bibinfo {author} {\bibfnamefont {S.}~\bibnamefont
  {Deswal}}, \bibinfo {author} {\bibfnamefont {D.}~\bibnamefont {Kumar}},
  \bibinfo {author} {\bibfnamefont {D.}~\bibnamefont {Rout}}, \bibinfo {author}
  {\bibfnamefont {S.}~\bibnamefont {Singh}},\ and\ \bibinfo {author}
  {\bibfnamefont {P.}~\bibnamefont {Kumar}},\ }\bibfield  {title} {\bibinfo
  {title} {Dynamics of electron--electron correlation and electron--phonon
  coupled phase progression in trilayer nickelate {La$_4$Ni$_3$O$_{10}$}},\
  }\href@noop {} {\bibfield  {journal} {\bibinfo  {journal} {Applied Physics
  Letters}\ }\textbf {\bibinfo {volume} {127}} (\bibinfo {year}
  {2025})}\BibitemShut {NoStop}%
\bibitem [{\citenamefont {Li}\ \emph {et~al.}(2024)\citenamefont {Li},
  \citenamefont {Wang}, \citenamefont {Ma}, \citenamefont {Wang}, \citenamefont
  {Liu}, \citenamefont {Fan}, \citenamefont {Han}, \citenamefont {Wang},
  \citenamefont {Tao},\ and\ \citenamefont {Zhang}}]{Li2024a}%
  \BibitemOpen
  \bibfield  {author} {\bibinfo {author} {\bibfnamefont {F.}~\bibnamefont
  {Li}}, \bibinfo {author} {\bibfnamefont {S.}~\bibnamefont {Wang}}, \bibinfo
  {author} {\bibfnamefont {C.}~\bibnamefont {Ma}}, \bibinfo {author}
  {\bibfnamefont {X.}~\bibnamefont {Wang}}, \bibinfo {author} {\bibfnamefont
  {C.}~\bibnamefont {Liu}}, \bibinfo {author} {\bibfnamefont {C.}~\bibnamefont
  {Fan}}, \bibinfo {author} {\bibfnamefont {L.}~\bibnamefont {Han}}, \bibinfo
  {author} {\bibfnamefont {S.}~\bibnamefont {Wang}}, \bibinfo {author}
  {\bibfnamefont {X.}~\bibnamefont {Tao}},\ and\ \bibinfo {author}
  {\bibfnamefont {J.}~\bibnamefont {Zhang}},\ }\bibfield  {title} {\bibinfo
  {title} {Flux growth of trilayer {La$_4$Ni$_3$O$_{10}$} single crystals at
  ambient pressure},\ }\href@noop {} {\bibfield  {journal} {\bibinfo  {journal}
  {Crystal Growth \& Design}\ }\textbf {\bibinfo {volume} {24}},\ \bibinfo
  {pages} {347} (\bibinfo {year} {2024})}\BibitemShut {NoStop}%
\bibitem [{\citenamefont {Dyadkin}\ \emph {et~al.}(2016)\citenamefont
  {Dyadkin}, \citenamefont {Pattison}, \citenamefont {Dmitriev},\ and\
  \citenamefont {Chernyshov}}]{Dyadkin:ie5157}%
  \BibitemOpen
  \bibfield  {author} {\bibinfo {author} {\bibfnamefont {V.}~\bibnamefont
  {Dyadkin}}, \bibinfo {author} {\bibfnamefont {P.}~\bibnamefont {Pattison}},
  \bibinfo {author} {\bibfnamefont {V.}~\bibnamefont {Dmitriev}},\ and\
  \bibinfo {author} {\bibfnamefont {D.}~\bibnamefont {Chernyshov}},\ }\bibfield
   {title} {\bibinfo {title} {{A new multipurpose diffractometer
  PILATUS@SNBL}},\ }\href@noop {} {\bibfield  {journal} {\bibinfo  {journal}
  {Journal of Synchrotron Radiation}\ }\textbf {\bibinfo {volume} {23}},\
  \bibinfo {pages} {825} (\bibinfo {year} {2016})}\BibitemShut {NoStop}%
\bibitem [{\citenamefont {Marshall}\ \emph {et~al.}(2023)\citenamefont
  {Marshall}, \citenamefont {Emerich}, \citenamefont {McMonagle}, \citenamefont
  {Fuller}, \citenamefont {Dyadkin}, \citenamefont {Chernyshov},\ and\
  \citenamefont {van Beek}}]{Marshall:ok5082}%
  \BibitemOpen
  \bibfield  {author} {\bibinfo {author} {\bibfnamefont {K.~P.}\ \bibnamefont
  {Marshall}}, \bibinfo {author} {\bibfnamefont {H.}~\bibnamefont {Emerich}},
  \bibinfo {author} {\bibfnamefont {C.~J.}\ \bibnamefont {McMonagle}}, \bibinfo
  {author} {\bibfnamefont {C.~A.}\ \bibnamefont {Fuller}}, \bibinfo {author}
  {\bibfnamefont {V.}~\bibnamefont {Dyadkin}}, \bibinfo {author} {\bibfnamefont
  {D.}~\bibnamefont {Chernyshov}},\ and\ \bibinfo {author} {\bibfnamefont
  {W.}~\bibnamefont {van Beek}},\ }\bibfield  {title} {\bibinfo {title} {{A new
  high temperature, high heating rate, low axial gradient capillary heater}},\
  }\href@noop {} {\bibfield  {journal} {\bibinfo  {journal} {Journal of
  Synchrotron Radiation}\ }\textbf {\bibinfo {volume} {30}},\ \bibinfo {pages}
  {267} (\bibinfo {year} {2023})}\BibitemShut {NoStop}%
\bibitem [{\citenamefont
  {Rodríguez-Carvajal}(1993)}]{RODRIGUEZCARVAJAL199355}%
  \BibitemOpen
  \bibfield  {author} {\bibinfo {author} {\bibfnamefont {J.}~\bibnamefont
  {Rodríguez-Carvajal}},\ }\bibfield  {title} {\bibinfo {title} {Recent
  advances in magnetic structure determination by neutron powder diffraction},\
  }\href@noop {} {\bibfield  {journal} {\bibinfo  {journal} {Physica B:
  Condensed Matter}\ }\textbf {\bibinfo {volume} {192}},\ \bibinfo {pages} {55}
  (\bibinfo {year} {1993})}\BibitemShut {NoStop}%
\bibitem [{\citenamefont {Garbarino}\ \emph {et~al.}(2024)\citenamefont
  {Garbarino}, \citenamefont {Hanfland}, \citenamefont {Gallego-Parra},
  \citenamefont {Rosa}, \citenamefont {Mezouar}, \citenamefont {Duran},
  \citenamefont {Martel}, \citenamefont {Papillon}, \citenamefont {Roth},
  \citenamefont {Got},\ and\ \citenamefont {Jacobs}}]{Garbarino2024-ID15B}%
  \BibitemOpen
  \bibfield  {author} {\bibinfo {author} {\bibfnamefont {G.}~\bibnamefont
  {Garbarino}}, \bibinfo {author} {\bibfnamefont {M.~E.}\ \bibnamefont
  {Hanfland}}, \bibinfo {author} {\bibfnamefont {S.}~\bibnamefont
  {Gallego-Parra}}, \bibinfo {author} {\bibfnamefont {A.~D.}\ \bibnamefont
  {Rosa}}, \bibinfo {author} {\bibfnamefont {M.}~\bibnamefont {Mezouar}},
  \bibinfo {author} {\bibfnamefont {D.}~\bibnamefont {Duran}}, \bibinfo
  {author} {\bibfnamefont {K.}~\bibnamefont {Martel}}, \bibinfo {author}
  {\bibfnamefont {E.}~\bibnamefont {Papillon}}, \bibinfo {author}
  {\bibfnamefont {T.}~\bibnamefont {Roth}}, \bibinfo {author} {\bibfnamefont
  {P.}~\bibnamefont {Got}},\ and\ \bibinfo {author} {\bibfnamefont
  {J.}~\bibnamefont {Jacobs}},\ }\bibfield  {title} {\bibinfo {title} {{Extreme
  conditions X-ray diffraction and imaging beamline ID15B on the ESRF extremely
  brilliant source}},\ }\href {https://doi.org/10.1080/08957959.2024.2379294}
  {\bibfield  {journal} {\bibinfo  {journal} {High Press. Res.}\ }\textbf
  {\bibinfo {volume} {44}},\ \bibinfo {pages} {199} (\bibinfo {year}
  {2024})}\BibitemShut {NoStop}%
\bibitem [{\citenamefont {Shen}\ \emph {et~al.}(2020)\citenamefont {Shen},
  \citenamefont {Wang}, \citenamefont {Dewaele}, \citenamefont {Wu},
  \citenamefont {Fratanduono}, \citenamefont {Eggert}, \citenamefont {Klotz},
  \citenamefont {Dziubek}, \citenamefont {Loubeyre}, \citenamefont
  {Fat’yanov}, \citenamefont {Asimow}, \citenamefont {Mashimo},\ and\
  \citenamefont {Wentzcovitch}}]{Shen2020}%
  \BibitemOpen
  \bibfield  {author} {\bibinfo {author} {\bibfnamefont {G.}~\bibnamefont
  {Shen}}, \bibinfo {author} {\bibfnamefont {Y.}~\bibnamefont {Wang}}, \bibinfo
  {author} {\bibfnamefont {A.}~\bibnamefont {Dewaele}}, \bibinfo {author}
  {\bibfnamefont {C.}~\bibnamefont {Wu}}, \bibinfo {author} {\bibfnamefont
  {D.~E.}\ \bibnamefont {Fratanduono}}, \bibinfo {author} {\bibfnamefont
  {J.}~\bibnamefont {Eggert}}, \bibinfo {author} {\bibfnamefont
  {S.}~\bibnamefont {Klotz}}, \bibinfo {author} {\bibfnamefont {K.~F.}\
  \bibnamefont {Dziubek}}, \bibinfo {author} {\bibfnamefont {P.}~\bibnamefont
  {Loubeyre}}, \bibinfo {author} {\bibfnamefont {O.~V.}\ \bibnamefont
  {Fat’yanov}}, \bibinfo {author} {\bibfnamefont {P.~D.}\ \bibnamefont
  {Asimow}}, \bibinfo {author} {\bibfnamefont {T.}~\bibnamefont {Mashimo}},\
  and\ \bibinfo {author} {\bibfnamefont {R.~M.~M.}\ \bibnamefont
  {Wentzcovitch}},\ }\bibfield  {title} {\bibinfo {title} {Toward an
  international practical pressure scale: A proposal for an ipps ruby gauge
  {(IPPS-Ruby2020)}},\ }\href@noop {} {\bibfield  {journal} {\bibinfo
  {journal} {High Pressure Research}\ }\textbf {\bibinfo {volume} {40}},\
  \bibinfo {pages} {299} (\bibinfo {year} {2020})}\BibitemShut {NoStop}%
\bibitem [{\citenamefont {Takemura}(2001)}]{Takemura2001}%
  \BibitemOpen
  \bibfield  {author} {\bibinfo {author} {\bibfnamefont {K.}~\bibnamefont
  {Takemura}},\ }\bibfield  {title} {\bibinfo {title} {{Evaluation of the
  hydrostaticity of a helium-pressure medium with powder x-ray diffraction
  techniques}},\ }\href@noop {} {\bibfield  {journal} {\bibinfo  {journal}
  {Journal of Applied Physics}\ }\textbf {\bibinfo {volume} {89}},\ \bibinfo
  {pages} {662} (\bibinfo {year} {2001})}\BibitemShut {NoStop}%
\bibitem [{\citenamefont {Dewaele}\ and\ \citenamefont
  {Loubeyre}(2007)}]{Dewaele2007}%
  \BibitemOpen
  \bibfield  {author} {\bibinfo {author} {\bibfnamefont {A.}~\bibnamefont
  {Dewaele}}\ and\ \bibinfo {author} {\bibfnamefont {P.}~\bibnamefont
  {Loubeyre}},\ }\bibfield  {title} {\bibinfo {title} {Pressurizing conditions
  in helium-pressure-transmitting medium},\ }\href
  {https://doi.org/10.1080/08957950701659627} {\bibfield  {journal} {\bibinfo
  {journal} {High Pressure Research}\ }\textbf {\bibinfo {volume} {27}},\
  \bibinfo {pages} {419} (\bibinfo {year} {2007})}\BibitemShut {NoStop}%
\bibitem [{\citenamefont {Paulmann}(2022)}]{paulmann2022a}%
  \BibitemOpen
  \bibfield  {author} {\bibinfo {author} {\bibfnamefont {C.}~\bibnamefont
  {Paulmann}},\ }\href@noop {} {\emph {\bibinfo {title} {P24ToolsCP, v1.12,
  software tools for postprocessing area detector, x-ray diffraction data}}}\
  (\bibinfo {year} {2022})\BibitemShut {NoStop}%
\bibitem [{\citenamefont {Schreurs}\ \emph {et~al.}(2010)\citenamefont
  {Schreurs}, \citenamefont {Xian},\ and\ \citenamefont
  {Kroon-Batenburg}}]{schreursamm2010a}%
  \BibitemOpen
  \bibfield  {author} {\bibinfo {author} {\bibfnamefont {A.~M.~M.}\
  \bibnamefont {Schreurs}}, \bibinfo {author} {\bibfnamefont {X.}~\bibnamefont
  {Xian}},\ and\ \bibinfo {author} {\bibfnamefont {L.~M.~J.}\ \bibnamefont
  {Kroon-Batenburg}},\ }\bibfield  {title} {\bibinfo {title} {{EVAL15:} a
  diffraction data integration method based on ab initio predicted profiles},\
  }\href {https://doi.org/10.1107/S0021889809043234} {\bibfield  {journal}
  {\bibinfo  {journal} {J. Appl. Crystallogr.}\ }\textbf {\bibinfo {volume}
  {43}},\ \bibinfo {pages} {70} (\bibinfo {year} {2010})}\BibitemShut {NoStop}%
\bibitem [{\citenamefont {Duisenberg}\ \emph {et~al.}(2003)\citenamefont
  {Duisenberg}, \citenamefont {Kroon-Batenburg},\ and\ \citenamefont
  {Schreurs}}]{duisenberga2003a}%
  \BibitemOpen
  \bibfield  {author} {\bibinfo {author} {\bibfnamefont {A.~J.~M.}\
  \bibnamefont {Duisenberg}}, \bibinfo {author} {\bibfnamefont {L.~M.~J.}\
  \bibnamefont {Kroon-Batenburg}},\ and\ \bibinfo {author} {\bibfnamefont
  {A.~M.~M.}\ \bibnamefont {Schreurs}},\ }\bibfield  {title} {\bibinfo {title}
  {{An intensity evaluation method: {\it EVAL}-14}},\ }\href
  {https://doi.org/10.1107/S0021889802022628} {\bibfield  {journal} {\bibinfo
  {journal} {Journal of Applied Crystallography}\ }\textbf {\bibinfo {volume}
  {36}},\ \bibinfo {pages} {220} (\bibinfo {year} {2003})}\BibitemShut
  {NoStop}%
\bibitem [{\citenamefont {Sheldrick}(2008)}]{sheldrick2008}%
  \BibitemOpen
  \bibfield  {author} {\bibinfo {author} {\bibfnamefont {G.~M.}\ \bibnamefont
  {Sheldrick}},\ }\href@noop {} {\emph {\bibinfo {title} {{SADABS,} Version
  2008/1}}}\ (\bibinfo  {publisher} {G\"{o}ttingen: University of
  G\"{o}ttingen},\ \bibinfo {year} {2008})\BibitemShut {NoStop}%
\bibitem [{\citenamefont {Zeman}\ \emph {et~al.}(2025)\citenamefont {Zeman},
  \citenamefont {Lefevre}, \citenamefont {Dang}, \citenamefont {Trollier},
  \citenamefont {Camus},\ and\ \citenamefont
  {Chaneli{\`e}re}}]{zeman2025design}%
  \BibitemOpen
  \bibfield  {author} {\bibinfo {author} {\bibfnamefont {M.}~\bibnamefont
  {Zeman}}, \bibinfo {author} {\bibfnamefont {B.}~\bibnamefont {Lefevre}},
  \bibinfo {author} {\bibfnamefont {S.}~\bibnamefont {Dang}}, \bibinfo {author}
  {\bibfnamefont {T.}~\bibnamefont {Trollier}}, \bibinfo {author}
  {\bibfnamefont {P.}~\bibnamefont {Camus}},\ and\ \bibinfo {author}
  {\bibfnamefont {T.}~\bibnamefont {Chaneli{\`e}re}},\ }\bibfield  {title}
  {\bibinfo {title} {Design and performances of a dry table-top optical
  cryostat at 2{K}},\ }\href@noop {} {\bibfield  {journal} {\bibinfo  {journal}
  {IOP Conference Series: Materials Science and Engineering}\ }\textbf
  {\bibinfo {volume} {1327}},\ \bibinfo {pages} {012153} (\bibinfo {year}
  {2025})}\BibitemShut {NoStop}%
\end{thebibliography}
\end{document}